\documentclass{emulateapj}
\usepackage{lscape}

\newcommand{\Ifg}{\ensuremath{I_\mathrm{fg}}}
\newcommand{\Ihalo}{\ensuremath{I_\mathrm{halo}}}

\newcommand{\Iobs}{\ensuremath{I_\mathrm{obs}}}
\newcommand{\Kalpha}{K$\alpha$}
\newcommand{\Kbeta}{K$\beta$}

\newcommand{\Lyalpha}{Ly$\alpha$}

\newcommand{\NH}{\ensuremath{N_{\mathrm{H}}}}

\newcommand{\RE}{\ensuremath{R_\mathrm{E}}}

\newcommand{\nm}{\ensuremath{\mbox{\nm}}}
\newcommand{\cm}{\ensuremath{\mbox{cm}}}

\newcommand{\km}{\ensuremath{\mbox{km}}}
\newcommand{\AU}{\ensuremath{\mbox{AU}}}
\newcommand{\pc}{\ensuremath{\mbox{pc}}}

\newcommand{\s}{\ensuremath{\mbox{s}}}

\newcommand{\ev}{\ensuremath{\mbox{eV}}}
\newcommand{\kev}{\ensuremath{\mbox{keV}}}

\newcommand{\erg}{\ensuremath{\mbox{erg}}}

\newcommand{\sr}{\ensuremath{\mbox{sr}}}

\newcommand{\K}{\ensuremath{\mbox{K}}}

\newcommand{\ph}{\ensuremath{\mbox{photons}}}

\newcommand{\cmsq}{\ensuremath{\cm^2}}

\newcommand{\pcc}{\ensuremath{\cm^{-3}}}
\newcommand{\pcmsq}{\ensuremath{\cm^{-2}}}

\newcommand{\ps}{\ensuremath{\s^{-1}}}
\newcommand{\psr}{\ensuremath{\sr^{-1}}}

\newcommand{\flux}{\erg\ \pcmsq\ \ps}
\newcommand{\lineunit}{\ph\ \pcmsq\ \ps\ \psr}

\newcommand{\LU}{\ensuremath{\mbox{L.U.}}}
\newcommand{\kmps}{\km\ \ps}

\newcommand{\pownorm}{\ph\ \pcmsq\ \ps\ \psr\ \ensuremath{\kev^{-1}}}

\newcommand{\HI}{H~\textsc{i}}

\newcommand{\OVII}{O~\textsc{vii}}
\newcommand{\OVIII}{O~\textsc{viii}}

\newcommand{\Iovii}{\ensuremath{I_\mathrm{O\,VII}}}
\newcommand{\Eovii}{\ensuremath{E_\mathrm{O\,VII}}}
\newcommand{\Ioviii}{\ensuremath{I_\mathrm{O\,VIII}}}
\newcommand{\Eoviii}{\ensuremath{E_\mathrm{O\,VIII}}}

\newcommand{\ace}{\textit{ACE}}
\newcommand{\asca}{\textit{ASCA}}

\newcommand{\chandra}{\textit{Chandra}}

\newcommand{\rosat}{\textit{ROSAT}}

\newcommand{\suzaku}{\textit{Suzaku}}

\newcommand{\wind}{\textit{Wind}}
\newcommand{\xmm}{\textit{XMM-Newton}}

\newcommand{\e}{\ensuremath{\mathrm{e}}}
\newcommand{\chisq}{\ensuremath{\chi^2}}

\newcommand{\raymondsmith}{\citeauthor{raymond77} (\citeyear{raymond77} and updates)}
\newcommand{\eqref}[1]{equation~(\ref{#1})}

\newcommand{\cosec}{\ensuremath{\mathrm{cosec}}}
\newcommand{\fsw}{\ensuremath{f_\mathrm{sw}}}
\newcommand{\Ftotal}{\ensuremath{F_\mathrm{total}^{2-5}}}
\newcommand{\Fexgal}{\ensuremath{F_\mathrm{exgal}^{2-5}}}
\newcommand{\Fsource}{\ensuremath{F_\mathrm{X}^{0.5-2.0}}}
\newcommand{\thetamin}{\ensuremath{\theta_\mathrm{min}}}

\shorttitle{\textit{XMM-NEWTON} SURVEY OF SOFT X-RAY BACKGROUND}
\shortauthors{HENLEY AND SHELTON}

\begin{document}

\title{An \textit{XMM-Newton} Survey of the Soft X-ray Background. I.
       The \OVII\ and \OVIII\ Lines Between $\lowercase{l}=120\degr$ and $\lowercase{l}=240\degr$}
\author{David B. Henley and Robin L. Shelton}
\affil{Department of Physics and Astronomy, University of Georgia, Athens, GA 30602}
\email{dbh@physast.uga.edu}

\begin{abstract}
We present measurements of the soft X-ray background (SXRB) \OVII\ and \OVIII\ intensity between
$l = 120\degr$ and $l = 240\degr$, the first results of a survey of the SXRB using archival \xmm\
observations. We do not restrict ourselves to blank-sky observations, but instead
use as many observations as possible, removing bright or extended sources by hand if necessary. In
an attempt to minimize contamination from near-Earth solar wind charge exchange (SWCX) emission, we
remove times of high solar wind proton flux from the data. Without this filtering we are able to extract
measurements from 586 \xmm\ observations. With this filtering, $\sim$1/2 of the observations are rendered
unusable, and we are able to extract measurements from 303 observations.
The oxygen intensities are typically $\sim$0.5--10~\lineunit\ (line units, \LU) for \OVII\ and
$\sim$0--5~\LU\ for \OVIII. The proton flux filtering does not systematically reduce the oxygen
intensities measured from a given observation.  However, the filtering does preferentially remove
the observations with higher oxygen intensities.
Our dataset includes 69 directions with multiple observations, whose oxygen intensity variations
can be used to constrain SWCX models. One observation exhibits an \OVII\ enhancement of
$\sim$25~\LU\ over 2 other observations of the same direction, although most SWCX enhancements are
$\la$4~\LU\ for \OVII\ and $\la$2~\LU\ for \OVIII. We find no clear tendency for the \OVII\ centroid to
shift toward the forbidden line energy in observations with bright SWCX enhancements. There is also
no universal association between enhanced SWCX emission and increased solar wind flux or the closeness of
the sightline to the sub-solar region of the magnetosheath.
After removing observations likely to be contaminated by heliospheric SWCX emission, we use our
results to examine the Galactic halo. There is some scatter in the halo intensity about the
predictions of a simple plane-parallel model, indicating a patchiness to the halo emission. The
\OVII/\OVIII\ intensity ratio implies a halo temperature of $\sim$2.0--$2.5 \times 10^6$~\K,
in good agreement with previous studies.
\end{abstract}


\section{INTRODUCTION}
\label{sec:Introduction}

The soft X-ray background (SXRB) below 1~\kev\ is dominated by line emission from within the Galaxy
\citep[e.g.,][]{sanders01,mccammon02}. For many years, this emission was thought to be produced by
$\sim$million-degree gas in the interstellar medium (ISM), including unabsorbed emission from the Local
Bubble (LB, a cavity in the local ISM of radius $\sim$100~\pc\ in which the Solar System resides,
thought to be filled with $\sim$$1 \times 10^6~\K$ plasma), and emission from $\sim$1--$3 \times
10^6~\K$ plasma in the Galactic halo, which is attenuated by the Galaxy's \HI\ \citep[e.g.,][and
references therein]{kuntz00}.  However, in recent years it has become apparent that X-ray line
emission can also originate within the Solar System, from charge exchange reactions between highly
ionized metals in the solar wind and neutral hydrogen and helium atoms in the heliosphere or in the
outer reaches of the Earth's atmosphere
\citep[e.g.,][]{cravens00,robertson03a,robertson03b,koutroumpa06}. From the point of view of someone
interested in studying the Galaxy's hot ISM, this solar wind charge exchange (SWCX) emission is a
time-varying contaminant of the soft X-ray emission.

Our current picture of the Galaxy's hot ISM is largely derived from maps of the SXRB obtained with
rocket-borne and satellite-borne proportional counters
\citep{mccammon83,marshall84,garmire92,snowden97}, which are presented in a few broad bands between
$\sim$0.1 and a few \kev. Higher-resolution spectra of the SXRB have been obtained with the CCD
cameras on board \chandra, \xmm, and \suzaku\
\citep{snowden04,smith05,smith07a,fujimoto07,galeazzi07,henley07,henley08a,kuntz08a,kuntz08b,masui09,yao09,lei09,yoshino09,gupta09b}.
CCD spectrometers can resolve some emission features in the SXRB spectrum, allowing the temperature
of the X-ray--emitting plasma to be measured more accurately; it may also be possible to measure the
ionization state and relative abundances of the plasma. CCD-resolution spectra help us address questions of
the origin and evolution of the hot Galactic gas, such as the possible contributions made by infall
and supernova explosions to the hot gas content of the halo \citep[e.g.,][]{lei09,henley09}.
We can also measure SWCX emission spectra, for comparison with SWCX models.

Currently, CCD-resolution spectra of the SXRB have been presented for only a few tens of directions.
The aim of the current project is to obtain CCD-resolution spectra for a large number ($\sim$1000)
of directions, using archival observations obtained with the EPIC cameras on
board \xmm\ \citep{jansen01}. An important innovation is that we do not concentrate only on
observations of blank-sky fields. If a target object does not take up too much of the field of view,
we can remove the region immediately surrounding the target from the dataset and extract a SXRB
spectrum from the periphery of the field of view. This technique
greatly increases the number of observations that we can use. The ultimate goal of this project is
to improve our global picture of the hot gas in the Galaxy, using higher--spectral-resolution data
than is available from existing all-sky datasets. More immediately, our survey includes many
directions that have been observed multiple times over spans of time from days to years.
Multiple observations of the same direction are useful because the differences between the spectra
taken at different times can be used to constrain models of SWCX emission. Such models are essential
for obtaining an accurate picture of the hot ISM.

In this paper, we present the first results from this survey. We present intensities of the
\OVII\ and \OVIII\ lines at 0.57 and 0.65~\kev\ obtained from 590 \xmm\ observations from a third of the sky,
between $l = 120\degr$ and $l = 240\degr$. These lines dominate the Galactic SXRB emission in the
energy range covered by \xmm\ \citep{mccammon02}, and are important for a number of reasons. If SWCX
contamination is reduced or removed (for example, by applying the selection criteria
described in Section~\ref{subsec:ReducingSWCXContamination}), the variation in these lines' intensities across the sky
provides information on the distribution of the hot ISM, while the intensity ratio provides
temperature information. In addition, the line intensities can be used in joint emission/absorption
analyses with \chandra\ grating measurements \citep[e.g.,][]{yao09} to constrain the electron density
of the hot gas.  Future papers will expand the survey to the whole sky, and include additional
spectral information (such as the results of fitting thermal plasma models to the spectra).

The remainder of this paper is arranged as follows. In Section~\ref{sec:SWCX} we briefly review the
properties of SWCX emission, and discuss methods for helping reduce the SWCX contamination. In
Section~\ref{sec:DataReduction} we describe our observation selection and data reduction. In
Section~\ref{sec:OxygenIntensities} we describe our method for measuring the \OVII\ and
\OVIII\ intensities (Section~\ref{subsec:OxygenIntensitiesMethod}), present the results
(Section~\ref{subsec:OxygenIntensitiesResults}), and also present results for directions that have been
observed multiple times (Section~\ref{subsec:ResultsMultiple}).

We discuss our measurements in Section~\ref{sec:Discussion}. In
Section~\ref{subsec:SystematicErrors} we discuss various systematic errors which could bias our
intensity measurements. In Section~\ref{subsec:DiscussionMultiple} we discuss the results from
directions with multiple observations, and the implications of these results for SWCX. In
Section~\ref{subsec:Halo} we examine the oxygen emission from the halo. To do this, we first apply
various filters to reduce the SWCX contamination, and model the remaining foreground emission using
\rosat\ shadowing data. We compare the latitude-dependence of the halo emission with the predictions
of a simple plane-parallel model, and also look at the \OVII/\OVIII\ ratio.  We conclude with a
summary in Section~\ref{sec:Summary}.

\section{SOLAR WIND CHARGE EXCHANGE EMISSION}
\label{sec:SWCX}

\subsection{Summary of Properties of SWCX}

As was noted in the Introduction, observations of the diffuse soft X-ray emission from
$\sim$1--$3 \times 10^6~\K$ gas in the Galaxy can be contaminated by SWCX emission. This emission is
from two sources within the solar system. Firstly, geocoronal SWCX reactions occur between solar
wind ions and neutral hydrogen atoms in the outer reaches of the Earth's atmosphere. For example,
\OVII\ emission is produced by the following charge exchange reaction:
\begin{eqnarray}
	\mathrm{O}^{+7} + \mathrm{H} & \rightarrow & \mathrm{O}^{+6\ast} + \mathrm{H}^+	\nonumber \\
                                     & \rightarrow & \mathrm{O}^{+6}     + \mathrm{H}^+ + h\nu,
\end{eqnarray}
where the $\ast$ indicates that the ion is in an excited state.  This emission is produced mainly in
the magnetosheath, between the magnetopause and the bowshock, and is brightest in the region between
the Earth and the Sun (the sub-solar region; \citealp{robertson03b}).  Secondly, heliospheric SWCX
reactions occur throughout the heliosphere between solar wind ions and neutral hydrogen and helium
atoms that have entered the solar system from the ISM \citep{cravens00}.

Enhancements in the geocoronal emission and/or the near-Earth heliospheric emission on a timescale
of $\sim$hours--days have been observed with \rosat\ (the so-called ``long-term enhancements'' in
the \rosat\ All-Sky Survey; \citealp{snowden95}), and with \xmm\ and \suzaku\
\citep{snowden04,fujimoto07,carter08}. These bursts of enhanced SWCX emission are often, but not always,
associated with times of increased solar wind proton flux \citep{cravens01,snowden04,fujimoto07,carter08,kuntz08a}.

The heliospheric SWCX emission is also expected to vary, but much more slowly, because its variation
is due to the 11-year cycle of the Sun from solar minimum to solar maximum back to solar
minimum.\footnote{Note that this is half of the full 22-year solar cycle.}
The variation of the heliospheric SWCX emission with time is particularly strong at high ecliptic latitudes
\citep{robertson03a,koutroumpa06}. This is due to a variation in the ionization state of the solar wind at
high latitudes -- at solar minimum there are fewer of the O$^\mathrm{+7}$ and O$^\mathrm{+8}$ ions
that produce \OVII\ and \OVIII\ SWCX emission than there are at solar maximum. As a result, the
heliospheric \OVII\ and \OVIII\ emission is fainter at high ecliptic latitudes at solar minimum than
at solar maximum.

\subsection{Reducing SWCX Contamination}
\label{subsec:ReducingSWCXContamination}

Our current knowledge of geocoronal and heliospheric emission, briefly summarized above, suggests
methods that can be used to reduce SWCX contamination. The association between bursts of enhanced
geocoronal and/or near-Earth heliospheric SWCX emission and increased solar wind proton flux
suggests that removing times of high proton flux from one's X-ray data will help reduce SWCX
contamination. As we will describe in Section~\ref{subsec:ProtonFluxFiltering} below, we use such
filtering in our data reduction. However, it is important to note that this method will only help
eliminate enhancements in the SWCX emission produced near the Earth -- it will not eliminate the
quiescent geocoronal emission, nor will it eliminate heliospheric emission produced away from the
Earth.

\citet{carter08} have suggested using the time-variation in the 0.5--0.7~\kev\ band (the ``line
band'', which is dominated by \OVII\ and \OVIII\ emission) relative to the time-variation in a
continuum band as a way of identifying \xmm\ observations that are affected by SWCX
emission. Short-term variations in the line band that are uncorrelated with continuum-band
variations are indicative of SWCX contamination (the degree of correlation between the line-band and
continuum-bands is quantified by two parameters, $\chi^2_\mu$ and $R_\chi$). However, their method
is only sensitive to SWCX emission that varies during the course of an \xmm\ observation, and so it
too only deals with time-varying geocoronal emission and/or near-Earth heliospheric emission. In
addition, \citet{carter08} do not quote thresholds for their $\chi^2_\mu$ and $R_\chi$ parameters
for determining whether or not an observation is likely to be contaminated, although they intend to
address this issue in a future paper. We have therefore not used their method in the current paper,
but we intend to incorporate it into future extensions to this survey.

The above-described methods only help with geocoronal and near-Earth heliospheric emission. To
reduce the heliospheric emission as whole, we make use of its variation with the solar cycle.  In
particular, we can expect to reduce the heliospheric SWCX contamination by removing observations of
low ecliptic latitudes and observations taken during high solar activity. We will do this in
Section~\ref{subsec:Halo}, where we use our oxygen line measurements to study the Galactic halo.

\section{DATA REDUCTION}
\label{sec:DataReduction}

\subsection{Observation Selection and Initial Data Processing}
\label{subsec:InitialProcessing}

We began by selecting all \xmm\ observations between $l = 120\degr$ and $l = 240\degr$ that were publicly available
as of 2008 May 18 and that had at least some exposure with the EPIC-MOS cameras \citep{turner01}. This is a total of 1422 observations.
The data were downloaded from HEASARC.\footnote{ftp://legacy.gsfc.nasa.gov/xmm/data/rev1/}
We processed the data using SAS version 7.0.0\footnote{http://xmm2.esac.esa.int/sas/7.0.0/} and the
\xmm\ Extended Source Analysis Software\footnote{http://heasarc.gsfc.nasa.gov/docs/xmm/xmmhp\_xmmesas.html} (XMM-ESAS)
version 2\footnote{ftp://legacy.gsfc.nasa.gov/xmm/software/xmm-esas/xmm-esas-v2/} \citep{snowden07,kuntz08a}.
We used only the MOS data as the version of XMM-ESAS that we used cannot calculate the particle background for data from
the EPIC-pn camera \citep{struder01}. We intend to use EPIC-pn data in future versions of this catalog.

We initially processed and filtered each dataset using the XMM-ESAS \texttt{mos-filter} script. This
script first runs the SAS \texttt{emchain} script, which produces a calibrated events list for each
MOS exposure. These events lists were then further cleaned using the XMM-ESAS \texttt{clean-rel}
program, which identifies and removes times affected by soft-proton flaring.  For each events list,
a 2.5--12~\kev\ lightcurve was extracted from the whole field of view using 1-second bins, and a
histogram of count-rates was created. For an observation not badly affected by soft-proton flares,
this histogram should have an approximately Gaussian peak at the nominal count-rate. A Gaussian was
fitted to the peak, and all times whose count-rates differed from the mean of this Gaussian by more
than $1.5\sigma$ were removed from the data. The good time intervals resulting from this lightcurve
analysis were used to produce cleaned events lists.

\begin{figure}
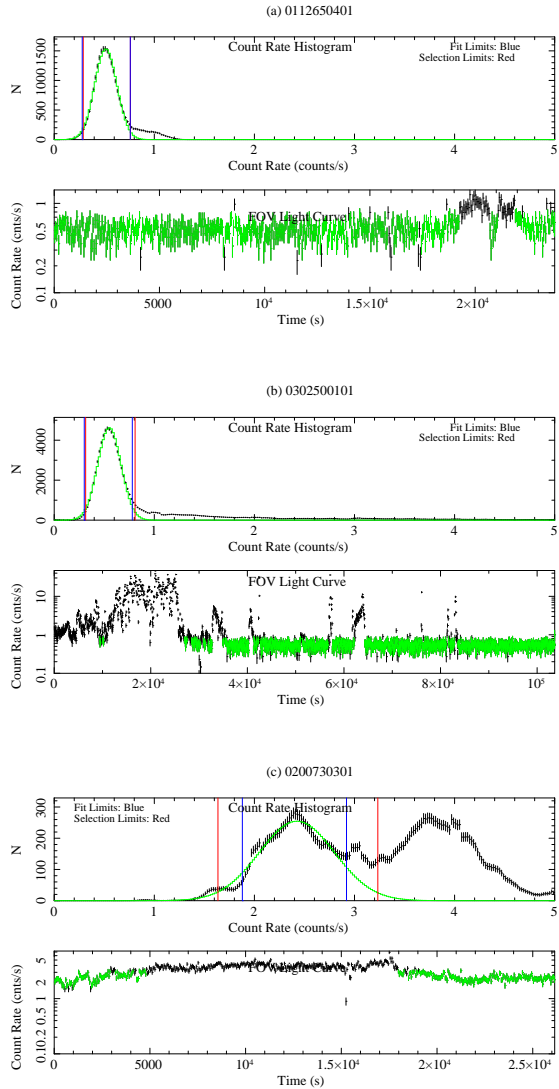

\centering
\includegraphics[angle=270,width=0.85\linewidth,clip]{f1a.eps} \\
\vspace{5mm}
\includegraphics[angle=270,width=0.85\linewidth,clip]{f1b.eps} \\
\vspace{5mm}
\includegraphics[angle=270,width=0.85\linewidth,clip]{f1c.eps} \\
\caption{Example MOS1 2.5--12~\kev\ count-rate histograms and lightcurves, illustrating different levels
of contamination by soft protons and the removal of contaminated portions of the data.
In the histogram panels, the black points show the data, and the green curve is the Gaussian that
was fitted to the peak (between the vertical blue lines). The vertical red lines show the mean of
the fitted Gaussian $\pm 1.5\sigma$. Times with count-rates outside that range were rejected. In the
lightcurve panels, the entire lightcurve is plotted in black, and the lightcurve for the accepted
times is overplotted in green.
(a) Obs.~0112650401. This is an example of an observation suffering from little or no flaring. Note that the
right vertical red line is obscured by the right vertical blue line.
(b) Obs.~0302500101. This is an example of an observation exhibiting several large flares, but which nevertheless yields a usable amount of good data.
(c) Obs.~0200730301. This is an example of an observation badly affected by flares -- such observations were rejected.
\label{fig:MOSlightcurves}}
\end{figure}

We inspected the lightcurve plots produced by \texttt{mos-filter}, in order to determine
whether or not each observation was badly contaminated by soft proton
flares. Figure~\ref{fig:MOSlightcurves} shows examples of the lightcurves and count-rate
histograms. Figure~\ref{fig:MOSlightcurves}(a) illustrates an observation suffering from little or
no flaring.  The observation shown in Figure~\ref{fig:MOSlightcurves}(b) suffers from a number of
bright flares.  However, after the removal of such flares, enough good time remains to yield a good
quality SXRB spectrum.  Figure~\ref{fig:MOSlightcurves}(c) illustrates an observation so badly
affected by flaring that it is unusable.

Our basic criterion for accepting an observation was that it had to have at least 1 MOS1 exposure
and 1 MOS2 exposure each with at least 5~ks of good time. Some observations had multiple exposures
from the MOS1 and/or MOS2 cameras -- we kept all exposures that had at least 5~ks of good time.  For
some observations that were badly contaminated by soft protons, the above filtering returned more than
5~ks of good time. Figure~\ref{fig:MOSlightcurves}(c) shows an example of this --
\texttt{mos-filter} identified 12~ks of good time for this MOS1 exposure. In most cases, it is clear from
a visual inspection of the count-rate histogram that the observation is contaminated, and such
observations were rejected. However, for some observations, some soft proton contamination remained in the
spectrum despite our filtering. We dealt with this by including an extra model component in our
spectral analysis to model this contamination (see Section~\ref{subsec:ParticleBackground}).

\begin{figure*}
\centering
\plottwo{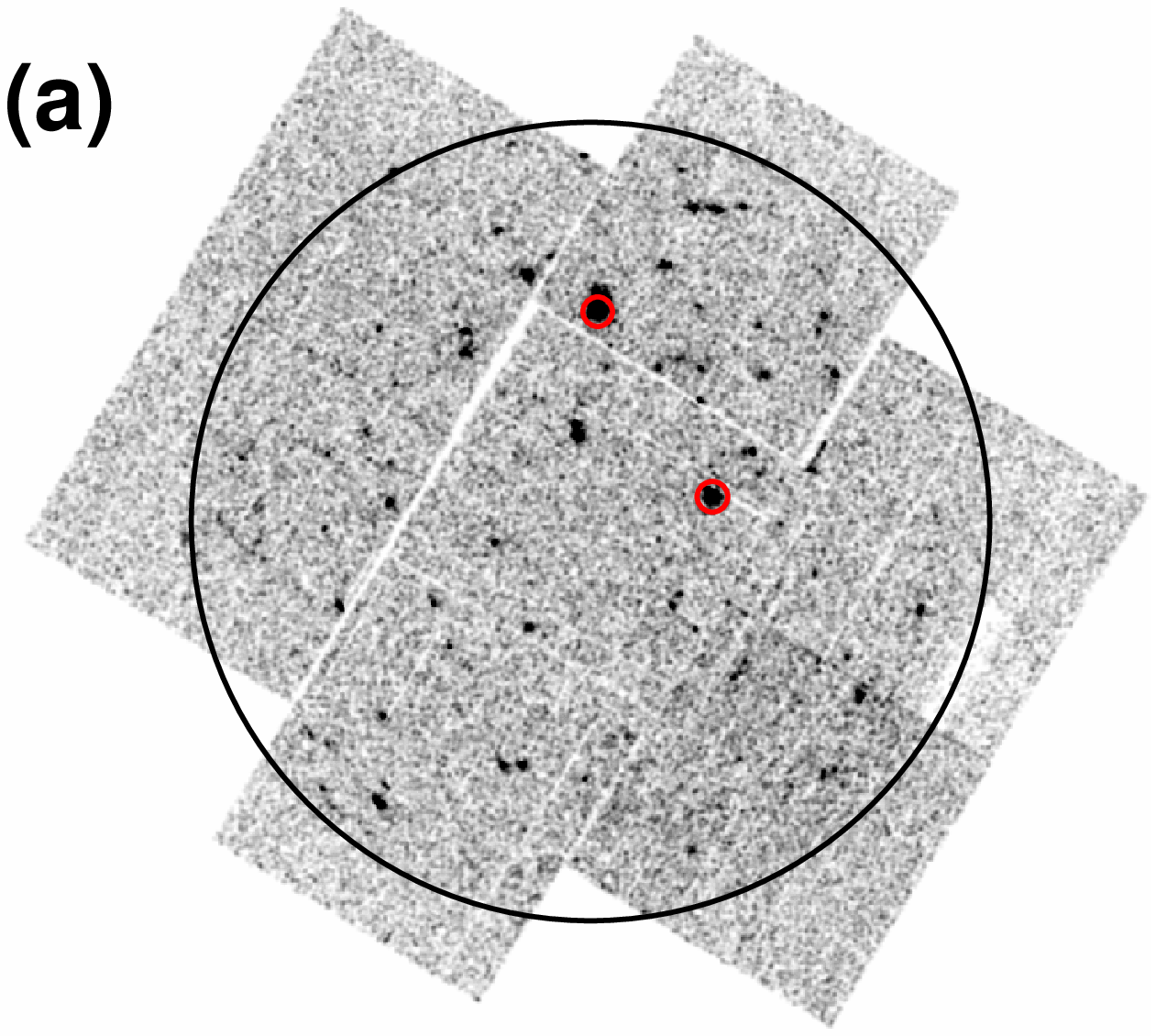}{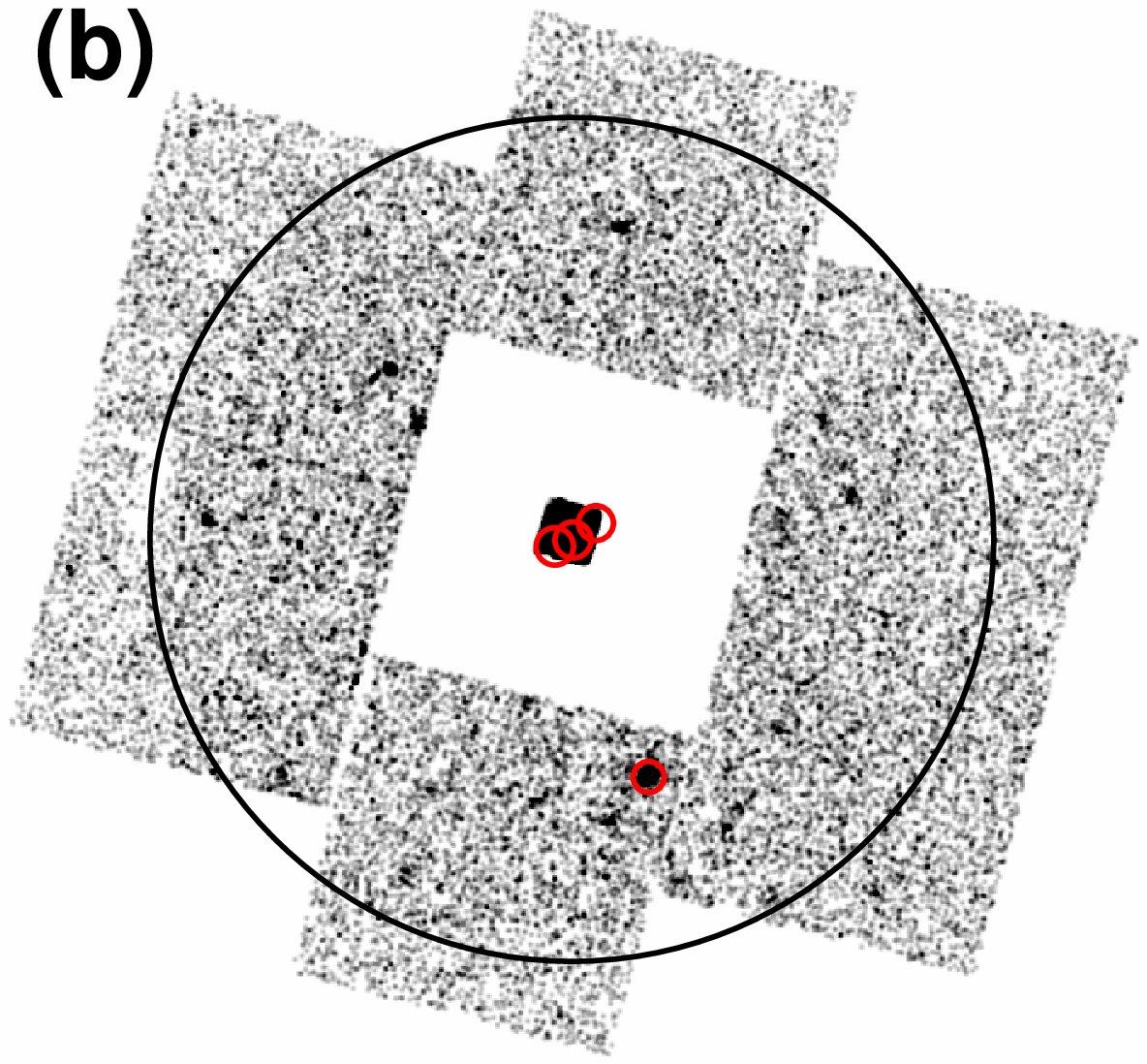}
\plottwo{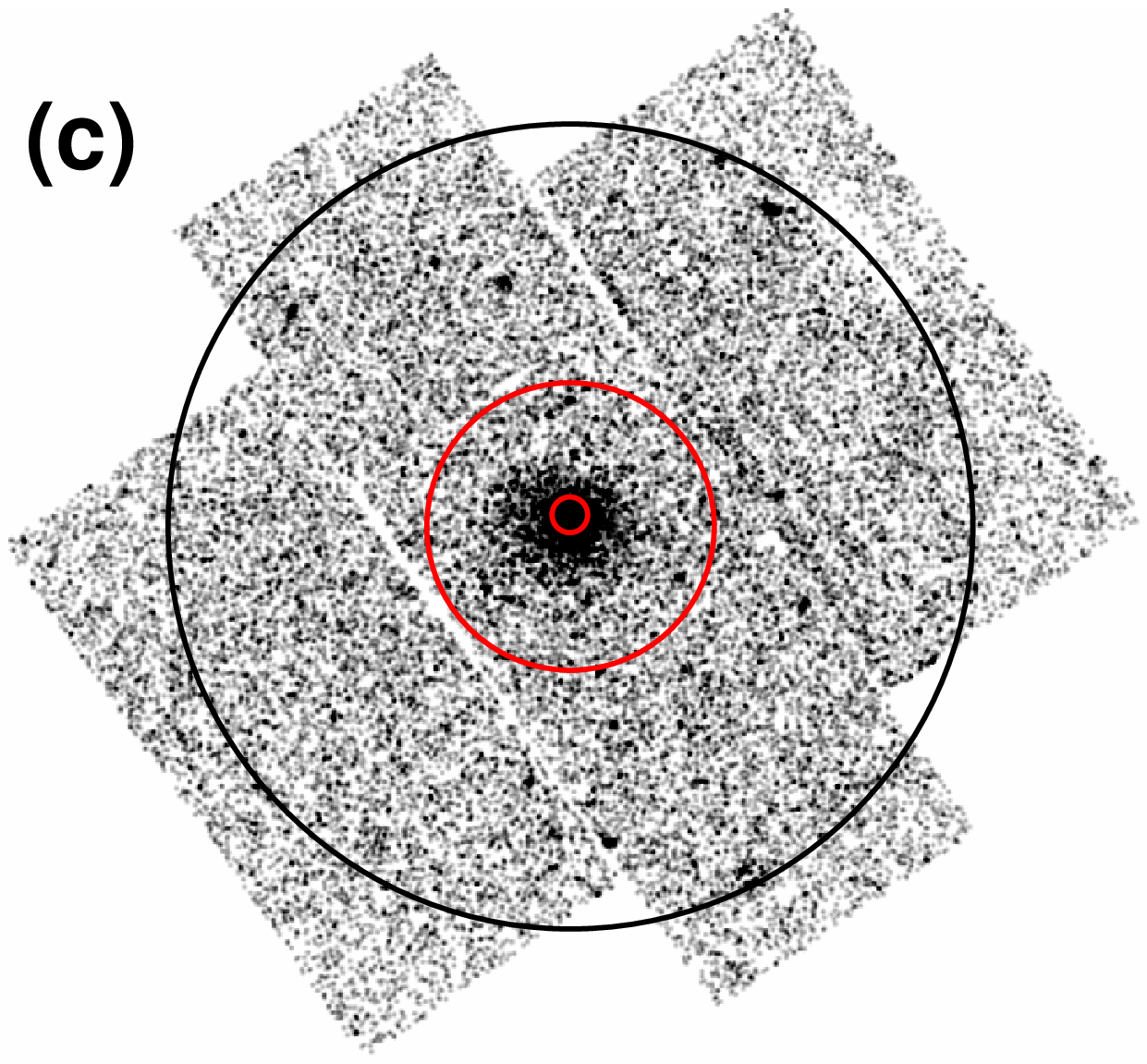}{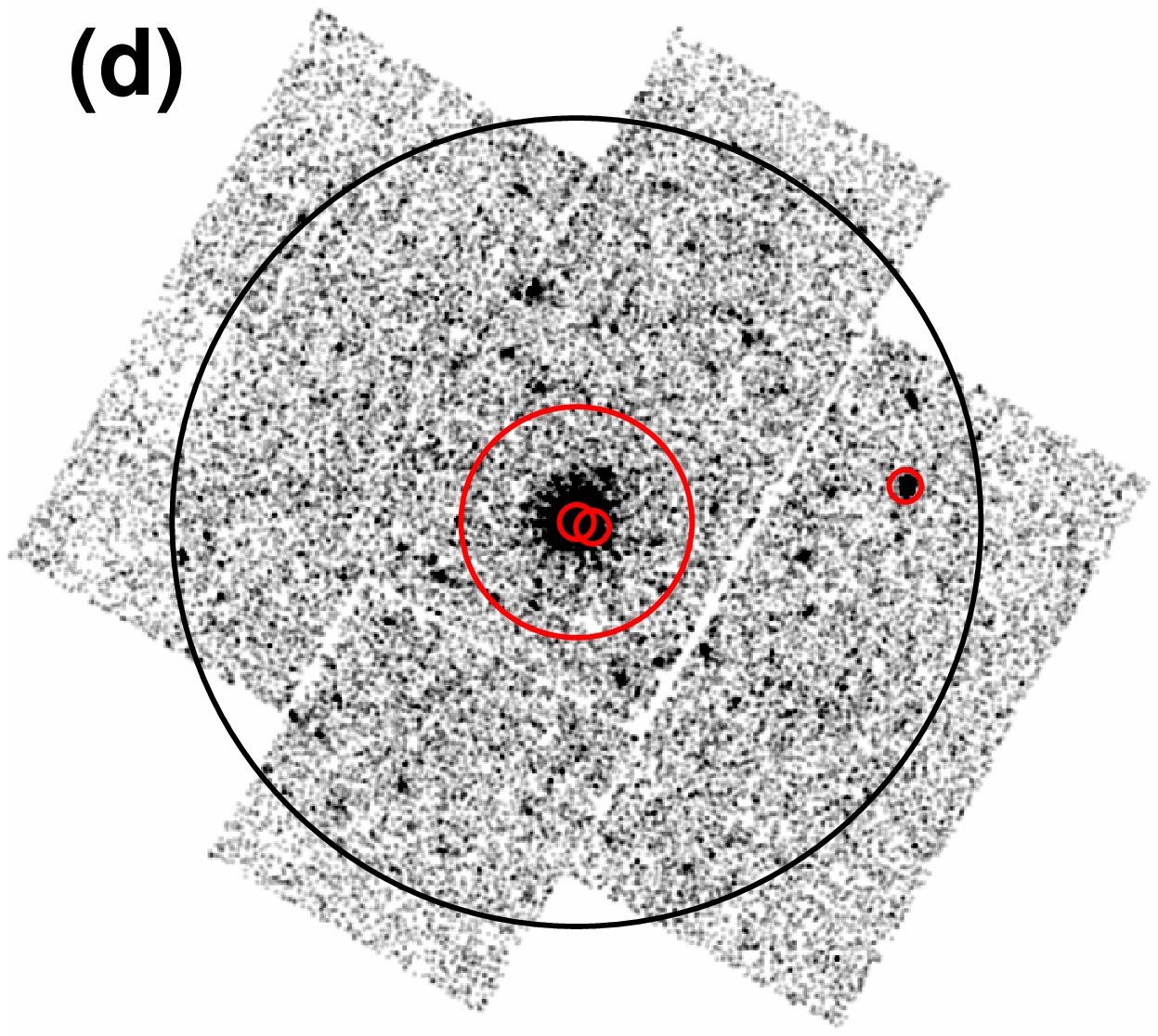}
\plottwo{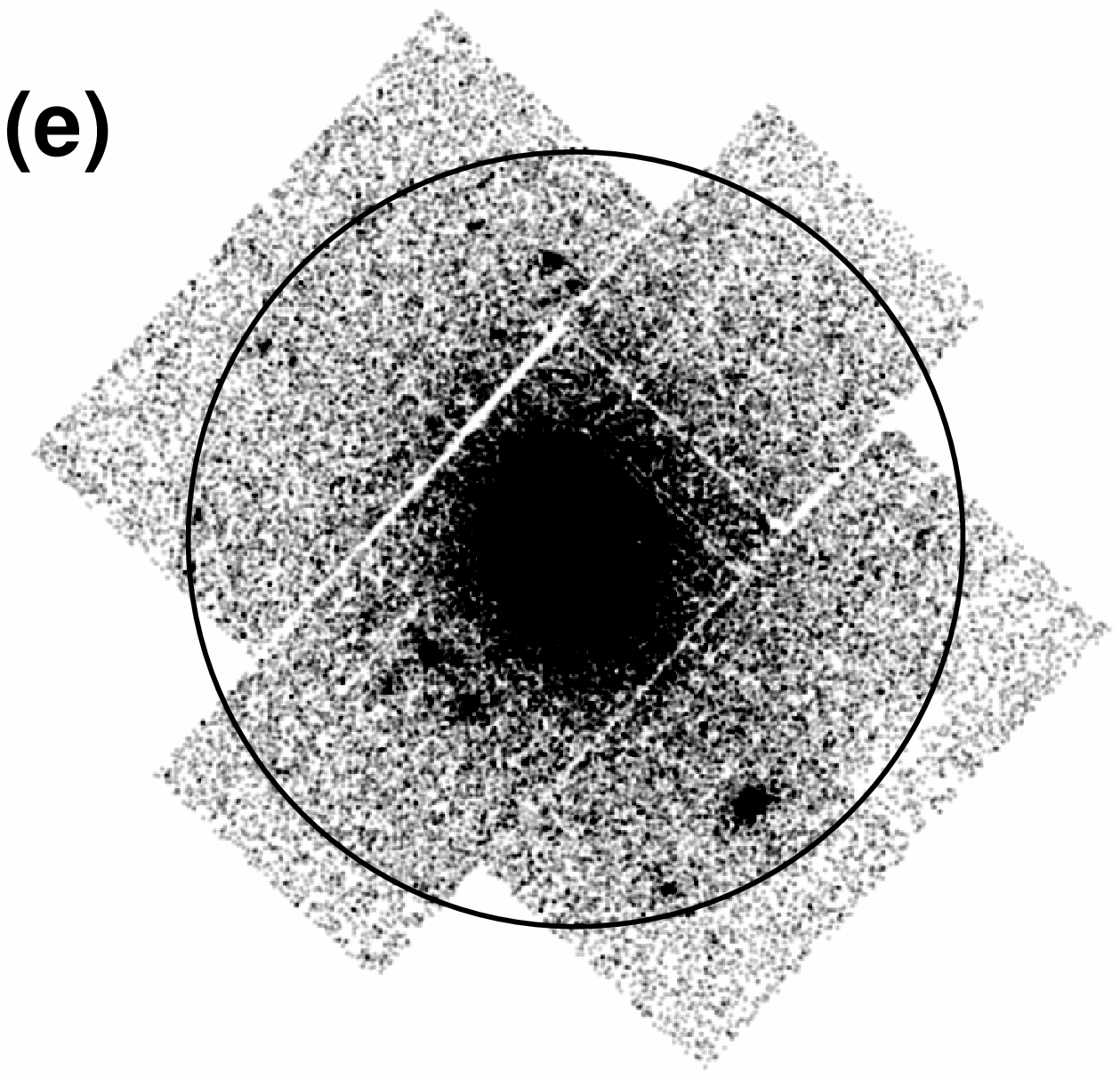}{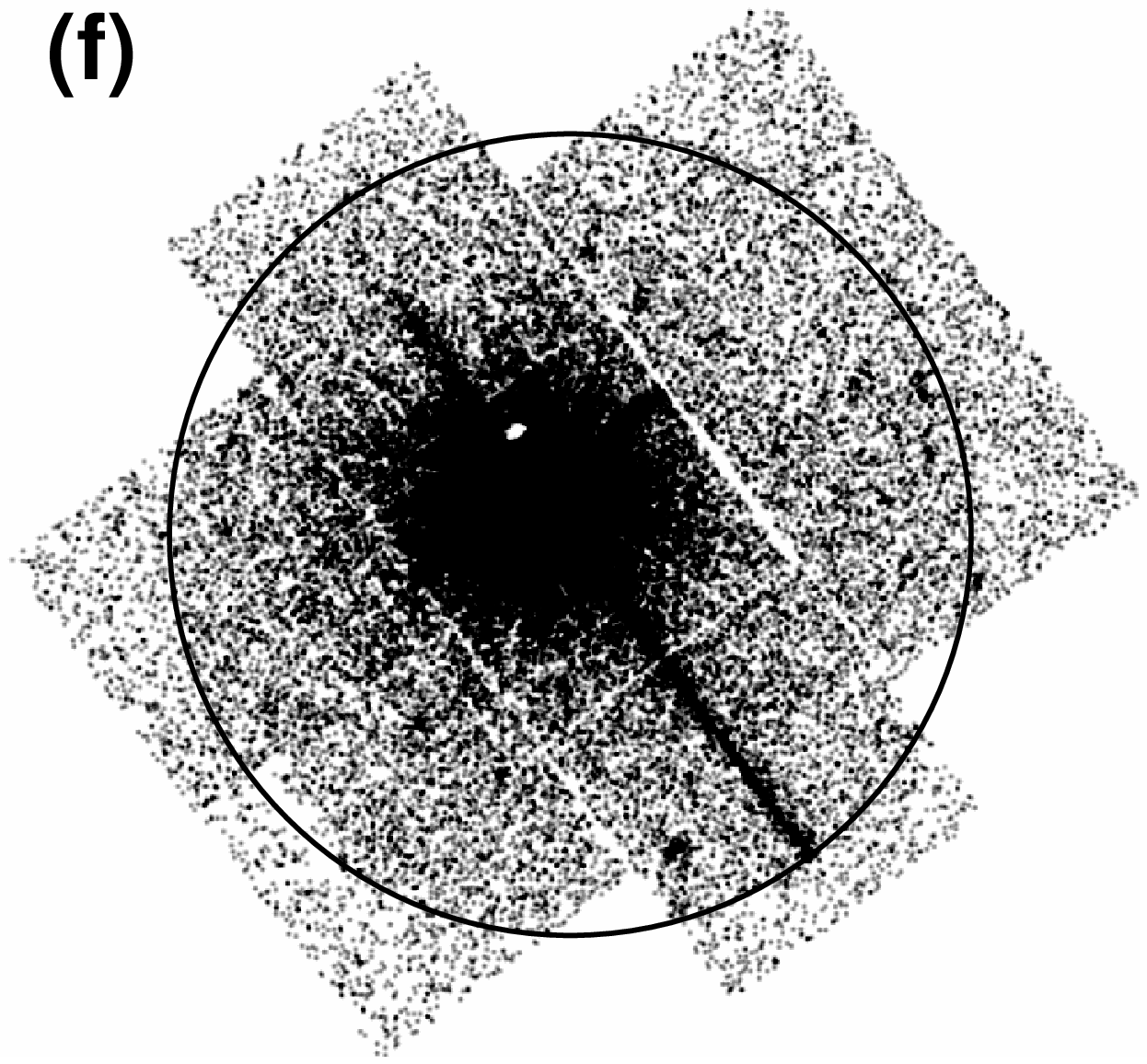}
\caption{Example smoothed MOS1 images, illustrating the different types of observations in the \xmm\ archive and the
removal of unusable portions of the data.
In all images, North is up, and East is to the left.
The large black circles ($\mathrm{radius} = 14\arcmin$) outline the approximate \xmm\ field of view.
Events from outside the field of view were used to calculate the quiescent particle background (see Section~\ref{subsec:ParticleBackground}).
The small red circles outline sources that were removed by the automated point source removal (see Section~\ref{subsec:PointSourceRemoval}).
The larger red circles in (c) and (d) outline bright regions that were removed by hand (see Section~\ref{subsec:BrightSourceRemoval}).
(a) the AXAF Ultra Deep Field (obs.~0108062301). This is an example of a blank-sky field.
(b) I~Zwicky~1 (obs.~0110890301). This is an example of an observation in which not all of the CCDs are usable.
In this observation, the central (\#1) CCD was operated in partial window mode.
This CCD was ignored, and a SXRB spectrum was extracted from the surrounding 6 CCDs.
(c) RX~J1241.5+3250 (obs.~0056020901). This is an example of an extended target source that does not fill the field of view.
The region around the target source was removed by hand.
(d) Markarian 1152 (obs.~0147920101). This is an example of a bright target source. The standard-sized exclusion region used in the automated point source removal
is inadequate, as there are a considerable number of photons from this source in the wings of the point spread function. As in (c),
a region around the target source was removed by hand.
(e) Abell~133 (obs.~0144310101). This is an example of a target source that was too extended for our purposes. Such observations were rejected.
(f) Markarian 421 (obs.~0136541101). This is an example of a target source that was too bright for our purposes. Again, such observations were rejected.
\label{fig:MOSimages}}
\end{figure*}

We next inspected the cleaned images produced by \texttt{mos-filter}. Example MOS1 sky images are
shown in Figure~\ref{fig:MOSimages}. This inspection had several purposes, exemplified by the images
in Figure~\ref{fig:MOSimages}:
\begin{itemize}

\item Figure~\ref{fig:MOSimages}(a) shows a simple blank-sky field, which did not require any special
treatment in the subsequent processing.

\item For some observations, not all of the CCDs were usable for our purposes.
Figure~\ref{fig:MOSimages}(b) shows an example of this -- for this observation the central (\#1)
MOS1 CCD was operated in partial window mode. This CCD was ignored in the subsequent processing, and
a SXRB spectrum was extracted from the surrounding 6 CCDs. In some observations, a CCD looked
significantly brighter than its neighbors. In such cases, the CCD was probably in an ``anomalous''
state (identified by \citealt{kuntz08a}), and it was ignored in the subsequent processing.
In other observations, data from some CCDs were missing altogether (for example, from MOS1-6
since its failure in 2005 March), and again these CCDs were ignored in the subsequent processing.

\item Some observations had bright and/or extended sources in the field. Such sources would
not be adequately removed by our automated source removal procedure (see Section~\ref{subsec:PointSourceRemoval},
below). Such observations were nevertheless usable for our purposes, as the target sources could be
removed by hand (see Section~\ref{subsec:BrightSourceRemoval}, below). Figures~\ref{fig:MOSimages}(c)
and \ref{fig:MOSimages}(d) show examples of observations of an extended source (a cluster of
galaxies) and a bright source (a Seyfert galaxy), respectively. The large red circles outline
the regions that were removed.

\item In some observations the target source either filled too much of the field of view, or was too bright.
Such observations were unusable, and were rejected. Examples are shown in
Figures~\ref{fig:MOSimages}(e) and \ref{fig:MOSimages}(f). We also
rejected a few observations whose fields were crowded with bright point sources.

\end{itemize}

The above-described rejections reduced our dataset from 1422 observations to 773 (a $\sim$45\%\
attrition rate). The following subsections describe our subsequent processing of the cleaned events
lists, culminating in the extraction of SXRB spectra.

\subsection{Point Source Removal}
\label{subsec:PointSourceRemoval}

We detected point sources using the standard SAS \texttt{edetect\_chain} script. Following the
Second \xmm\ Serendipitous Source Catalogue (2XMM; \citealt{watson09}), we carried out the source
detection simultaneously in five bands (0.2--0.5, 0.5--1.0, 1.0--2.0, 2.0--4.5, and
4.5--12.0~\kev). For observations with 1 good MOS1 exposure and 1 good MOS2 exposure, we used both
exposures simultaneously in the source detection. For observations with 2 or more good exposures
from either camera, we carried out the source detection on each exposure individually, as
\texttt{edetect\_chain} cannot handle more than one exposure from each MOS camera.

Since the extragalactic background is composed of resolved and unresolved point sources, the flux threshold
used in the point-source removal algorithm affects the strength of the remaining extragalactic background.
We removed point sources with fluxes down to $5 \times 10^{-14}~\flux$ in the 0.5--2.0~\kev\ band, for approximate
agreement with \citet{chen97}, whose model~A (fitted to their \rosat\ and \asca\ data) we use in
Section~\ref{subsec:OxygenIntensitiesMethod} to model the extragalactic background.

We used the energy conversion factors from the 2XMM
website\footnote{http://xmmssc-www.star.le.ac.uk/dev/Catalogue/2XMM/UserGuide\_xmmcat.html\#TabECFs}
(see also Table~4 in \citealt{watson09}) to convert the observed source count-rates to fluxes. The
region removed for each source was a circle whose radius enclosed 90\%\ of the source flux (assuming
that the source is point-like). This radius depended on the distance of the source from the optical
axis, and was typically 30\arcsec--40\arcsec.

Sources that were detected and removed in this way are shown by the small red circles in
Figures~\ref{fig:MOSimages}(a)--(d).  As can be seen, a number of sources that fall
below our flux threshold remain in the data, especially in Figure~\ref{fig:MOSimages}(a). In
order to evaluate the spectral effects of the chosen flux cut-off, we extracted and summed spectra
of sources with $5 \times 10^{-15}~\flux < \Fsource < 5 \times 10^{-14}~\flux$ for
a subset of our observations, where \Fsource\ is the 0.5--2.0~\kev\ source flux.
The 0.5--1.0~\kev\ count-rates of these summed point-source spectra
were typically $\la$20\%\ of the count-rates of the diffuse spectra in the same energy band
(extracted using a $5 \times 10^{-14}~\flux$ source removal threshold). More importantly, the summed
point-source spectra do not exhibit strong oxygen emission lines.
In addition, \citet{gupta09a} have examined the summed spectra of sources detected in several deep
\xmm\ observations. The combined spectrum of the sources with $\Fsource > 2 \times 10^{-15}~\flux$
does not exhibit any excess emission above a power-law spectrum. The combined
spectrum of the sources with $\Fsource < 2 \times 10^{-15}~\flux$ exhibits excess
counts below 0.7~\kev\ and could not be fitted with a simple power-law. \citet{gupta09a} fitted this
excess emission with a thermal plasma component. The oxygen emission from this extra thermal component
is 0.19~\LU\ for \OVII\ and 0.06~\LU\ for \OVIII. These intensities are smaller
than the typical errors on our measurements. We therefore think that the faint sources that
remain after the point source removal will not significantly affect our analysis
of the Galactic line emission.

\subsection{Removal of Bright and Extended Sources}
\label{subsec:BrightSourceRemoval}

As was mentioned in Section~\ref{subsec:InitialProcessing}, we removed extended sources and sources that were too bright to be
adequately removed by the above-described automated point source removal. In all cases we used
circular exclusion regions. If the source to be removed was the original observation target, we
centered the circle on the target position, which we extracted from the events list header. For
other sources, we centered the circle on the source by eye.

\begin{figure}
\centering
\plotone{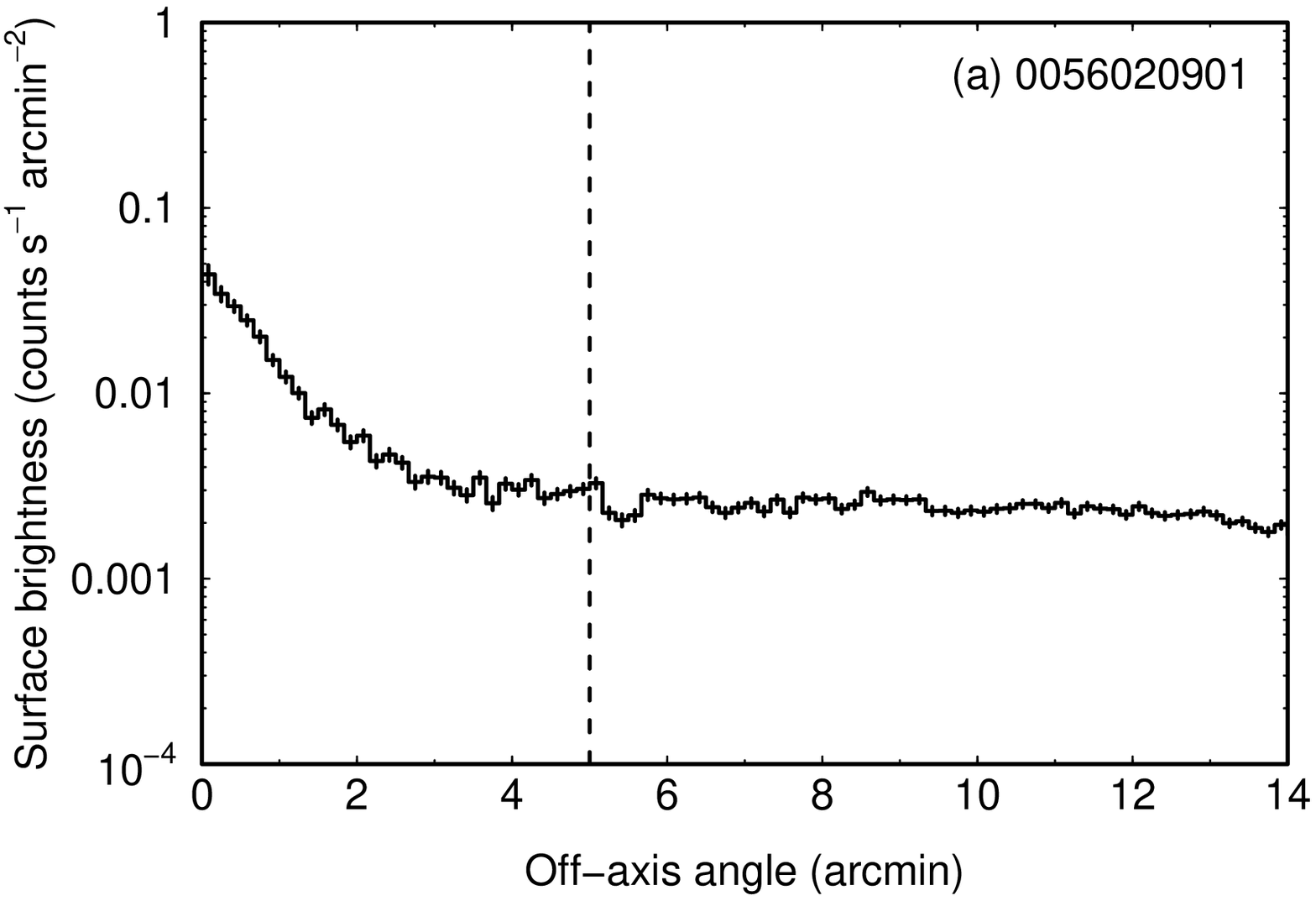}
\plotone{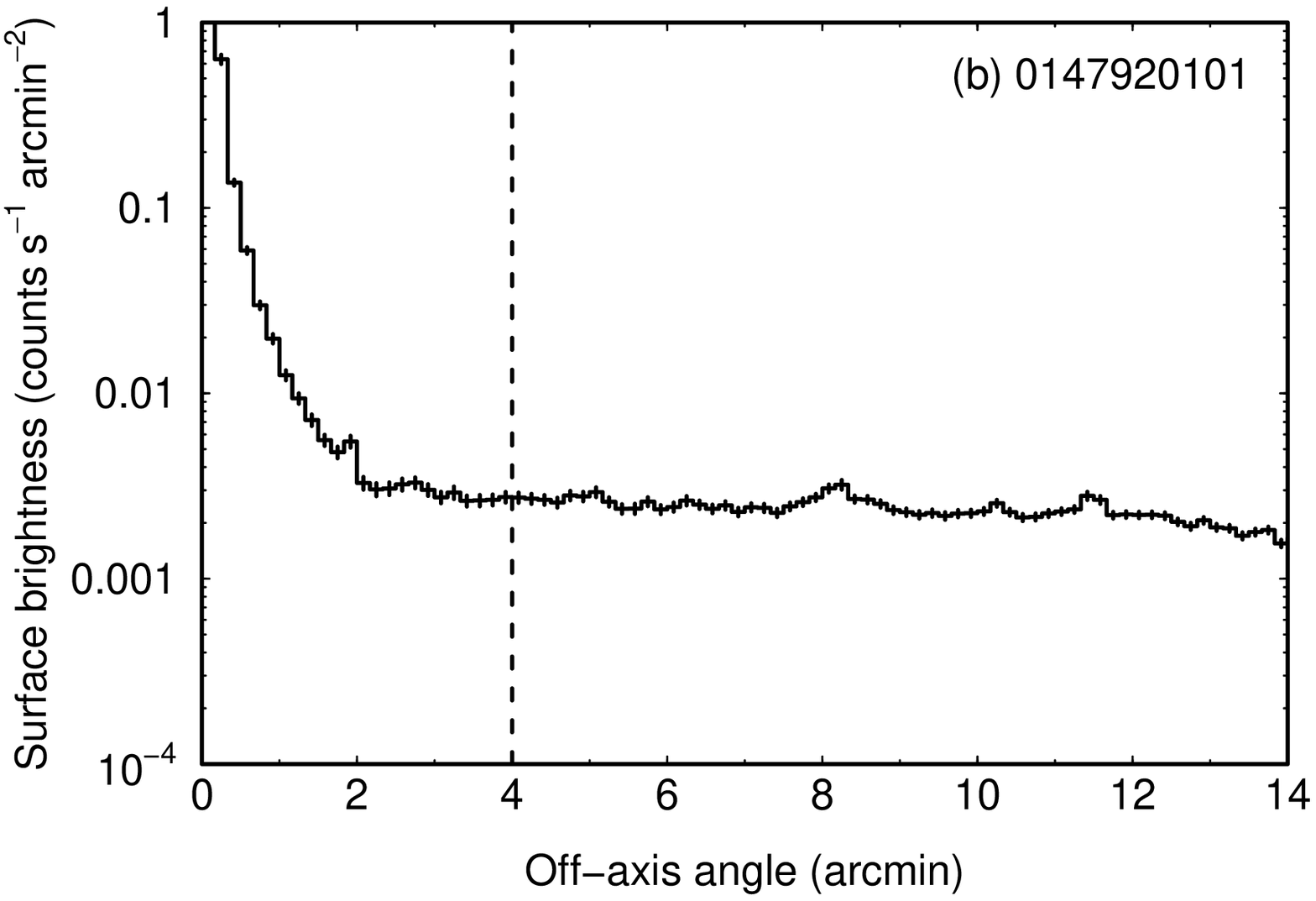}
\caption{Radial full-band MOS1 surface brightness profiles for the two observations shown in Figures~\ref{fig:MOSimages}(c) and
\ref{fig:MOSimages}(d). The vertical dashed lines indicate the radii of the source exclusion regions, shown by the
large red circles in Figure~\ref{fig:MOSimages}.
\label{fig:MOSprofiles}}
\end{figure}

The radii of the source exclusion regions were chosen by eye, although for some sources we used radial
surface brightness profiles to aid our selection of the source exclusion radius (see Figure~\ref{fig:MOSprofiles}).
We erred on the side of choosing larger source exclusion radii, at the expense of
reducing the number of photons from the SXRB. The source exclusion regions that we used typically
had radii $r = 1\arcmin$--4\arcmin. For a few observations we used larger exclusion regions: the
largest region that we used had $r = 10\arcmin$ (for 1 observation), and we also used exclusion
regions with $r = 8\arcmin$ for 4 observations and $r = 7\arcmin$ for 24 observations.

\subsection{Filtering by Solar Wind Proton Flux}
\label{subsec:ProtonFluxFiltering}

As was mentioned in Section~\ref{sec:SWCX}, we attempted to reduce the near-Earth SWCX contamination
of our data set by filtering out the portions of the \xmm\ data that were taken when the solar wind
proton flux was high.  The solar wind proton flux data were obtained from
OMNIWeb,\footnote{http://omniweb.gsfc.nasa.gov/} which combines \textit{in situ} solar wind
measurements from several satellites. The OMNIWeb solar wind proton flux data covering the
\xmm\ mission are mainly from the \textit{Advanced Composition Explorer} (\ace) and \wind\ -- these
data are time-shifted to the Earth, based on the relevant spacecraft's position and the observed solar wind
speed (this time-shifting is included in the OMNIWeb data). The OMNIWeb data
covering the \xmm\ mission also include data from \textit{IMP-8} and \textit{Geotail}. These data
are not time-shifted. However, the apogee of each satellite (35\RE\ for \textit{IMP-8}\footnote{http://spdf.gsfc.nasa.gov/imp8/project.html}
and 30\RE\ for \textit{Geotail}\footnote{http://www.isas.jaxa.jp/e/enterp/missions/geotail/},
where \RE\ is the Earth's radius) divided
by a typical solar wind speed of 400~\kmps\ is $<$10~min, which is much less than the 1-hour time
resolution of the OMNIWeb data. We therefore do not think that the lack of time-shifting in the
\textit{IMP-8} and \textit{Geotail} data will significantly affect our results.

In choosing a proton flux threshold for this filtering, we wanted to reduce the potential SWCX
contamination as much as possible without discarding too much \xmm\ data. We chose to remove times
when the measured proton flux exceeded $2 \times 10^8~\pcmsq~\ps$ from the \xmm\ data. This
threshold was chosen to be somewhat lower than the average proton flux at 1~\AU: an average solar
wind speed of 400~\kmps\ and an average density at 1~\AU\ of 7~\pcc\ \citep[e.g.,][]{wargelin04}
yield an average proton flux of $2.8 \times 10^8~\pcmsq~\ps$.  We also removed times for which no
proton flux data were available from OMNIWeb. This filtering reduced the total amount of usable time
over all observations by 55\%. For some observations, the good time that remained after this
filtering fell below our 5~ks acceptance threshold (see Section~\ref{subsec:InitialProcessing}).  As
a result, filtering on the basis of solar wind proton flux reduced the number of usable observations
from 773 to 412.  Because of this severe reduction in the number of usable observations, we measured
the oxygen line intensities both with and without this solar wind proton flux filtering.

\subsection{Source Spectra and Response Files}

The SXRB source spectra and the spectral response files were created using the XMM-ESAS
\texttt{mos-spectra} script. The SXRB spectra were extracted from the cleaned and filtered events lists
using the whole field of view, minus any CCDs that were ignored or missing and any sources that
were automatically removed (Section~\ref{subsec:PointSourceRemoval}) or removed by hand
(Section~\ref{subsec:BrightSourceRemoval}). The spectra were binned so that each bin contained at least 25
counts. Redistribution matrix files (RMFs) and ancillary response files (ARFs) were created with the
SAS \texttt{rmfgen} and \texttt{arfgen} programs.

\subsection{The MOS Particle Background}
\label{subsec:ParticleBackground}

The MOS particle background has two main components \citep[e.g.,][]{read03,kuntz08a}. Soft-proton flares are
caused by protons with $E \sim \mathrm{few} \times 100~\kev$ interacting directly with the
detector. These flares are largely removed by the lightcurve analysis described in
Section~\ref{subsec:InitialProcessing}.  However, even after this cleaning, some soft-proton contamination
may remain in the data. We modeled this residual contamination by adding a broken power-law component
to our spectral analysis model. This broken power-law was not convolved with the instrumental response,
and its break was fixed at 3.2~\kev. Such a model is a good description of the mean shape of the soft
proton contamination \citep{kuntz08a}.

The second particle background component is the so-called quiescent particle background
(QPB). This is produced by cosmic rays either interacting directly with the detector, or producing
fluorescent X-rays from the detector's surroundings. As the QPB varies with time, as well as across
the detector, its spectrum has to be calculated for each observation. We used the XMM-ESAS
\texttt{xmm-back} program to calculate QPB spectra for our observations. For each exposure from a
given observation, the QPB spectrum is constructed from a database of filter-wheel--closed data,
scaled using data from the unexposed corner pixels outside the field of view (see
Figure~\ref{fig:MOSimages}). This scaling is energy-dependent, and is based on the 0.3--10.0~\kev\
count-rate and the (2.5--5.0~\kev)/(0.4--0.8~\kev) hardness ratio from the unexposed corner pixels.
For more details of the modeling of the QPB spectrum, see \citet{kuntz08a}. The QPB spectra were
subtracted from the corresponding source spectra before we carried out our spectral analysis.

The QPB includes two bright fluorescent instrumental lines at 1.49 and 1.74~\kev, produced within
the telescope by aluminum and silicon, respectively. These lines cannot be adequately removed by the
above-described procedure, because small variations in the gain and the line strengths between the
source and background spectra can lead to large residuals in the spectral fitting
\citep{kuntz08a}. Instead, the continuum QPB spectrum was interpolated across the 1.2--1.9~\kev\
energy range, and we modeled these instrumental lines by adding two Gaussians to our spectral
analysis model.

\citet{kuntz08a} identified several periods in which certain MOS CCDs were in an ``anomalous''
state, characterized by a low hardness ratio and a high background count-rate in the unexposed
corner pixels. Because such data should not be used, after processing each observation we inspected
plots of the (2.5--5.0~\kev)/(0.4--0.8~\kev) hardness ratio against the 0.3--10.0~\kev\ count-rate
for the corner pixels. If any CCDs were found to have the hardness ratio and count-rate that
characterize the anomalous state, we excluded those CCDs, and then re-ran the spectral extraction
and QPB calculation for that observation.

\section{OXYGEN LINE INTENSITIES}
\label{sec:OxygenIntensities}

\subsection{Method}
\label{subsec:OxygenIntensitiesMethod}

In order to measure the diffuse \OVII\ and \OVIII\ intensities, we fitted a multicomponent spectral
model to the cleaned and QPB-subtracted MOS spectra extracted from each \xmm\ observation. The model
that we used is similar to that described in \citet{henley08a}, and consisted of Galactic ISM,
extragalactic, and instrumental components.

The Galactic ISM emission was modeled using a single-temperature APEC thermal plasma model
\citep{smith01a}, except for the \OVII\ and \OVIII\ \Kalpha\ lines, which were modeled separately using 2
$\delta$ functions.\footnote{Note that the \OVII\ \Kalpha\ ``line'' is actually a
forbidden-intercombination-resonance triplet.  However, as the energy resolution of the MOS cameras
($\sim$50~eV) is much larger than the splitting of the triplet ($\sim$10~eV), using a single
$\delta$ function to model the \OVII\ emission is a reasonable approximation.} In
\citet{henley08a} we disabled the oxygen emission from the APEC component by simply setting its
oxygen abundance to zero. The disadvantage of this method is that higher transitions (e.g., the
\Kbeta\ lines) and continuum emission (due to two-photon processes and radiative recombination) from
oxygen are also removed from the model. Here, we followed \citet{lei09}, and removed only the \OVII\
and \OVIII\ \Kalpha\ lines (and their satellite lines) from the APEC model.  We did this by setting
these lines' emissivities to zero in the APEC line emissivity data file (\texttt{apec\_v1.3.1\_line.fits}).

As the thermal diffuse emission in the \xmm\ band mainly originates in the Galactic halo, beyond the
majority of the Galaxy's \HI, the APEC component was attenuated by absorption. For each observation,
we fixed the column density \NH\ at the appropriate \HI\ column density from the LAB survey
\citep{kalberla05}. The oxygen lines were not subject to this absorption, so the intensities that we
report in Section~\ref{subsec:OxygenIntensitiesResults} below are observed intensities, not
intrinsic, deabsorbed intensities. As we are reporting the observed oxygen intensities, the fact
that our absorption model neglects the effects of molecular hydrogen and dust should not
significantly affect our intensity measurements.

The extragalactic background was modeled using a power-law. Because of possible residual
contamination from soft protons (see Section~\ref{subsec:ParticleBackground}), we could not
independently constrain the extragalactic background spectrum. We therefore fixed the extragalactic
background spectrum at $10.5(E/1~\kev)^{-1.46}~\pownorm$ \citep{chen97}. The extragalactic component
was assumed to be attenuated to the same extent as the APEC component.

As described in Section~\ref{subsec:ParticleBackground}, we included components to model parts of
the instrumental particle background.  We used two Gaussians to model the aluminum and silicon instrumental lines
at 1.49 and 1.74~\kev, respectively, and a broken power-law to model any residual soft-proton
contamination that may have remained after the cleaning described in
Section~\ref{subsec:InitialProcessing}.

We carried out our spectral analysis using
XSPEC\footnote{http://heasarc.gsfc.nasa.gov/docs/xanadu/xspec/} version 12.5.0.  For each observation, we
fitted the above-described model simultaneously to all the usable exposures (normally this was 1
MOS1 exposure and 1 MOS2 exposure, but some observations had more). The $\delta$ functions used to
model the oxygen lines were XSPEC \texttt{gauss} models with the widths fixed at 0. The energy of the \OVII\
\Kalpha\ feature was a free parameter, but that of the \OVIII\ \Lyalpha\ line was fixed at
0.6536~\kev\ (from APEC). The temperature and normalization of the APEC component were both free
parameters. We used the XSPEC \texttt{phabs} absorption model (\citealt{balucinska92}, with an
updated He cross-section from \citealt{yan98}) to attenuate the APEC and extragalactic
components. We used \citet{wilms00} interstellar abundances for the APEC and \texttt{phabs}
models. The parameters of the particle background components (the Gaussian instrumental lines and
the soft-proton broken power-law) were independent for each exposure.

\subsection{Measurements}
\label{subsec:OxygenIntensitiesResults}

The results of the above-described oxygen line measurements are presented in Tables~\ref{tab:OxygenIntensities1}
and \ref{tab:OxygenIntensities2}, sorted by increasing Galactic longitude, $l$. Note that the intensities are
observed intensities, not deabsorbed intrinsic intensities. Table~\ref{tab:OxygenIntensities1}
contains the results obtained without the proton flux filtering described in Section~\ref{subsec:ProtonFluxFiltering}, while
Table~\ref{tab:OxygenIntensities2} contains the results obtained with this additional filtering.

In both tables, column~1 contains the \xmm\ observation ID.
Column~2 contains the observation start date, in YYYY-MM-DD format.
Columns~3 and 4 contain the Galactic coordinates $(l,b)$ of the pointing direction.
Column~5 contains the usable MOS1 exposure, and column~6 contains the solid angle, $\Omega$, from which the MOS1 SXRB spectrum was extracted.
Columns~7 and 8 contain the corresponding data for MOS2.
For some observations (e.g., obs.~0144230101), there is more than one MOS1 and/or MOS2
exposure. For these observations, the data for the individual exposures are presented separately.
Column~9 contains the \OVII\ intensity, $\Iovii$, and column~10 contains the 68\%\ confidence interval.
Column~11 contains the photon energy, $\Eovii$, of the $\delta$ function used to model the
\OVII\ emission. Because we use a $\delta$ function, the fits are insensitive to shifts in the
\OVII\ energy within an RMF energy bin (these bins are 5~\ev\ wide). We have therefore
rounded $\Eovii$ to the central energy of the RMF bin in which $\Eovii$ lies.
Column~12 contains the \OVIII\ intensity, $\Ioviii$, and column~13 contains the 68\%\ confidence interval.
Column~14 contains the solar wind proton flux at the Earth, \fsw, averaged over the duration of the \xmm\ observation.
The proton flux data were obtained from OMNIWeb. If an \fsw\ value is missing, it means that there were no good solar wind data during the observation.
Column~15 contains the ratio of the total model 2--5~\kev\ count-rate, $\Ftotal$, to that expected from our extragalactic power-law
model, $\Fexgal$. This ratio was used as a measure of how much soft-proton contamination remained in the spectrum after the lightcurve analysis described
in Section~\ref{subsec:InitialProcessing}. Any observations for which this ratio exceeded 2 were rejected. We also rejected two
observations for which the soft-proton broken power-law component dominated the model at low energies and interfered with
the oxygen line measurements. These rejections further reduced the
number of good observations from 773 to 586 (without proton flux filtering) or from 412 to 303 (with proton flux filtering).
Column~16 in Table~\ref{tab:OxygenIntensities1} indicates whether or not the observation also appears in Table~\ref{tab:OxygenIntensities2}.
In Table~\ref{tab:OxygenIntensities2}, column~16 indicates whether or not the observation was affected by the proton flux filtering.

It should be noted that the 303 obserations in Table~\ref{tab:OxygenIntensities2} are not a strict subset of the 586 observations
in Table~\ref{tab:OxygenIntensities1}. Four observations (0083000101, 0111100101, 0200250101, 0202210301) fail the
requirement that $\Ftotal / \Fexgal \le 2$ without the proton flux filtering, but pass this requirement with the proton flux
filtering. These four observations therefore appear in Table~\ref{tab:OxygenIntensities2} but not in Table~\ref{tab:OxygenIntensities1},
and are indicated by a missing value in column~16 of Table~\ref{tab:OxygenIntensities2}.
The remaining 299 observations in Table~\ref{tab:OxygenIntensities2} are a subset of the observations in Table~\ref{tab:OxygenIntensities1}.
Of these 299 observations, 204 were completely unaffected by the proton flux filtering (i.e., the solar wind proton flux remained
below the threshold of $2 \times 10^8~\pcmsq~\ps$ throughout these observations). For the remaining 95 observations, the proton flux
filtering does not seem to have had a systematic effect on the measured oxygen intensities.

It should also be noted that our method gives the average \OVII\ and \OVIII\ intensities for each
observation. \citet{snowden04} found that the first part of obs.~0111550401 of the Hubble Deep Field
North (HDF-N), which is in our dataset, exhibited enhanced SWCX emission relative to the later part
of that observation, and to other \xmm\ observations of the HDF-N. Also, \citet{carter08} found
several \xmm\ observations in which the count-rate in the 0.5--0.7~\kev\ oxygen line band varied
independently of the count-rate in a continuum band, indicating variable SWCX emission. A detailed
examination of the oxygen line variability within each observation is beyond the scope of this
paper. However, future extensions to this survey may use this variability to identify
SWCX-contaminated observations \citep{carter08}.

\begin{figure}
\centering
\plotone{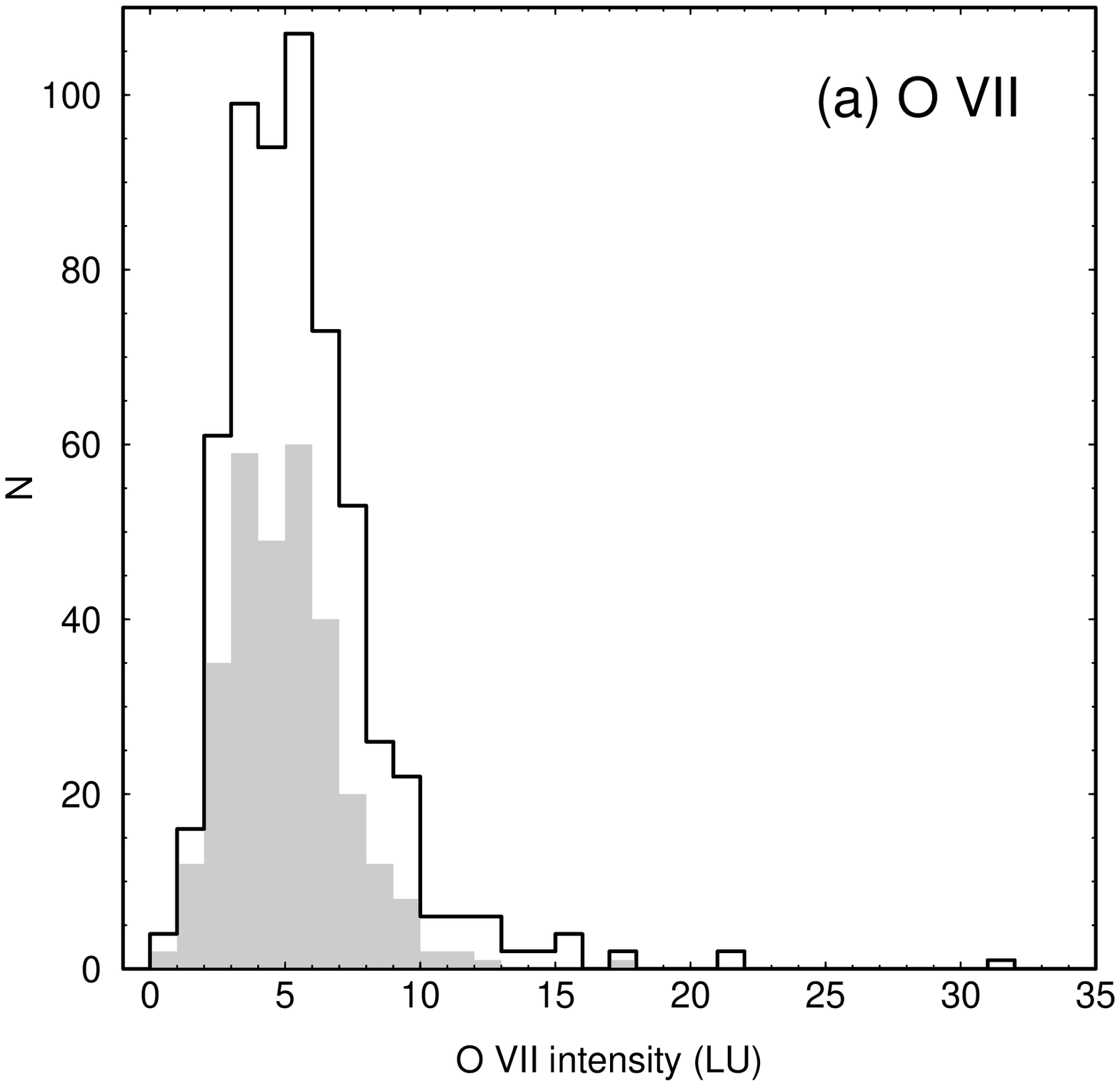}
\plotone{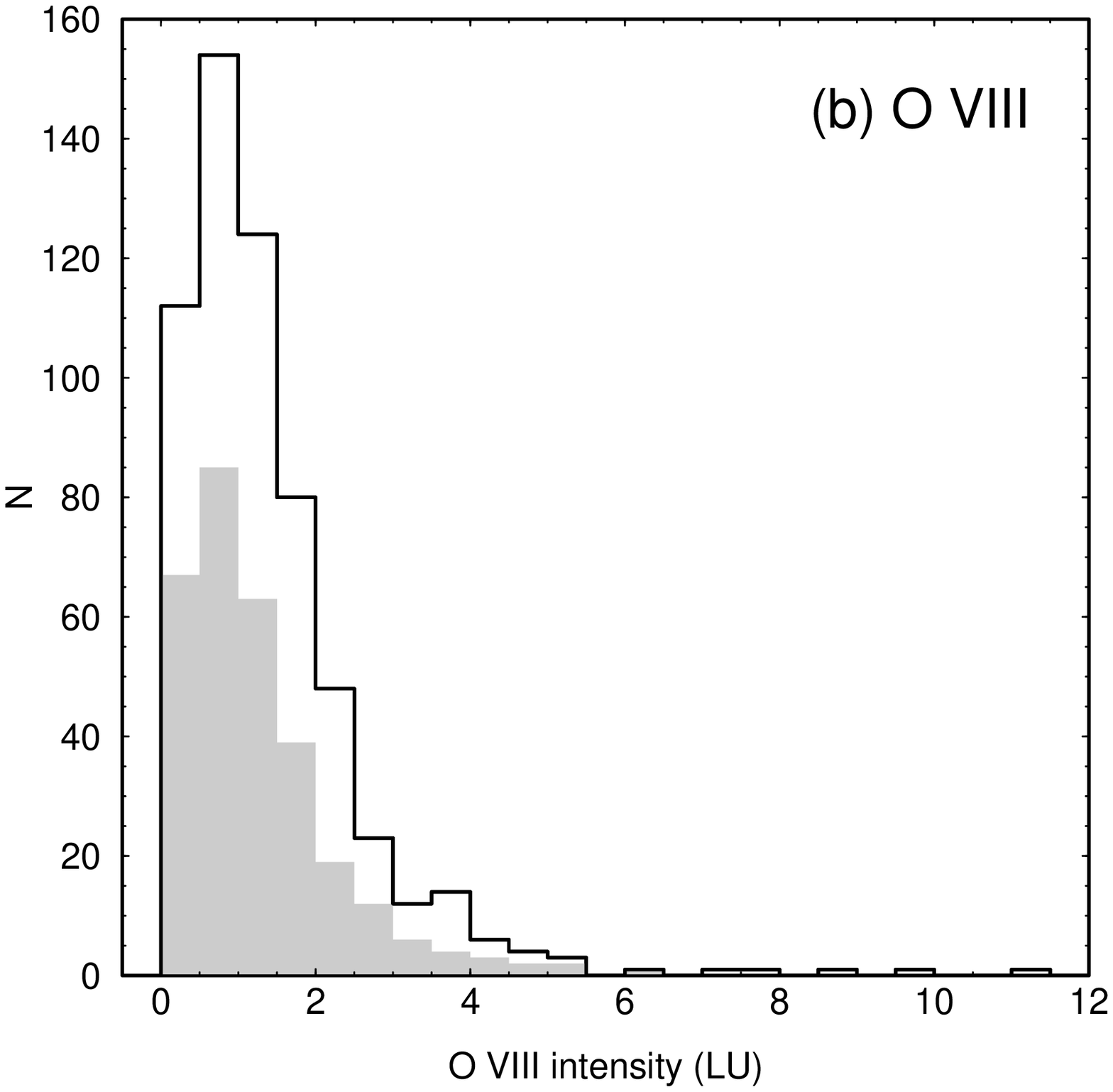}
\caption{Histograms of the (a) \OVII\ and (b) \OVIII\ intensities.
The solid black lines show the histograms of intensities obtained without the solar
wind proton flux filtering described in Section~\ref{subsec:ProtonFluxFiltering}, and the gray areas
show the histograms of intensities obtained with this filtering.
\label{fig:IntensityHistogram}}
\end{figure}

Figure~\ref{fig:IntensityHistogram} shows histograms of the oxygen intensities measured with and
without the proton flux filtering. The ranges and quartiles of the intensities are summarized in
Table~\ref{tab:RangeMedianMode}. The table also shows the 90\%\ confidence intervals on the
quartiles, calculated by bootstrapping.\footnote{All the statistical analysis in this paper was
  carried out using the R software package \citep{R}.} In general, the confidence intervals calculated with and
without proton flux filtering overlap, implying that the proton flux filtering does not cause a
significant, systematic shift toward lower average oxygen intensities. In addition, for the 95
observations that have at least some good time removed by the proton flux filtering and that yield
measurable oxygen line intensities after the filtering, the median difference between the
\OVII\ intensity obtained without proton flux filtering and the intensity obtained with proton flux
filtering is 0.04~\LU\ (90\%\ bootstrap confidence interval: $-0.05$ to 0.14~\LU). The corresponding
value for \OVIII\ is 0~\LU\ ($-0.03$ to 0.02~\LU). Again, these results show that proton flux
filtering does not cause a systematic shift to lower oxygen intensities.

Proton flux filtering does, however, preferentially remove the
observations with higher oxygen intensities (meaning that, during such observations, the solar wind
proton flux tended to be above the filtering threshold of $2 \times 10^8~\pcmsq~\ps$).  For example,
31 observations in Table~\ref{tab:OxygenIntensities1} have $\Iovii > 10~\LU$, obtained without the
proton flux filtering. Among those 31 observations, only 5 yield usable data after the proton flux
filtering. Similarly, among the 9 observations with $\Iovii > 15~\LU$ without proton flux filtering,
only 1 yields usable data after the proton flux filtering (this observation, 0103861001, has $\Iovii
= 17.6~\LU$, and is one of the observations unaffected by the proton flux filtering). In contrast,
53\%\ of the observations in Table~\ref{tab:OxygenIntensities1} with $\Iovii \le 10~\LU$ yield
usable data after the proton flux filtering. Among the \OVIII\ measurements, 9 observations have
$\Ioviii > 5~\LU$ without proton flux filtering, but only 2 of those yield usable data after the
proton flux filtering. In contrast, 51\%\ of the observations in Table~\ref{tab:OxygenIntensities1}
with $\Ioviii \le 5~\LU$ yield usable data after the proton flux filtering.

\begin{figure*}
\centering
\plotone{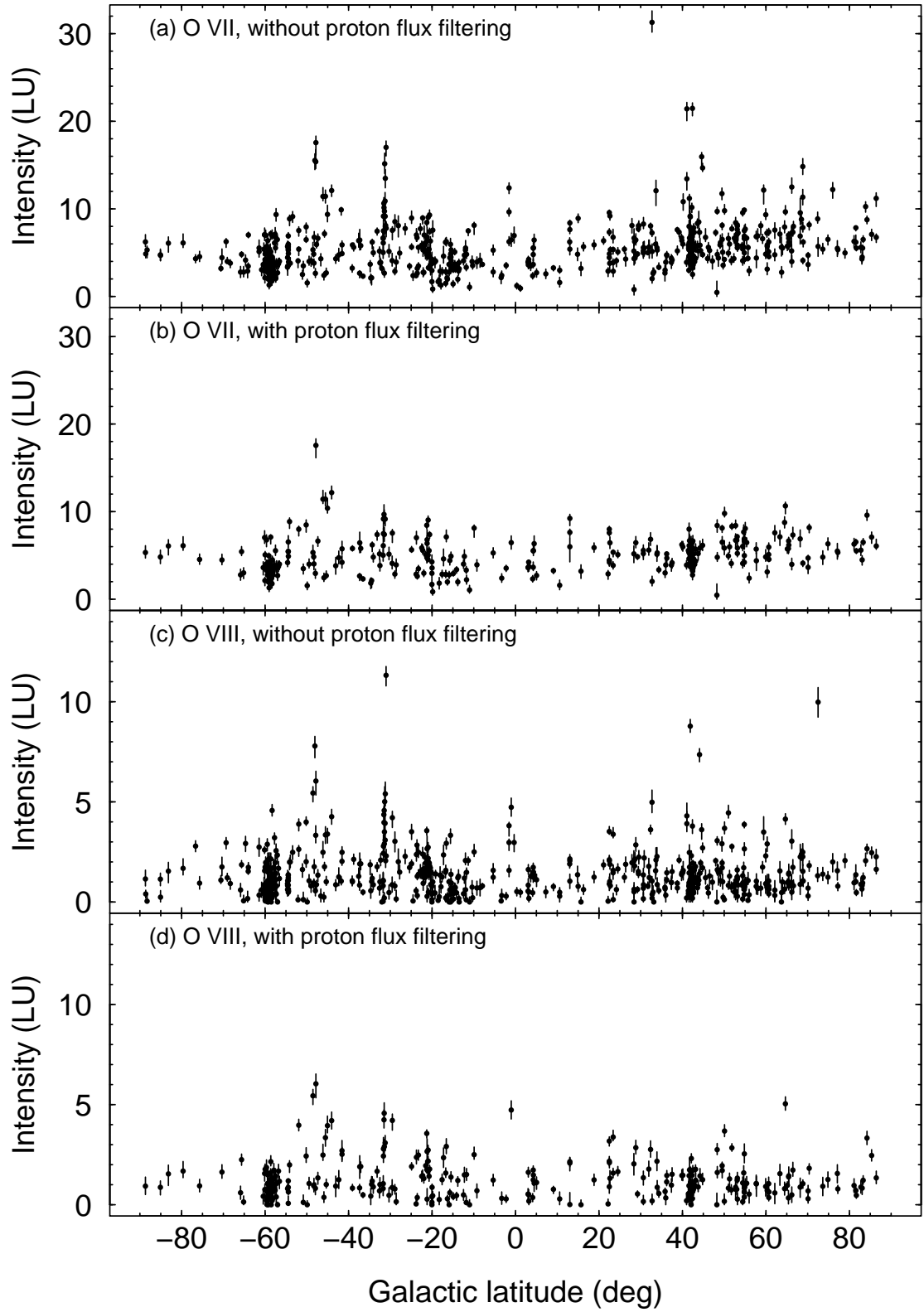}
\caption{Variation of the observed oxygen line intensities with Galactic latitude.
Panel (a) shows the \OVII\ intensities obtained without the solar
wind proton flux filtering described in Section~\ref{subsec:ProtonFluxFiltering},
and panel (b) shows the \OVII\ intensities obtained with this filtering.
Panels (c) and (d) show the corresponding \OVIII\ intensities.
\label{fig:Oxygen-vs-b}}
\end{figure*}

Figure~\ref{fig:Oxygen-vs-b} shows the measured oxygen intensities plotted against Galactic
latitude. We have looked for correlations between the measured intensities and Galactic latitude
using Kendall's $\tau$ \citep[e.g.,][]{press92}. The results are summarized in
Table~\ref{tab:Correlation}.  In the northern Galactic hemisphere, the \OVII\ intensity is
significantly correlated with Galactic latitude (i.e, the intensity tends to increase from the
Galactic plane to the north Galactic pole).  This correlation exists with or without the proton flux
filtering, and whether or not we exclude data from low Galactic latitudes. No such correlation
exists for \OVIII\ in the north. In the south, if we exclude the observations from low Galactic
latitudes, we find that both the \OVII\ and \OVIII\ intensities are significantly correlated with
latitude -- here the correlation implies a decrease from the Galactic plane to the south Galactic
pole.

\begin{figure*}
\centering
\plotone{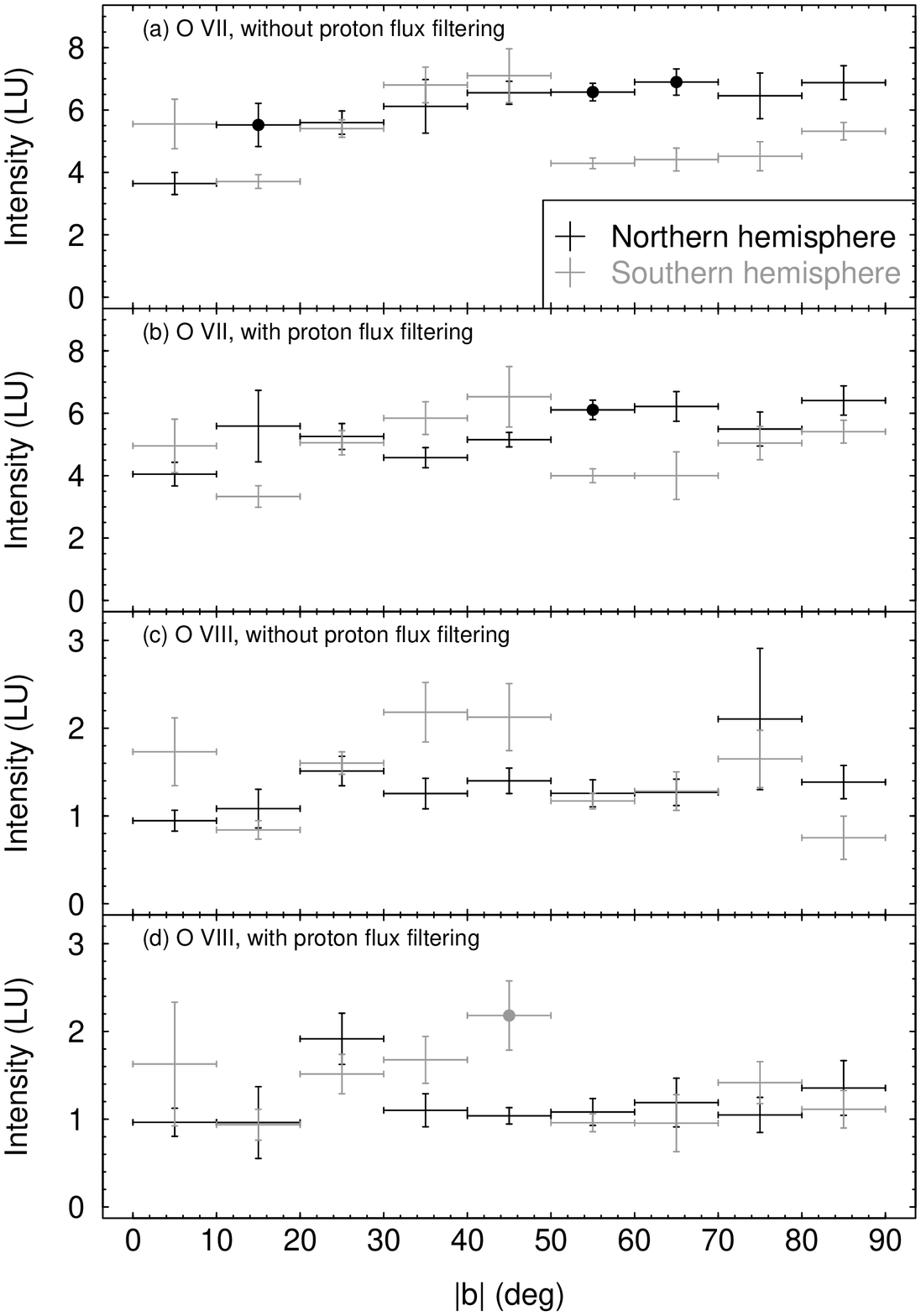}
\caption{Variation of the observed oxygen line intensities with Galactic latitude, grouped in
  10\degr\ bins. The vertical errorbars indicate the errors on the means.  Panel (a) shows the
  \OVII\ intensities obtained without the solar wind proton flux filtering described in
  Section~\ref{subsec:ProtonFluxFiltering}, and panel (b) shows the \OVII\ intensities obtained with
  this filtering.  Panels (c) and (d) show the corresponding \OVIII\ intensities. Datapoints marked
  with a solid circle are significantly brighter (at the 1\%\ level) than the corresponding datapoint
  from the other hemisphere.
  \label{fig:Oxygen-vs-b-binned}}
\end{figure*}

Figure~\ref{fig:Oxygen-vs-b-binned} compares the oxygen intensities, averaged over 10\degr\ bins,
for the two hemispheres. For each latitude bin, we compared the mean intensities from the two
hemispheres using the $t$ test -- mean intensities that are significantly brighter (at the
1\%\ level) than their counterparts from the opposite hemisphere are marked with solid circles.
At high latitudes, the northern hemisphere appears somewhat brighter in \OVII\ than the southern
hemisphere, whereas for \OVIII\ there is in general no difference between the hemispheres.

The correlations noted above could ultimately be due to variations in the SWCX intensity (although
this is unlikely to be correlated with Galactic latitude), LB intensity, observed halo intensity
(which in turn could be due to variations in intrinsic intensity or absorbing column), or any
combination thereof. The different correlations in the two hemispheres, and the results shown in
Figure~\ref{fig:Oxygen-vs-b-binned}, suggest at least some differences between the two hemispheres,
and also possible differences between the distributions of \OVII\ and \OVIII\ emission. Without an
accurate model for the SWCX emission in each observation, we cannot use the above-noted trends to
draw conclusions about the hot ISM. However, in Section~\ref{subsec:Halo} we apply various filters
to our dataset to remove observations likely to be contaminated by SWCX emission, in order to study
the Galactic halo emission.  Such filtering greatly reduces the number of usable observations, but
those that remain should give a more accurate picture of the halo than if we were to use the whole,
unfiltered dataset.

\subsection{Measurements for Directions with Multiple Observations}
\label{subsec:ResultsMultiple}

Many directions have been observed multiple times by \xmm. The separations in time between
observations of the same direction range from $\sim$1~day to several years. The contributions to
the SXRB from the LB and the Galactic halo are not expected to vary on such a short time scale. For
example, if the LB is filled with $10^6$-K plasma, then the sound-crossing time for crossing \xmm's
field of view ($\sim$0.5\degr) at a distance of 10~pc is $\sim$1000~yr. Variations in the oxygen
intensities measured in a given direction must therefore be due to SWCX. As a result, multiple
observations of a given direction are important as they can be used to constrain models of
SWCX emission, in particular the time-varying aspects of such models.

In practice, the pointing directions for \xmm\ observations of the same target are rarely
identical. We therefore searched Table~\ref{tab:OxygenIntensities1} for sets of observations whose
pointing directions were within 0.1\degr\ of each other (cf. the \xmm\ field of view is
$\sim$0.5\degr\ across). We found 69 such sets from the set of 586 good observations in
Table~\ref{tab:OxygenIntensities1}.

The oxygen line intensities from directions with multiple observations are shown in Table~\ref{tab:MultipleObs}.
Column~1 contains a unique number (1--69) which identifies each set of observations of nearby directions.
Column~2 contains the number of observations in each set.
Column~3 contains the \xmm\ observation IDs of these observations.
Columns~4 and 5 contain the Galactic coordinates $(l,b)$ of the pointing direction.
Columns~6 through 9 contain the \OVII\ intensity, the 68\%\ confidence interval on the \OVII\ intensity,
the \OVIII\ intensity, and the 68\%\ confidence interval on the \OVIII\ intensity, all obtained
without the proton flux filtering described in Section~\ref{subsec:ProtonFluxFiltering}.
Columns~10 through 13 contain the corresponding values obtained with the proton flux filtering.
Missing values in columns~10 through 13 indicate observations that were unusable after the proton flux filtering.

For each of the 69 sets of observations in Table~\ref{tab:MultipleObs}, we
find the minimum intensity, $\min(\Iovii)$ or $\min(\Ioviii)$. Then, for each observation
in Table~\ref{tab:MultipleObs}, we calculate $\Iovii - \min(\Iovii)$ and $\Ioviii - \min(\Ioviii)$,
where $\min(\Iovii)$ or $\min(\Ioviii)$ is the minimum measured intensity \textit{for that direction}.
The difference $I - \min(I)$ can be attributed to SWCX. This SWCX intensity is actually a lower limit,
because SWCX may have contributed photons to the dimmest observations, as well as too the brighter
observations.

\begin{figure}
\centering
\plotone{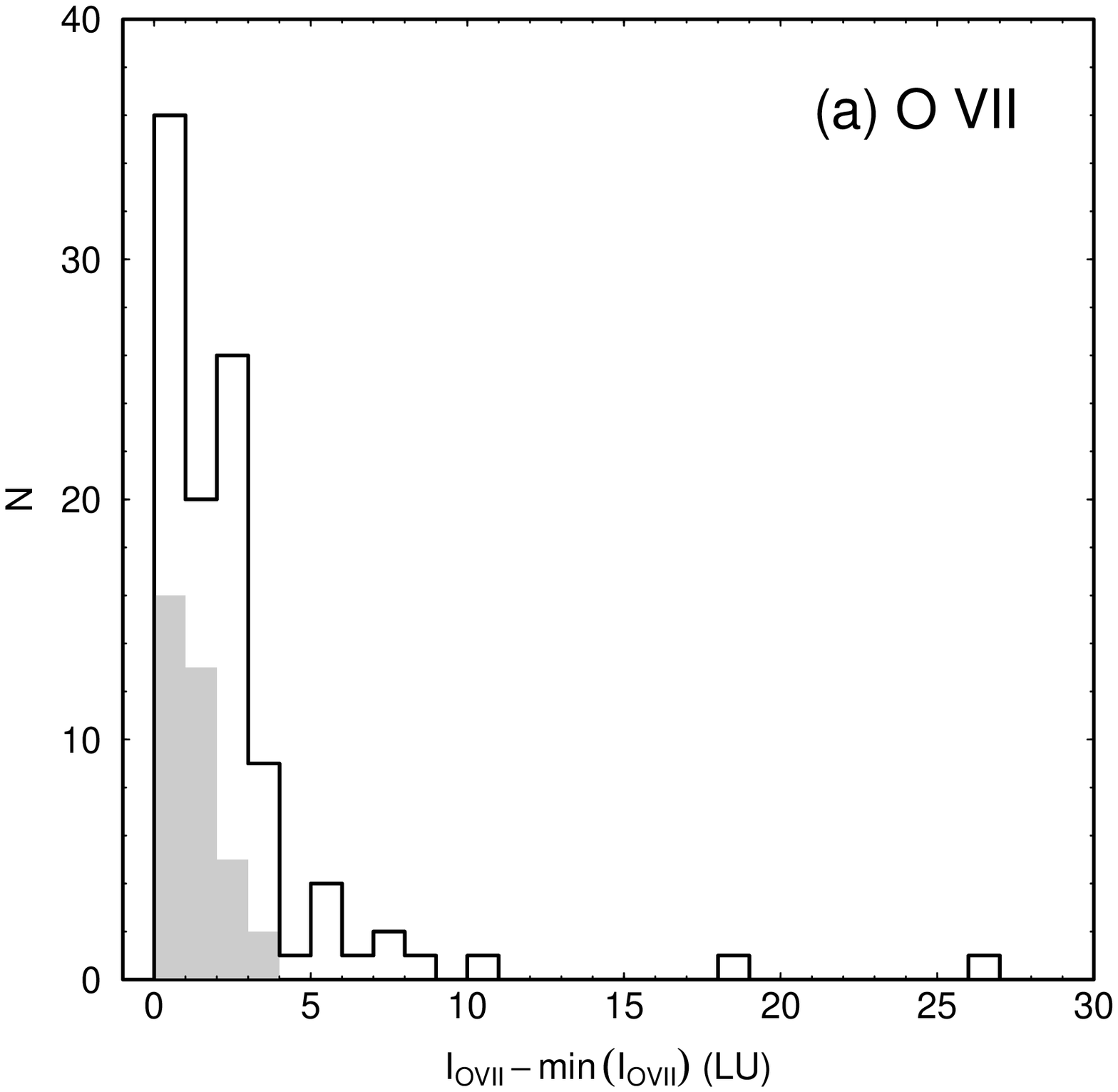}
\plotone{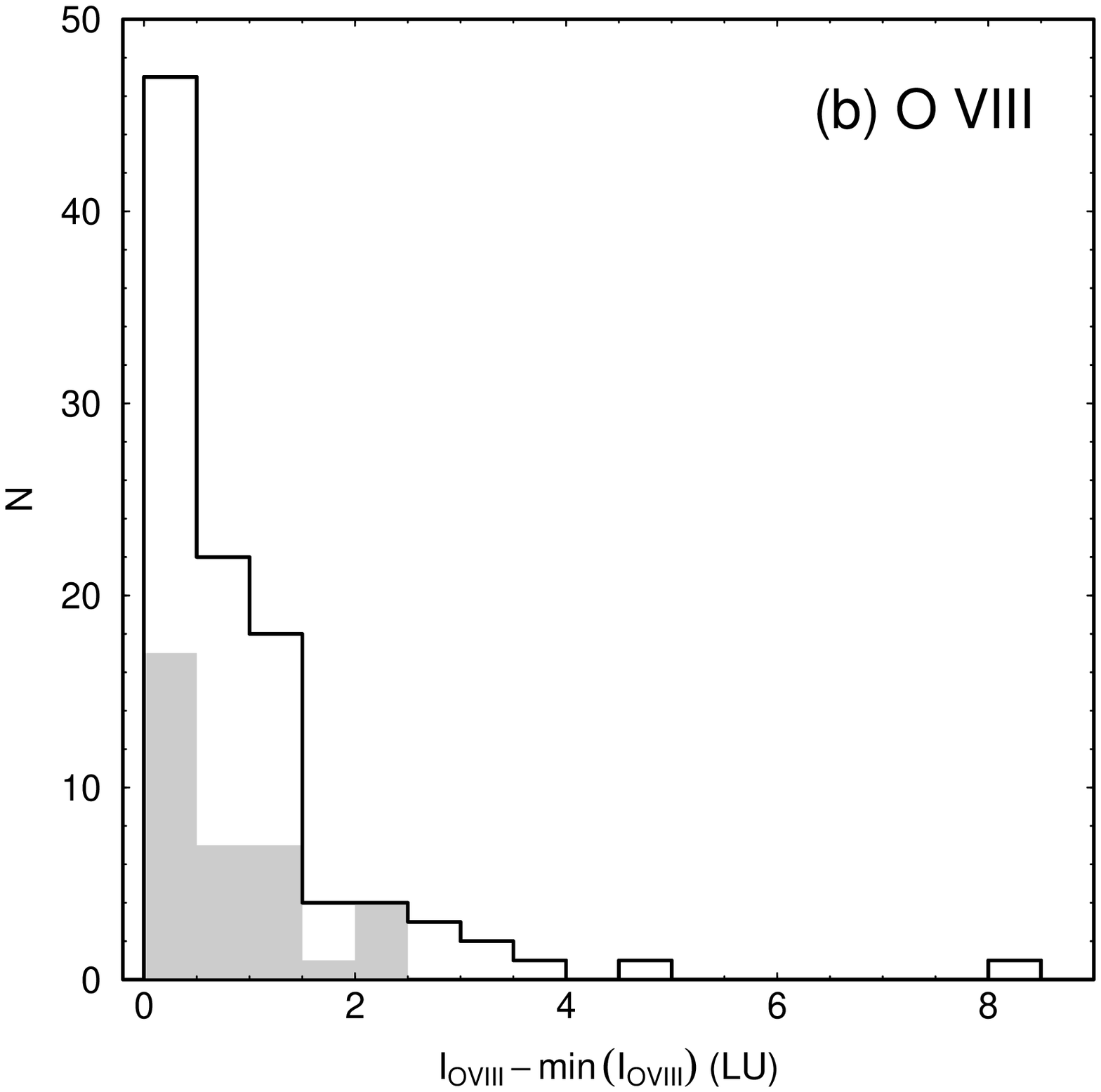}
\caption{Histograms of $I - \min(I)$ for (a) \OVII\ and (b) \OVIII, where $\min(I)$
is the minimum measured intensity in the same direction as the $I$ measurement. These histograms were
constructed from the data in Table~\ref{tab:MultipleObs}. The solid black lines
show the results obtained without the solar wind proton flux filtering described in
Section~\ref{subsec:ProtonFluxFiltering}, and the gray areas show the results obtained with this filtering.
For each of the 69 sets of observations in Table~\ref{tab:MultipleObs} there is, by definition,
an observation with $I - \min(I) = 0$; these observations are omitted.
\label{fig:I-Imin}}
\end{figure}

Figure~\ref{fig:I-Imin} shows histograms of $\Iovii - \min(\Iovii)$ and $\Ioviii - \min(\Ioviii)$,
obtained with and without the solar wind proton flux filtering described in
Section~\ref{subsec:ProtonFluxFiltering}.  For each of the 69 sets of observations in
Table~\ref{tab:MultipleObs} there is, by definition, an observation with $I - \min(I) = 0$. These
observations are omitted from the histograms in Figure~\ref{fig:I-Imin}.

\begin{figure*}
\centering
\plottwo{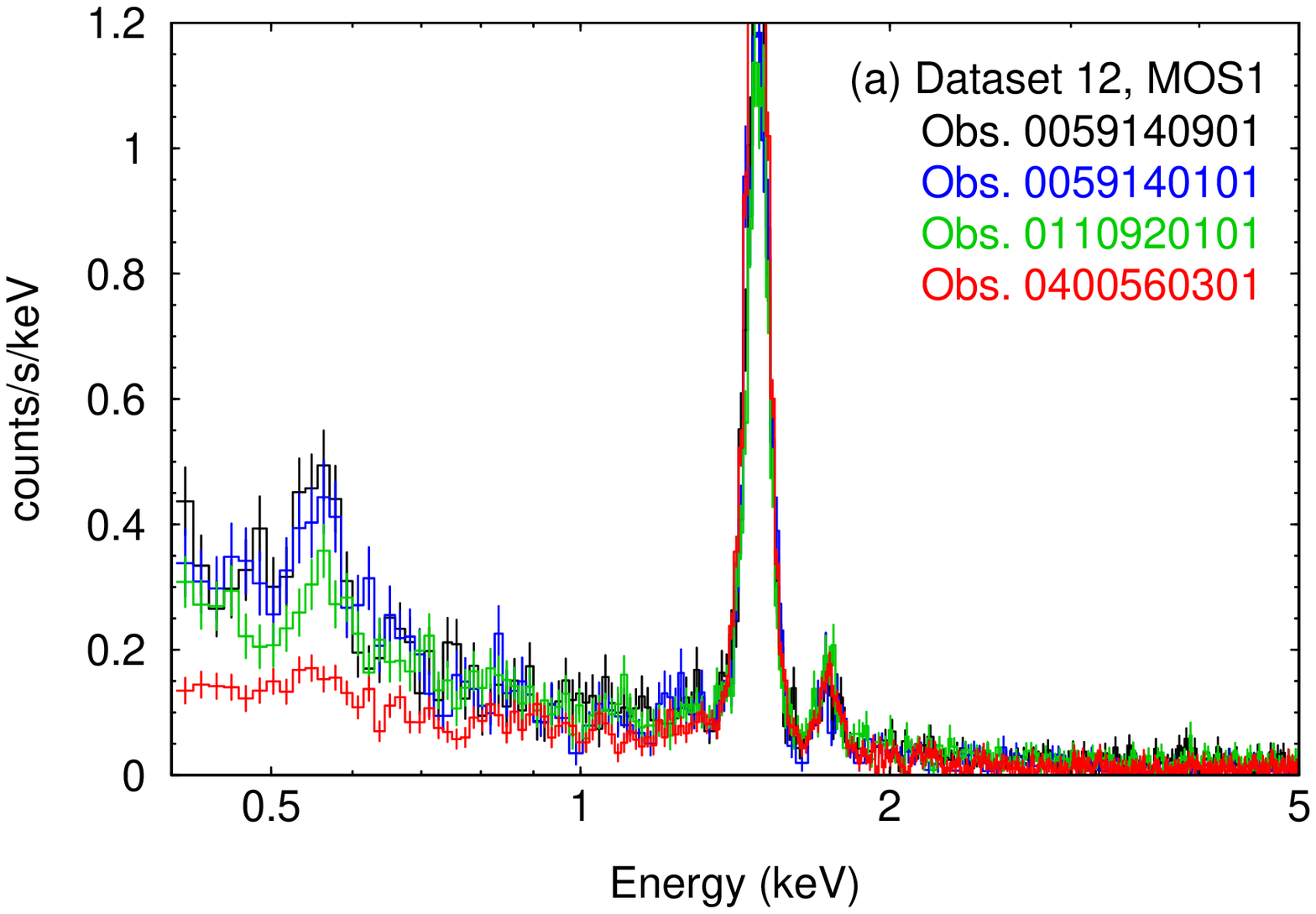}{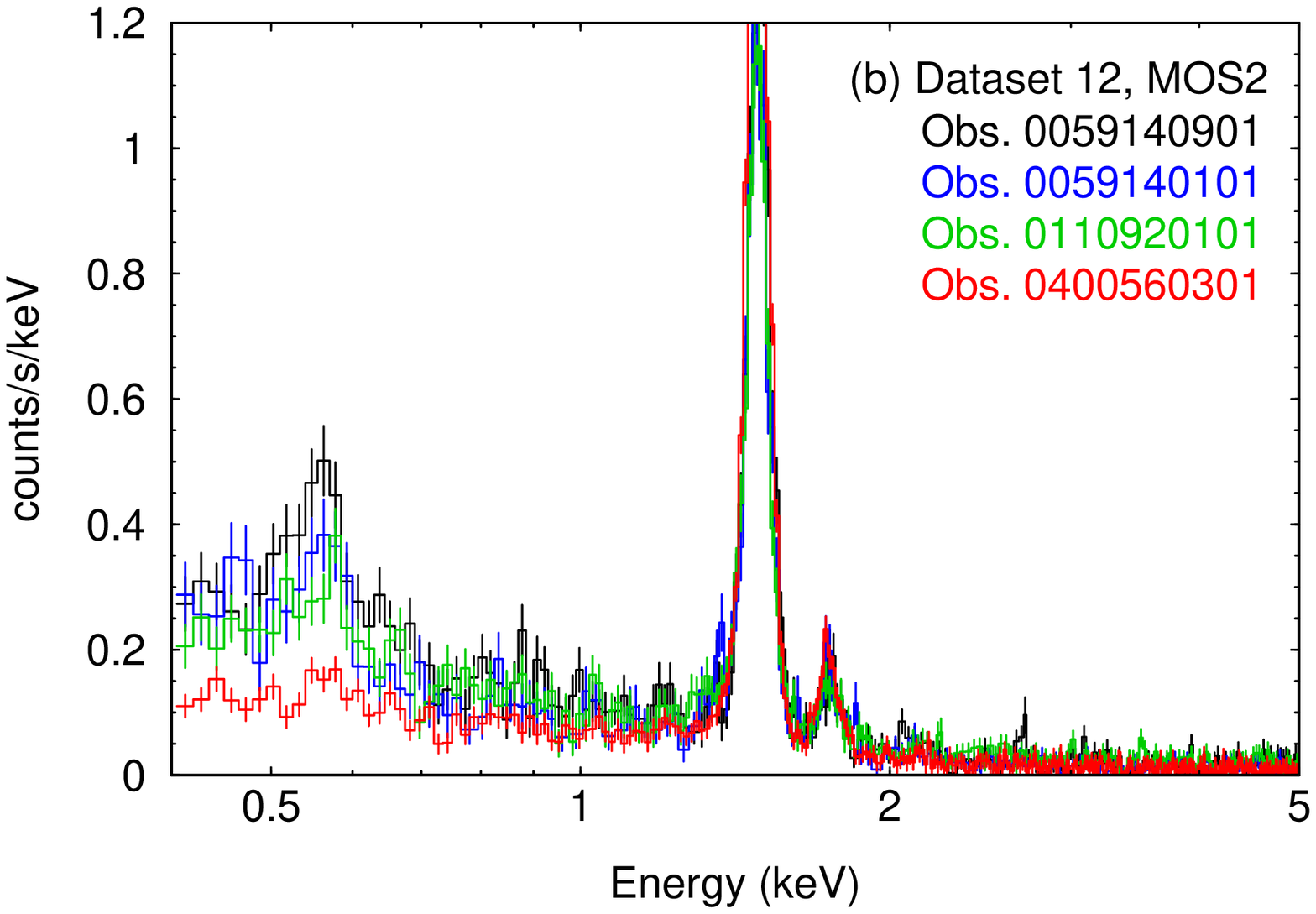}
\plottwo{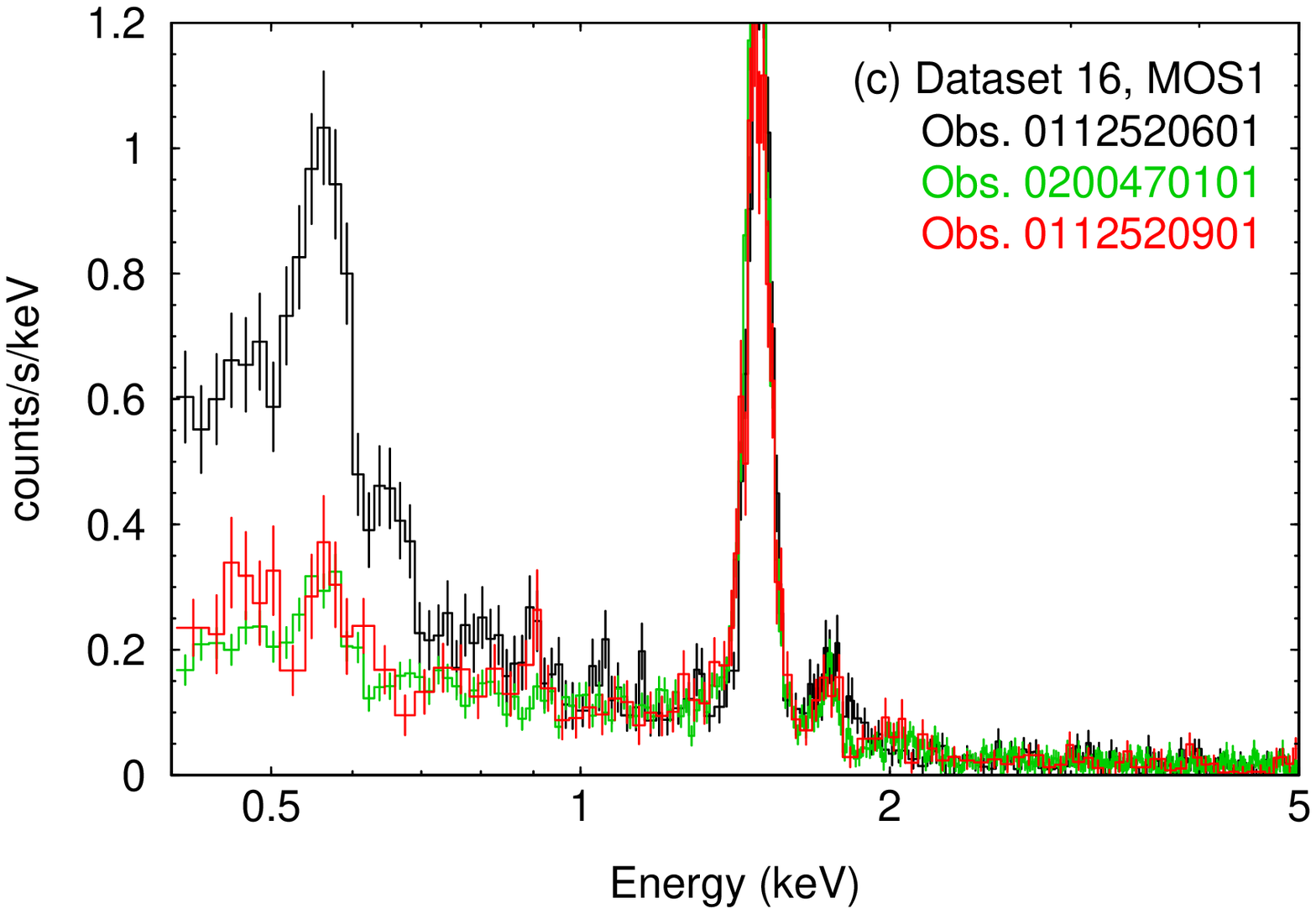}{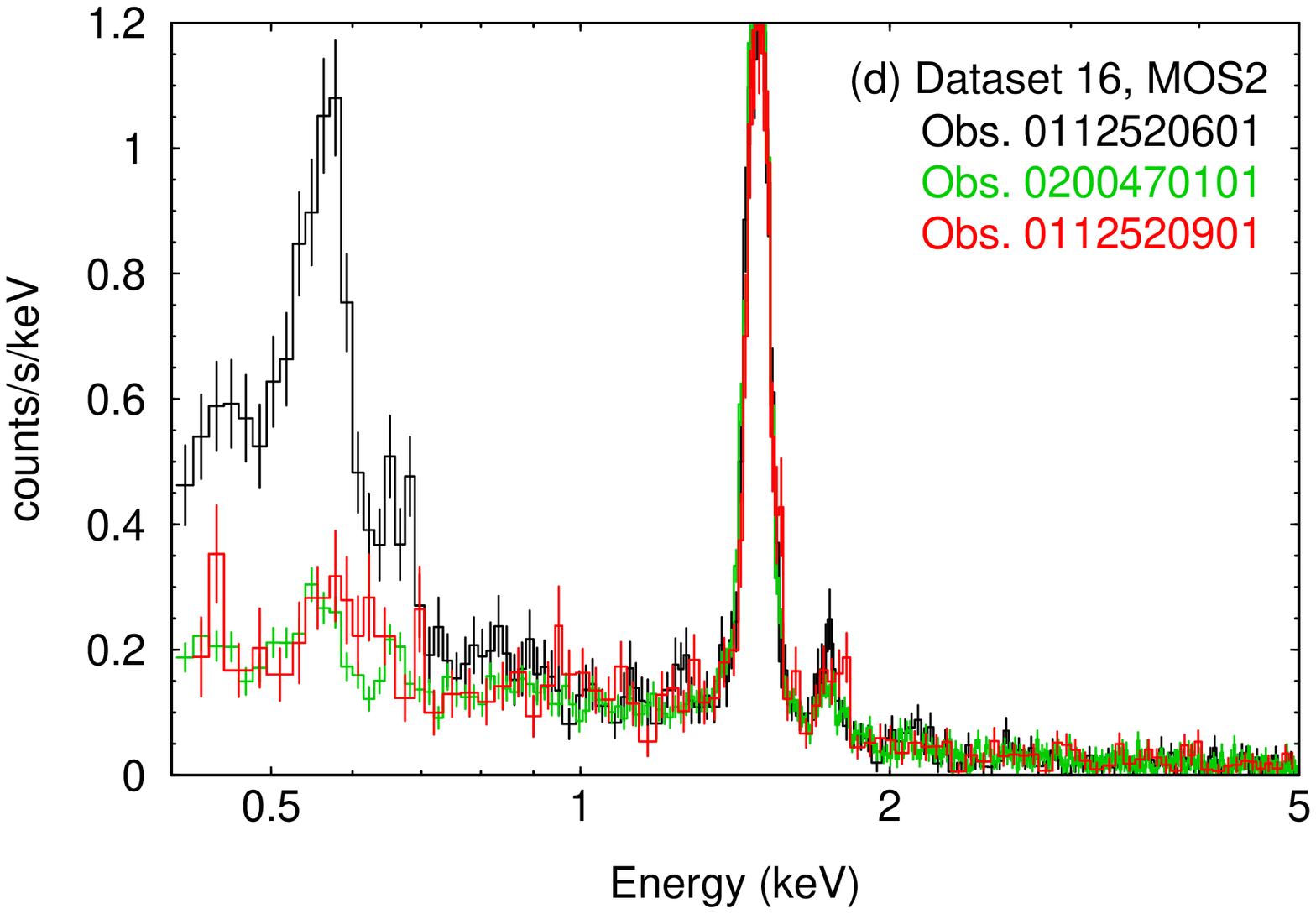}
\plottwo{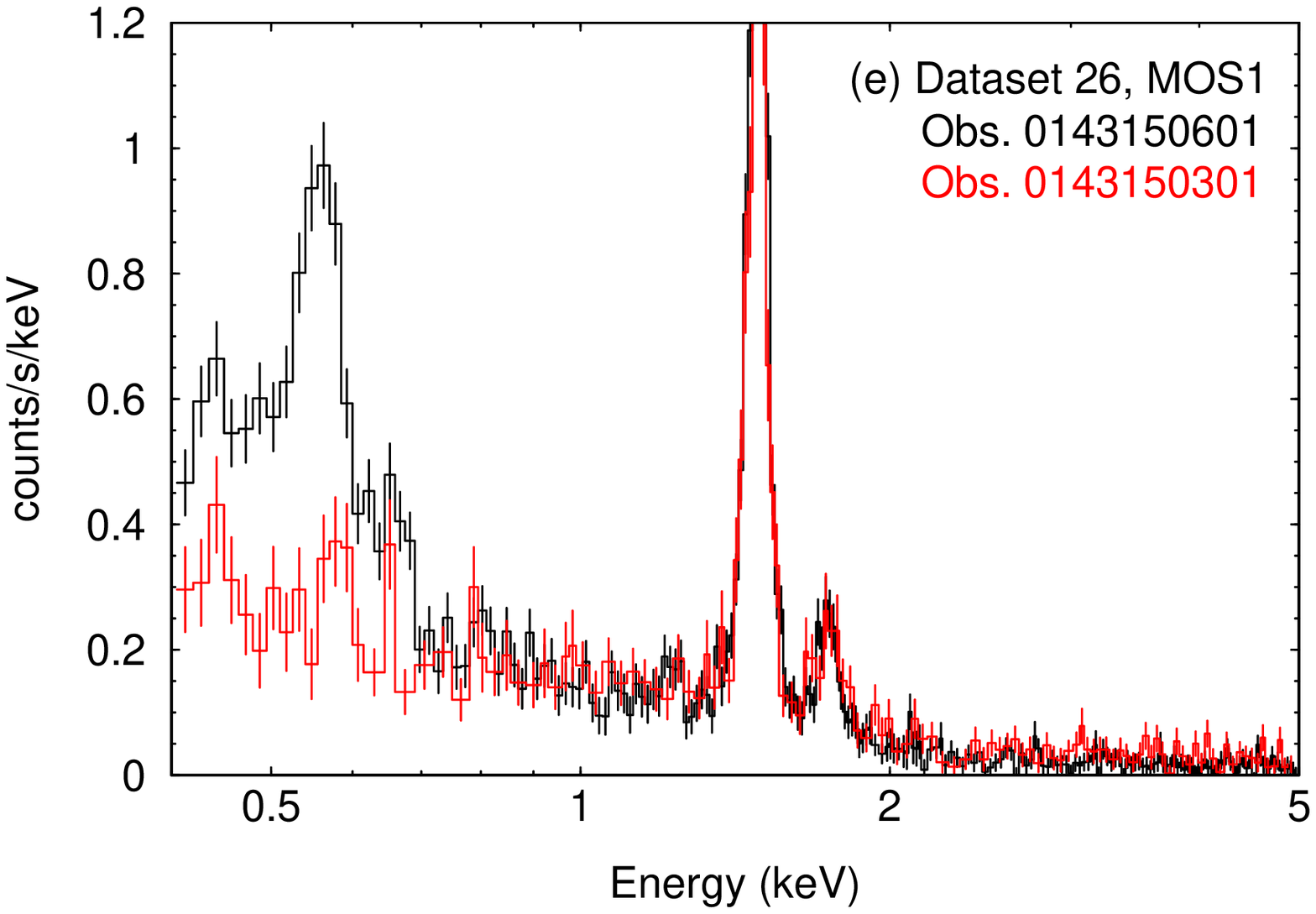}{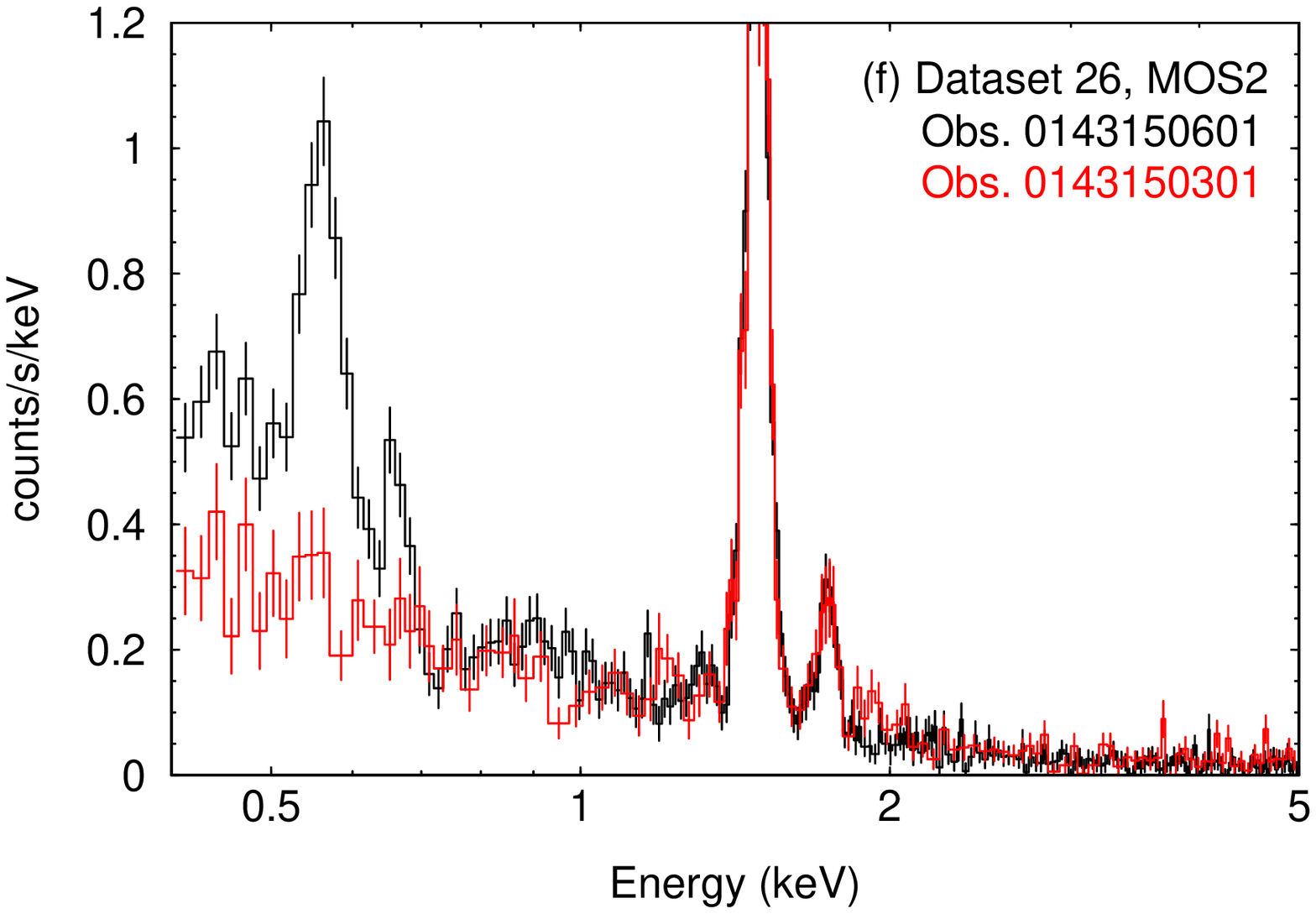}
\plottwo{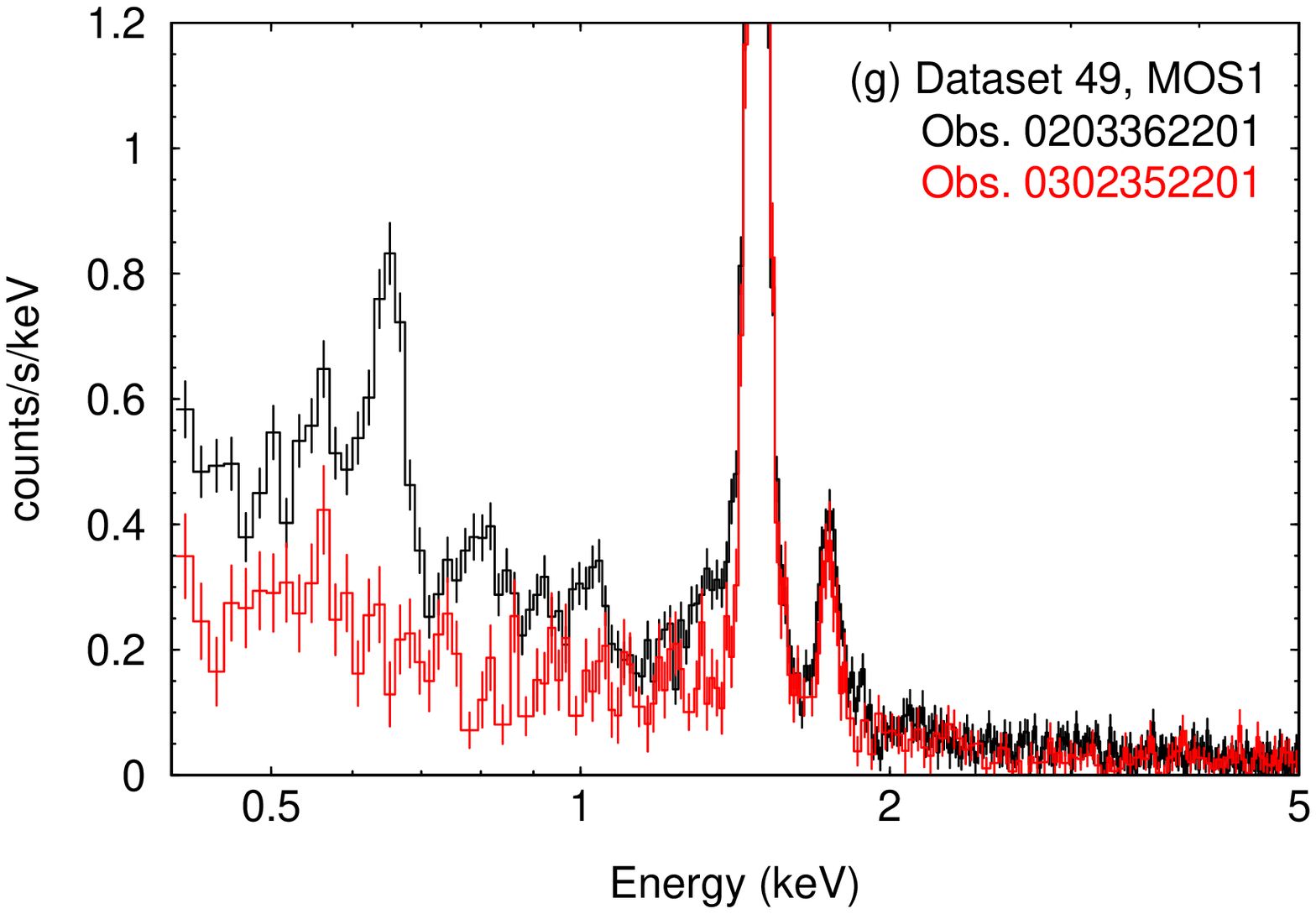}{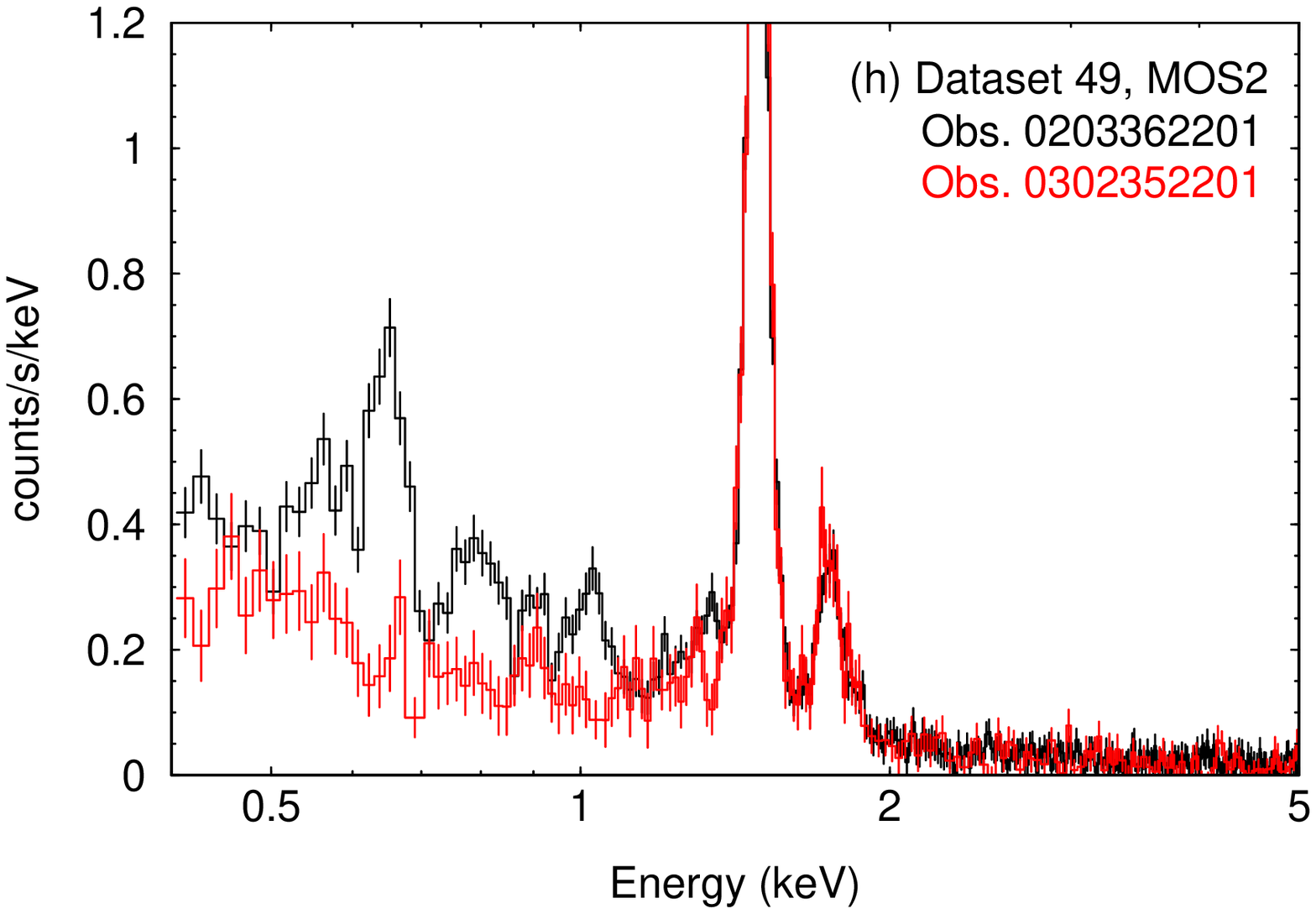}
\caption{MOS1 (left) and MOS2 (right) spectra from our survey that exhibit the strongest SWCX emission.
The spectra were extracted without the solar wind proton flux filtering described in Section~\ref{subsec:ProtonFluxFiltering}
(the observations exhibiting the brightest SWCX emission were unusable after this filtering).
The dataset number refers to the dataset numbers in Table~\ref{tab:MultipleObs}.
In each panel, the black spectrum exhibits the brightest \OVII\ (a--f) or \OVIII\
(g--h) emission, and the red spectrum the faintest \OVII\ or \OVIII\ emission. The
blue and green spectra, where plotted, are intermediate.
The \OVII\ and \OVIII\ lines are at $\sim$0.57 and $\sim$0.65~\kev. The bright lines at 1.49 and
1.75~\kev\ are the Al and Si instrumental lines.
\label{fig:SWCXSpectra}}
\end{figure*}

The histograms in Figure~\ref{fig:I-Imin} show that the measured enhancements due to SWCX are
typically $\la$4~\LU\ for \OVII\ and $\la$2~\LU\ for \OVIII. However, there are more extreme
enhancements. The largest measured enhancement for \OVII\ is 26~\LU, and there are two other
observations with enhancements that exceed 10~\LU.  These observations are from datasets 12, 16, and
26 in Table~\ref{tab:MultipleObs}. The spectra from these sets of observations are shown in
Figure~\ref{fig:SWCXSpectra}(a)--(f), and the brightest and faintest \OVII\ intensities for these
directions are shown in the first three rows of Table~\ref{tab:BrightSWCX}.  The brightest
\OVIII\ enhancement is 8~\LU, for dataset 49. This is the only \OVIII\ enhancement in our survey
that exceeds 5~\LU. The spectra are shown in Figure~\ref{fig:SWCXSpectra}(g)--(h), and the
intensities are in the final row of Table~\ref{tab:BrightSWCX}. It should be noted that these
extreme enhancements are only seen when we do not apply the proton flux filtering described in
Section~\ref{subsec:ProtonFluxFiltering}. When this filtering is applied, we find $\Iovii -
\min(\Iovii) < 4~\LU$ and $\Ioviii - \min(\Ioviii) < 2.5~\LU$.

The extreme \OVII\ enhancements shown in Figure~\ref{fig:SWCXSpectra} and Table~\ref{tab:BrightSWCX}
are particularly noteworthy, as they are much larger than most previously reported \OVII\ SWCX
enhancements ($\sim$3--7~\LU; \citealt{snowden04,fujimoto07,henley08a}). \citet{koutroumpa07} have
reported \OVII\ SWCX enhancements of up to 10~\LU\ for some observations of the Lockman Hole.
However, we do not present results for the three observations of the Lockman Hole exhibiting the
brightest \OVII\ enhancements, because we find that these observations are badly contaminated by
soft protons.  In particular, for obs.~0147510901 the XMM-ESAS software yielded no good time at all
for the MOS1 exposure, and obs.~0147510801 and 0147511101 failed our $\Ftotal / \Fexgal \le 2$
requirement (see Section~\ref{subsec:OxygenIntensitiesResults}).

\section{DISCUSSION}
\label{sec:Discussion}

The main purpose of this paper is to present the first measurements from our \xmm\ survey of the
SXRB. In Section~\ref{subsec:SystematicErrors} we discuss possible systematic errors which could be
affecting these measurements. We also discuss some of the implications of our results.
In Section~\ref{subsec:DiscussionMultiple} we discuss the results
obtained from directions with multiple observations, and the implications of these results for SWCX.
In Section~\ref{subsec:Halo} we look at the oxygen emission from the Galactic halo. We apply various
filters to our measurements in an attempt to minimize the SWCX contamination. However, because more
sophisticated methods for removing SWCX contamination are unavailable, and because we only have data
for one third of the sky, the results for the halo must be considered preliminary.

\subsection{Possible Systematic Errors}
\label{subsec:SystematicErrors}

In this section, we discuss possible systematic errors which could bias our intensity measurements.
In particular, in Section~\ref{subsubsec:CentralSource} we investigate possible contamination of our
SXRB spectra by photons in the wings of the point spread functions of bright sources. In
Section~\ref{subsubsec:SoftProton} we investigate if the residual soft proton contamination has a
systematic effect on our measurements. Because of this soft proton contamination, we had to fix the
normalization of the extragalactic background in our spectral analysis. In
Section~\ref{subsubsec:ExtragalacticNormalization} we investigate if the value we used for this
normalization (10.5~\pownorm\ at 1~\kev) significantly affects our results.

\subsubsection{Contamination from Bright Sources}
\label{subsubsec:CentralSource}

Although bright sources were removed from the \xmm\ observations, and although we tended to err on
the side of choosing larger source exclusion radii, it is possible that photons in the wings of the
\xmm\ point spread function could be contaminating our spectra of the SXRB. Bright sources with
non-thermal spectra should not be a problem. However, bright sources with thermal spectra, such as
stars, could contribute line emission photons to our SXRB spectra, potentially biasing our SXRB line
intensity measurements.

To investigate if thermal emission from bright sources affects our SXRB measurements, we selected observations
of stars for which the central source had been removed by hand (we chose only observations for which the
central source was the only object removed by hand). For these observations, we increased the radius of the
source exclusion region from its original value, and measured the oxygen intensities as a function of the
source exclusion radius. If contamination from the central source were a problem, we would expect the SXRB intensities
to decrease with increasing source exclusion radius.

\begin{figure*}
\centering
\plottwo{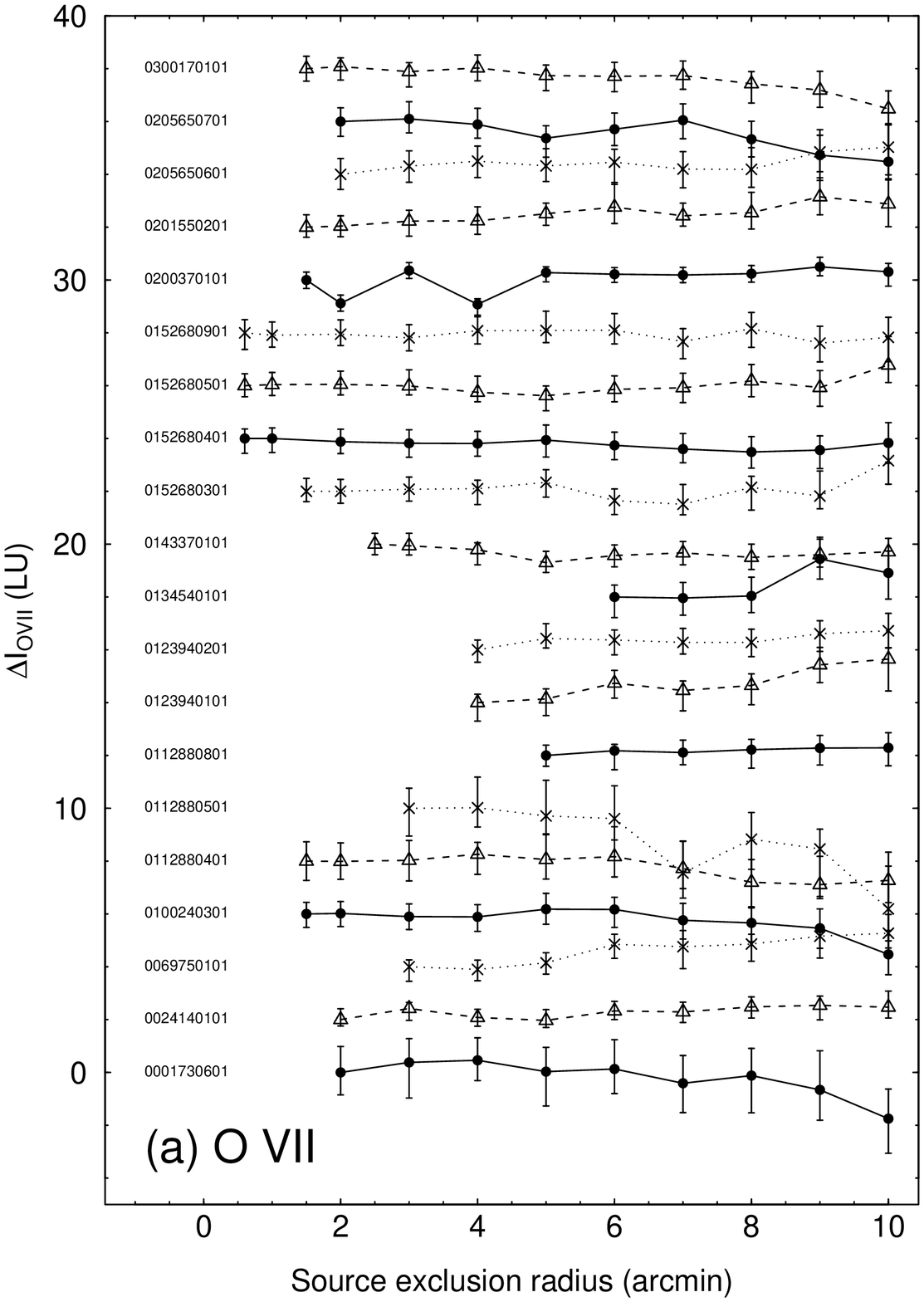}{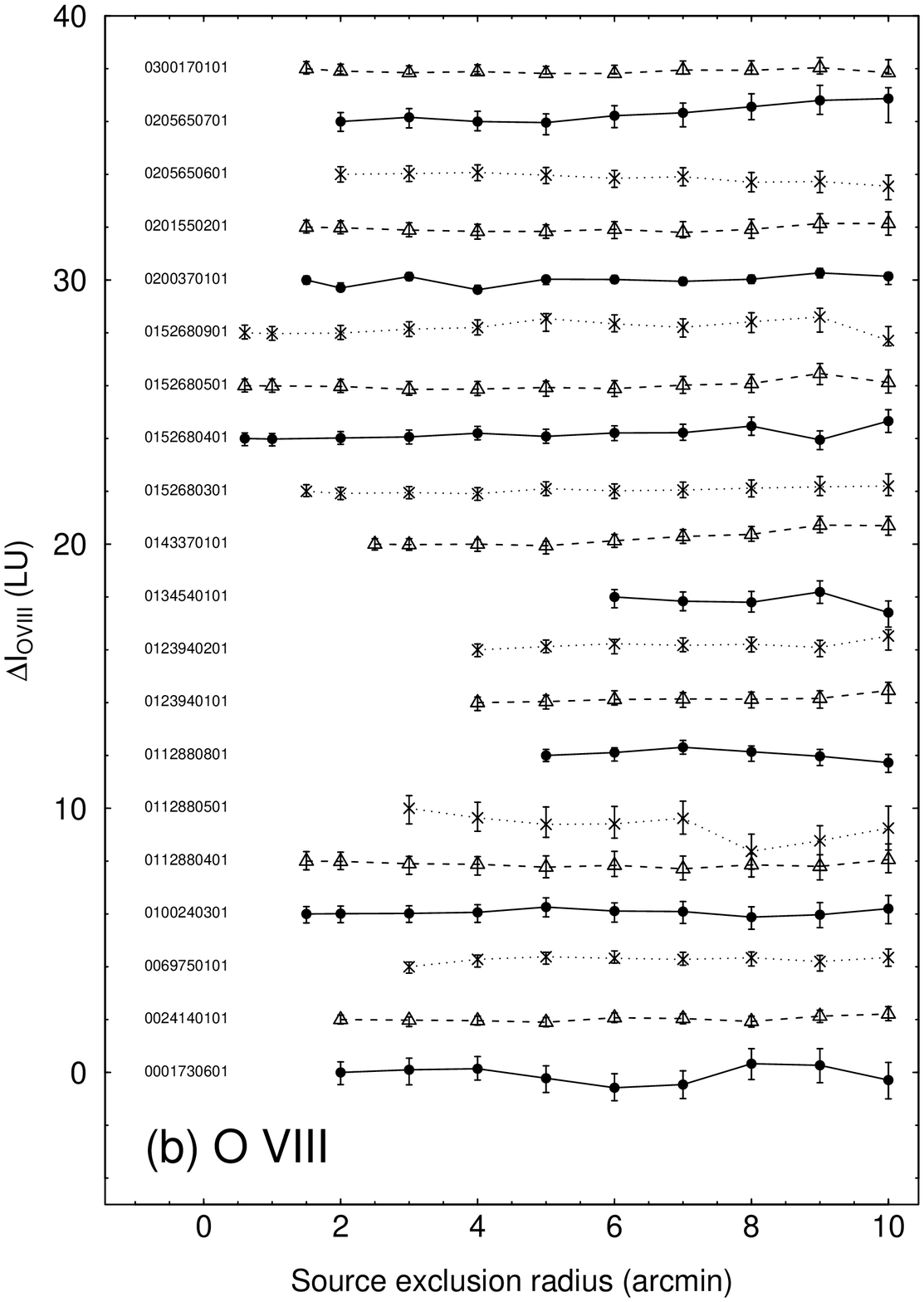}
\caption{(a) \OVII\ and (b) \OVIII\ intensities as a function of the radius used to exclude
  the central source.  For each observation, the intensities are plotted as the differences from the
  intensity measured using the original source exclusion radius. The curves have been shifted
  upwards by 0, 2, 4, ...~\LU\ for clarity.
\label{fig:ExclusionRadius}}
\end{figure*}

The results of this experiment are shown in Figure~\ref{fig:ExclusionRadius}. Although there is some
variation in the SXRB oxygen intensities as we increase the source exclusion radii, the intensities
do not systematically decrease as the source exclusion radii are increased. For each observation, and
for each of the two lines, we have used \chisq\ to test if the measured intensities as a function of
source exclusion radius are consistent with no variation from the intensity measured with the original
source exclusion radius. At the 5\%\ level, only the \OVII\ intensity from obs.~0200370101 shows significant
variation with source exclusion radius. However, as can be seen in Figure~\ref{fig:ExclusionRadius}(a), the variation
is non-monotonic, which is not what we would expect if the central source were contaminating the SXRB spectrum.
Furthermore, for this particular observation the central source is not bright. We therefore conclude that
our SXRB spectra are not significantly contaminated by thermal emission from bright sources.

\subsubsection{Soft Proton Contamination}
\label{subsubsec:SoftProton}

Despite the cleaning of the data described in Section~\ref{subsec:InitialProcessing}, some soft
proton contamination may remain in the spectra. We modeled this residual contamination as a broken
power-law in our spectral analysis. Here we wish to examine whether or not this contamination significantly
affects our intensity measurements.

To investigate the extent to which the soft proton contamination affects our measurements, we used
the results from directions with multiple observations. For a given direction, the variation in the
intensity is expected to be due to SWCX. However, if the presence of soft proton contamination
biases the intensity measurements, we would expect correlations between measures of the soft proton
contamination and $I - \min(I)$, where $\min(I)$ is the minimum intensity measured in the same direction
as the $I$ measurement (see Section~\ref{subsec:ResultsMultiple}).

\begin{figure}
\centering
\plotone{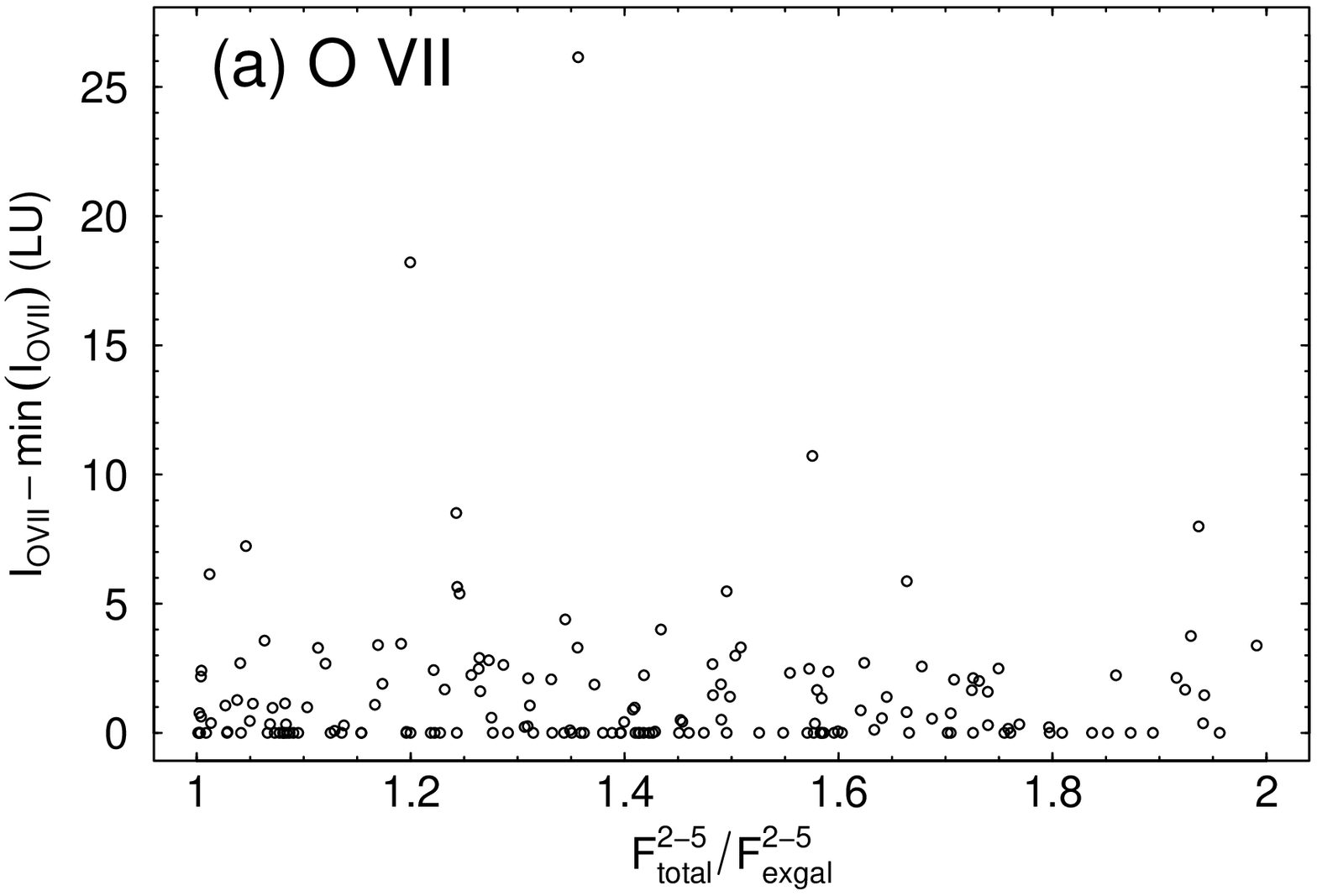}
\plotone{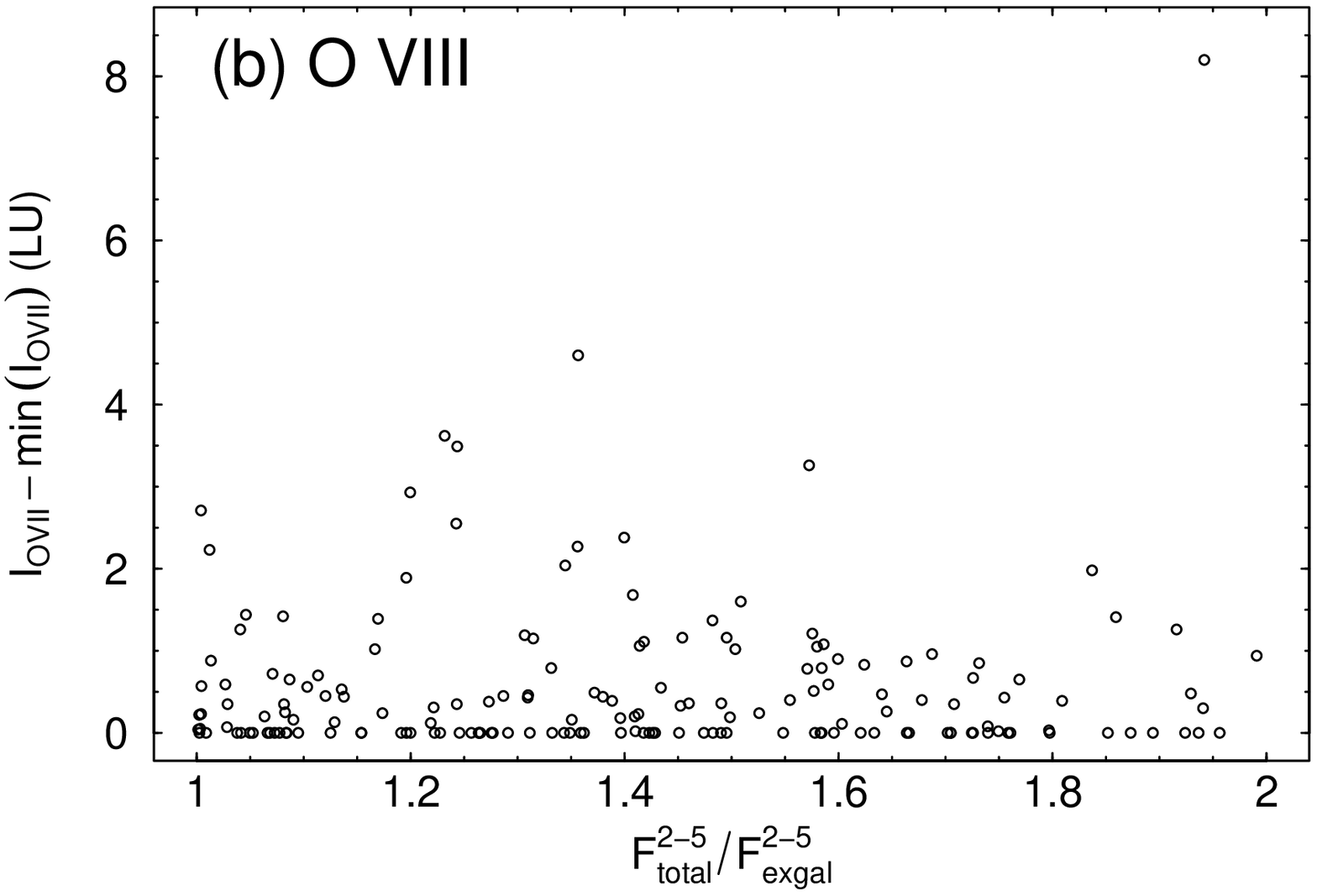}
\caption{(a) $\Iovii - \min(\Iovii)$ and (b) $\Ioviii - \min(\Ioviii)$ against $\Ftotal / \Fexgal$,
  where $\min(I)$ is the minimum measured intensity in the same direction as the $I$ measurement,
  and $\Ftotal / \Fexgal$ is a measure of the soft proton contamination.
  \label{fig:SoftProton}}
\end{figure}

Figure~\ref{fig:SoftProton} shows $\Iovii - \min(\Iovii)$ and $\Ioviii - \min(\Ioviii)$ against
$\Ftotal / \Fexgal$, which is a measure of the soft-proton contamination (see
Section~\ref{subsec:OxygenIntensitiesResults}). Using Kendall's $\tau$ \citep[e.g.,][]{press92} we find
there is no significant correlation between the oxygen intensity and the amount of soft proton
contamination. This statement is also true if we use other measures of the soft proton contamination, such
as the normalization of the broken power-law (at 1~\kev), or its spectral index below the break at 3.2~\kev.
Therefore, soft proton contamination does not seem to have a systematic effect on our intensity measurements.

\subsubsection{The Normalization of the Extragalactic Background}
\label{subsubsec:ExtragalacticNormalization}

As mentioned in Section~\ref{subsec:OxygenIntensitiesMethod}, we were unable to independently
constrain the normalization of the extragalactic background, because of the broken power-law that we
used to model the residual soft-proton contamination. We therefore had to assume a normalization -- we
used 10.5~\pownorm\ at 1~\kev\ \citep{chen97}. \citet{chen97} obtained this value
after removing a few bright sources with $\Fsource \ga 5 \times 10^{-14}~\flux$, and so we too
removed sources down to this flux level.

\citet{moretti03} present X-ray source counts in the 0.5--2.0~\kev\ band obtained from shallow wide-field
and deep pencil-beam surveys carried out with \rosat, \chandra, and \xmm. Using their results for the source
flux distribution, we find that sources with $\Fsource < 5 \times 10^{-14}~\flux$ contribute a total
0.5--2.0~\kev\ flux of $5.46 \times 10^{-12}~\flux~\mathrm{deg}^{-2}$. Assuming a power-law index of
1.46 \citep{chen97}, this corresponds to a normalization of 7.9~\pownorm\ at 1~\kev, in contrast to the
value of 10.5~\pownorm\ that we used.

\begin{figure}
\centering
\plotone{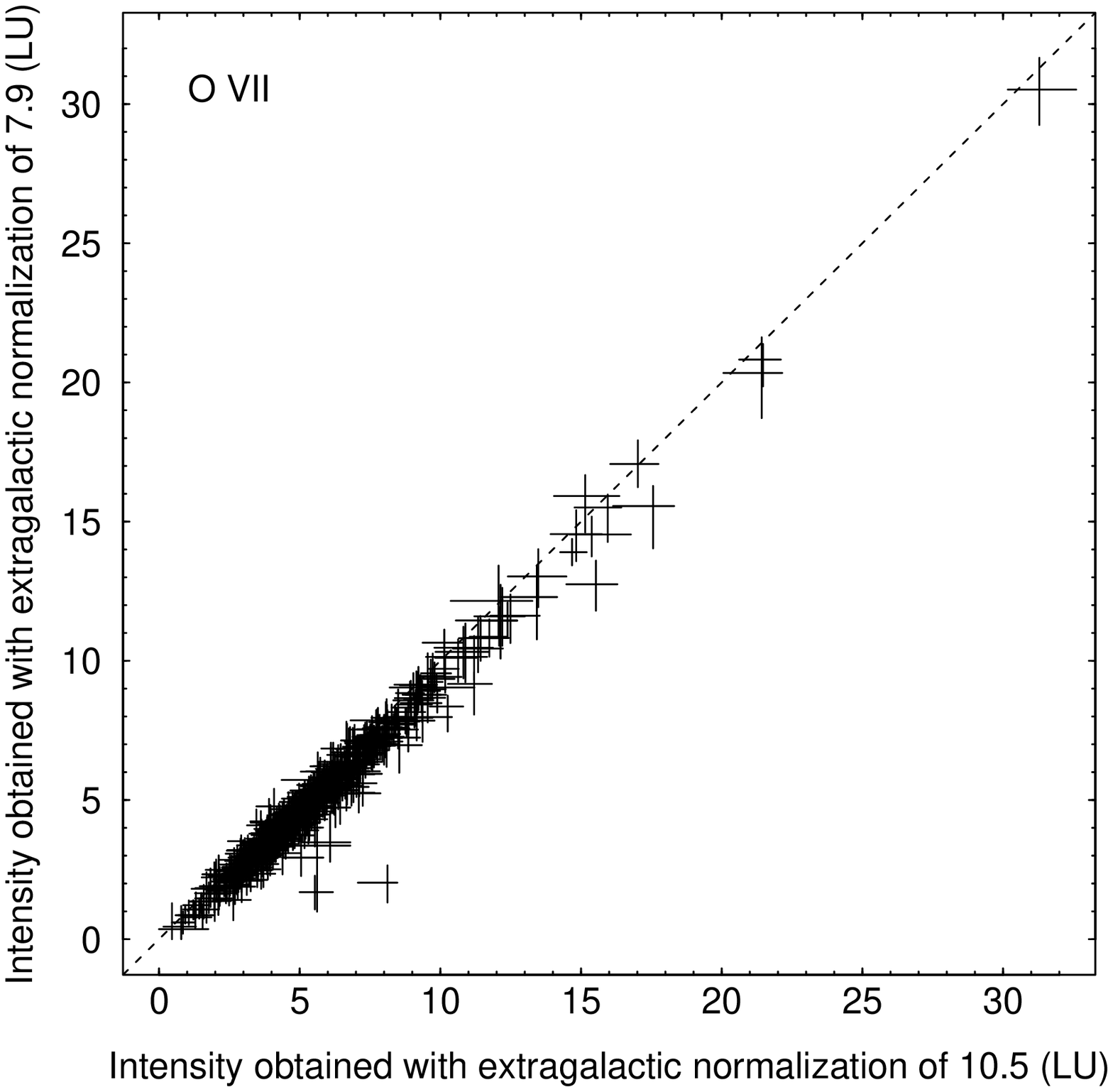}
\plotone{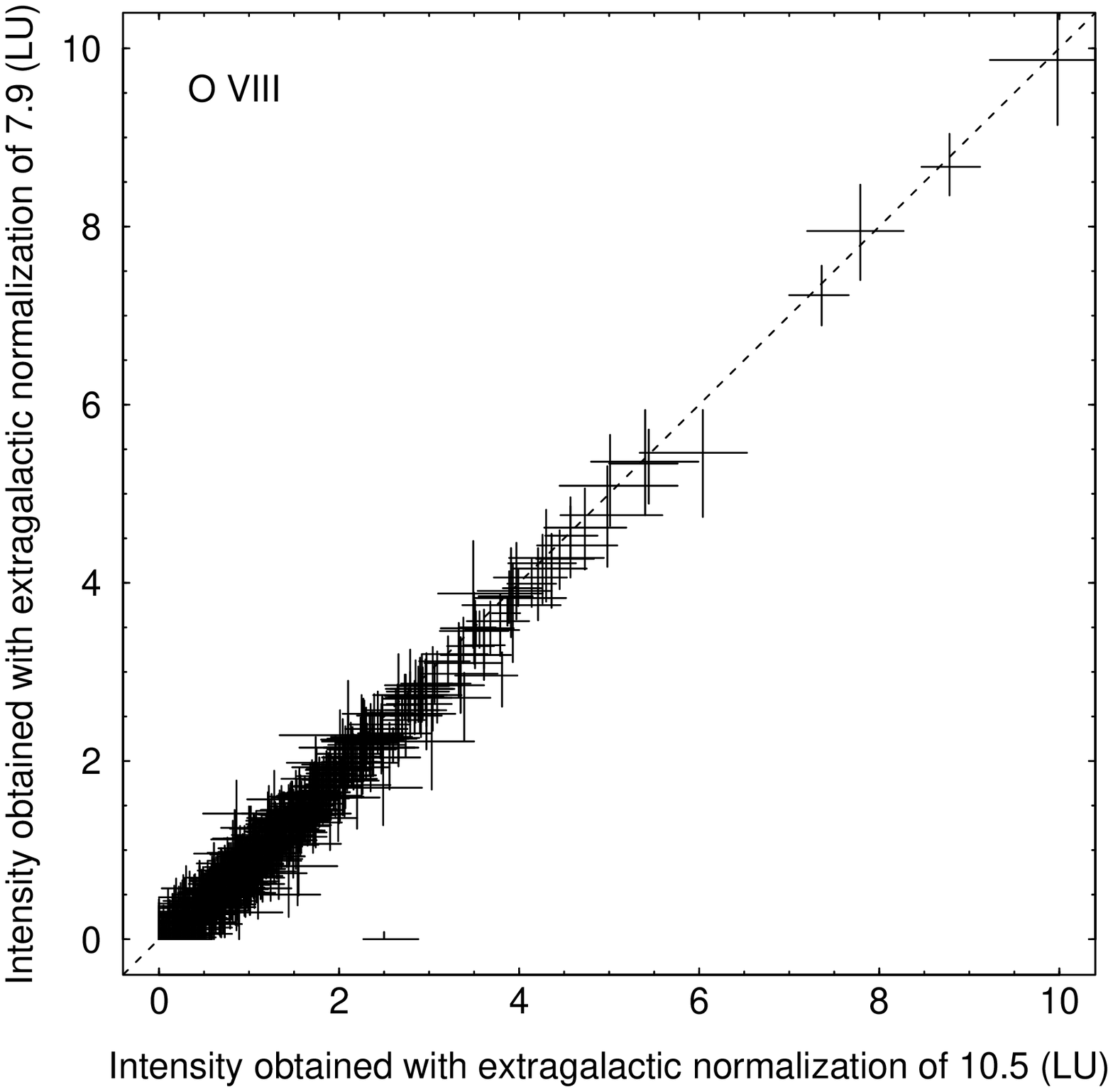}
\caption{Comparison of the oxygen intensities obtained with extragalactic normalizations of
10.5~\pownorm\ (\citealp{chen97}; abscissae) and 7.9~\pownorm\ (calculated using X-ray source counts from
\citealp{moretti03}; ordinates). The top panel shows the results for \OVII, and the bottom panel for
\OVIII. The dashed lines indicate equality.
\label{fig:CompareExgalNorm}}
\end{figure}

We examined the effect of our assumed extragalactic background normalization on our results by
repeating the measurements described in Section~\ref{subsec:OxygenIntensitiesMethod}, but this time
using an extragalactic background normalization of 7.9~\pownorm. The results obtained with the
two different extragalactic background normalizations are compared in Figure~\ref{fig:CompareExgalNorm}.

In general, using the lower normalization for the extragalactic background results in slightly lower oxygen intensities.
Lowering the extragalactic background normalization increases the normalization of the soft-proton broken power-law,
which in turn tends to decrease the intensities of the thermal emission components, including those of the oxygen
lines. However, it should be noted that the differences are generally not significant within the errorbars.
In addition, our general conclusions are not affected by these differences.

\subsection{Directions with Multiple Observations: Implications for Solar Wind Charge Exchange}
\label{subsec:DiscussionMultiple}

Multiple observations of individual directions are useful because they allow us to study the
variation in SWCX X-ray intensity. Such observed variations can be used to constrain models of SWCX
emission.  As was noted in Section~\ref{subsec:ResultsMultiple}, the measured \OVII\ and
\OVIII\ enhancements due to SWCX are typically $\la$4 and $\la$2~\LU, respectively (see
Figure~\ref{fig:I-Imin}) although some observations exhibit much larger intensity enhancements (see
Figure~\ref{fig:SWCXSpectra}). In this section, we discuss the variations in the oxygen intensities
seen in directions that have been observed multiple times, and the implications of these variations
for SWCX.

In a collisionally excited plasma, the brightest component of the \Kalpha\ emission from an He-like
ion is the resonance $\mbox{1s2p}\,^1\mathrm{P}_1 \rightarrow \mbox{1s}^2\,^1\mathrm{S}_0$
line. However, if the \Kalpha\ emission is produced by charge exchange, the lower-energy forbidden
$\mbox{1s2s}\,^3\mathrm{S}_1 \rightarrow \mbox{1s}^2\,^1\mathrm{S}_0$ line dominates
\citep[e.g.,][]{wargelin08}. For example, in a plasma in collisional ionization equilibrium with $T
\sim \mbox{few} \times 10^6~\K$, the \OVII\ forbidden line is roughly half as bright as the
resonance line (using line emissivity data from APEC), whereas the \OVII\ forbidden line yield from
charge exchange between O$^{+7}$ and He is $\sim$5 times the resonance line yield
\citep{krasnopolsky04}.  (Note that a recombining interstellar plasma would also produce a bright
forbidden line; e.g., see Figure 26 in \citealt{shelton99}.)  As the splitting between the
\OVII\ resonance and forbidden lines is 12.8~\ev\ (from APEC) and the energy bin size in the
\xmm\ RMF is 5~\ev, we might expect to see a shift in the \OVII\ centroid toward lower energies in
observations with brighter SWCX emission. Such a shift could potentially be used as a diagnostic of
SWCX contamination.

\begin{figure}
\centering
\plotone{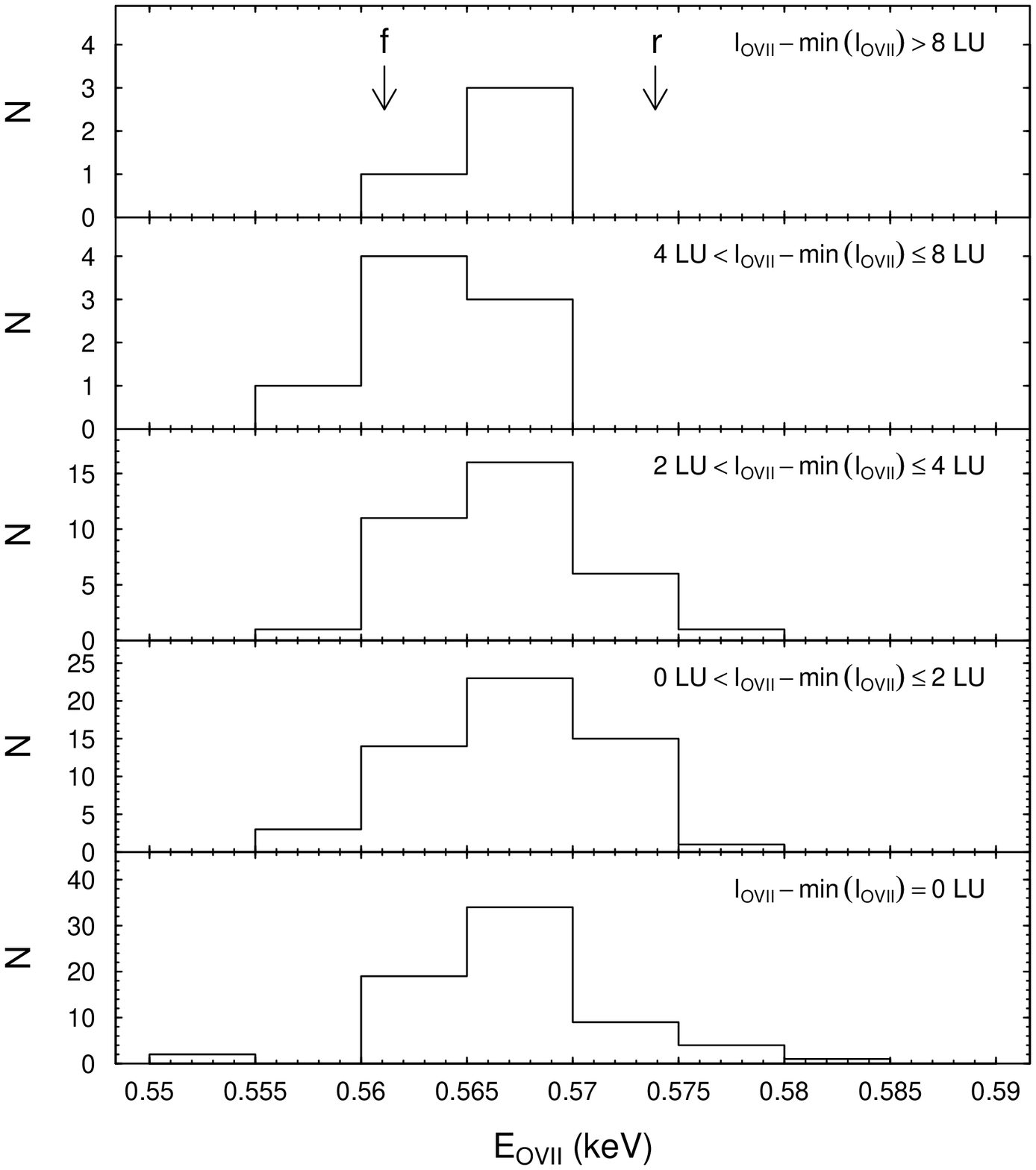}
\caption{Histograms of \Eovii\ (the energy of the \OVII\ centroid) for different ranges of $\Iovii -
  \min(\Iovii)$, where $\min(\Iovii)$ is the minimum measured intensity in the same direction as a
  given \Iovii\ measurement. The arrows mark the energies of the forbidden and resonance lines.
  \label{fig:Intensity-vs-Energy}}
\end{figure}

Figure~\ref{fig:Intensity-vs-Energy} shows histograms of the centroid energy of the \OVII\ emission,
\Eovii, for different ranges of $\Iovii - \min(\Iovii)$ (i.e., for different levels of enhanced SWCX
emission).  We take the instrumental gain shift between observations to have unobservably small
effects on the apparent line centroids, because we find no measurable variation in the \OVIII\ line
centroid energy in a sample of observations (probably partly due to the insensitivity of our
analysis to line shifts small than a few \ev; see below). There appears to be a shift in
\Eovii\ toward the energy of the forbidden line for $4~\LU < \Iovii - \min(\Iovii) \le 8~\LU$, but
not for $\Iovii - \min(\Iovii) > 8~\LU$. Therefore, enhancements in the \OVII\ intensity are not
clearly associated with shifts in the centroid energy toward that of the forbidden line, at least
not to the extent that \Eovii\ could be used as a diagnostic of SWCX contamination.  In fact,
\chisq\ tests show that all the histograms in Figure~\ref{fig:Intensity-vs-Energy}, except for the
$\Iovii - \min(\Iovii) = 0~\LU$ histogram, are consistent with a Gaussian distribution centered on
$\Eovii = 0.5675~\kev$ (roughly midway between the energies of the forbidden and resonance lines)
with a standard deviation of 5~\ev.

The lack of an observable shift toward the forbidden line energy in observations with enhanced
\OVII\ emission is most likely due to the uncertainty in \Eovii. Because we used a $\delta$ function
for the \OVII\ emission, the fit statistic (\chisq) is insensitive to changes in \Eovii\ within
an RMF energy bin (each of which is 5~\ev\ wide); i.e., the model intensity integrated between, say,
$E_1 = 0.565~\kev$ and $E_2 = 0.570~\kev$, and hence the corresponding model count spectrum, will be the
same no matter where \Eovii\ lies between $E_1$ and $E_2$, and so \chisq\ only changes when \Eovii\
moves from one RMF bin to the next. As a result, plotting \chisq\ as a function
of \Eovii\ does not result in a smooth parabola, but instead results in a stepped function with the
steps at the boundaries of the RMF energy bins. With such a \chisq\ curve, we find that the XSPEC
\texttt{error} command is generally unable to calculate the
uncertainty on \Eovii\ (which is why we do not quote errors for \Eovii\ in
Tables~\ref{tab:OxygenIntensities1} and \ref{tab:OxygenIntensities2}). However, using XSPEC's
\texttt{steppar} command to estimate the uncertainty on \Eovii, we find that the 90\%\ confidence
interval typically spans $\ga$10~\ev\ (i.e., similar to or greater than the splitting between the
\OVII\ resonance and forbidden lines). Therefore, it appears that we cannot measure \Eovii\ with
\xmm\ with sufficient precision to use \Eovii\ as a diagnostic of SWCX contamination. However, the
XIS cameras on \suzaku, which have a higher spectral resolution than \xmm's EPIC-MOS cameras, may be
able to detect a shift in the \OVII\ centroid toward the forbidden line energy in SWCX-contaminated
observations.

\begin{figure*}
\centering
\plotone{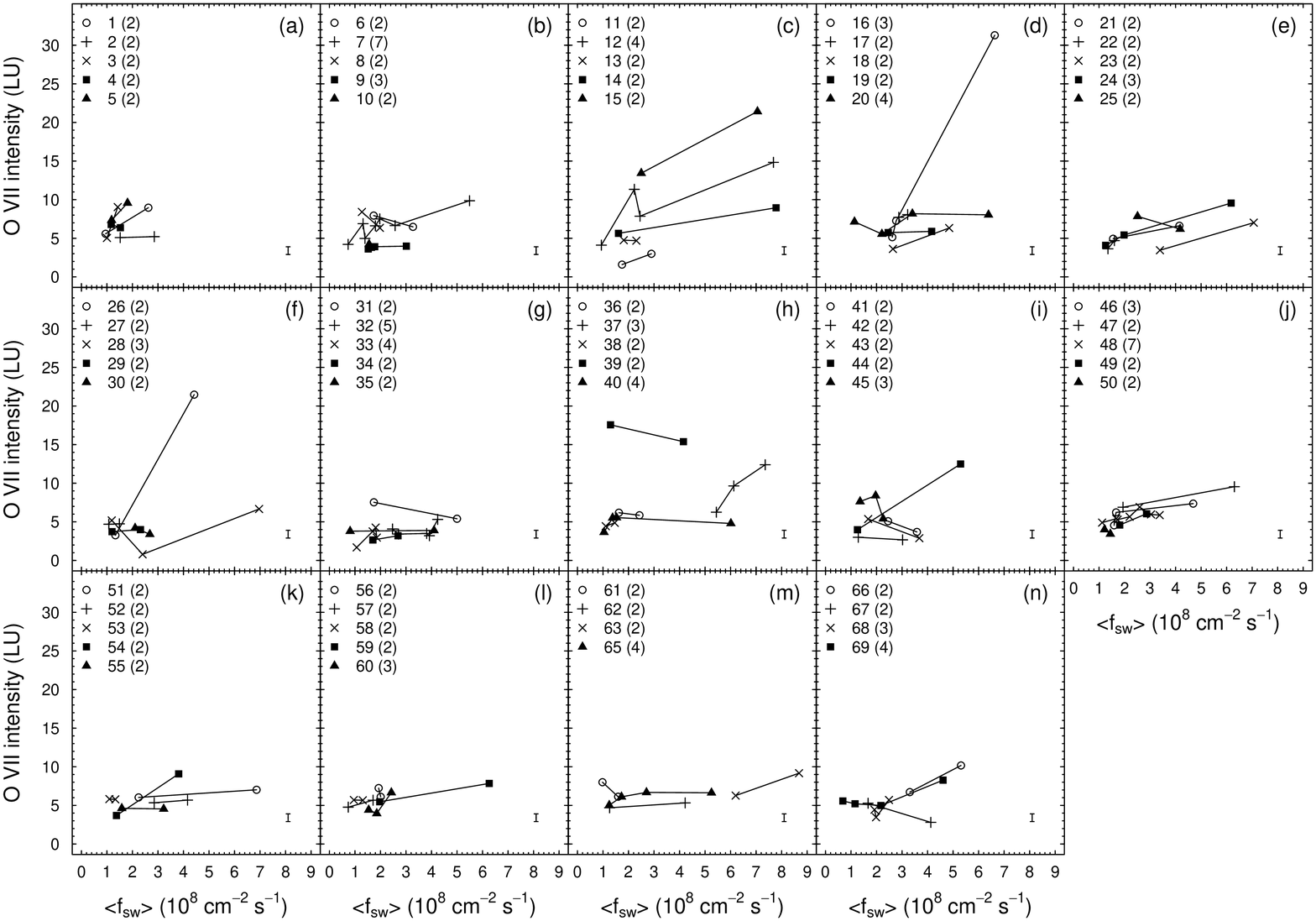}
\plotone{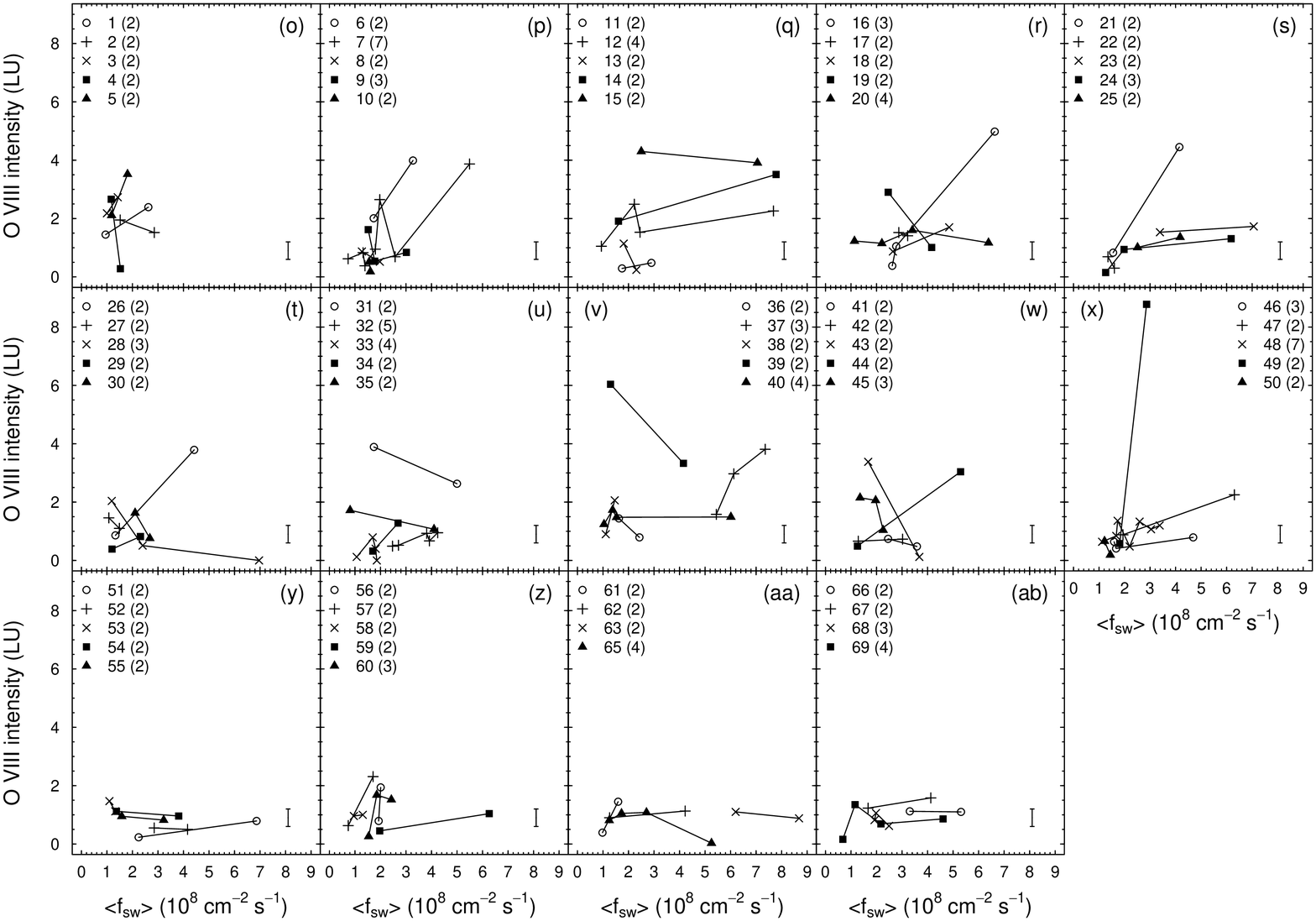}
\caption{Oxygen intensity versus average solar wind proton flux, for directions with multiple observations.
Each panels shows 4 or 5 sets of observations, in which each set consists of multiple coincident observations.
Panels~(a)--(n) show the \OVII\ intensity, and panels~(o)--(ab) the \OVIII\ intensity.
The intensities were obtained without the solar wind proton flux filtering described in Section~\ref{subsec:ProtonFluxFiltering}.
The bar in the lower-right corner of most panels indicates the typical error bar.
The numbers in the legends indicate the dataset number from Table~\ref{tab:MultipleObs} and, in parentheses, the number
of observations in that dataset with solar wind proton flux measurements.
The lines are used to join observations from the same dataset.
Observations without proton flux measurements are not plotted.
Note that dataset~64 is not plotted, as only 1 of the 2 observations has a proton flux measurement.
\label{fig:Oxygen-vs-Protonflux}}
\end{figure*}

In Section~\ref{sec:SWCX} we noted that times of enhanced SWCX emission have been observationally
associated with times of enhanced solar wind proton flux
\citep{cravens01,snowden04,fujimoto07,carter08,kuntz08a}. Such an association is at least partly
expected, as an increase in the solar wind proton flux striking the Earth will increase the SWCX
emission from the geocorona. Figure~\ref{fig:Oxygen-vs-Protonflux} shows how the oxygen intensities
vary with the average solar wind proton flux for 68 of the 69 sets of observations in
Table~\ref{tab:MultipleObs} (dataset~64 is not plotted, as only 1 of the 2 observations has a proton
flux measurement).

For many of the sets of observations plotted in Figure~\ref{fig:Oxygen-vs-Protonflux}, there is a
general trend that the oxygen intensity increases with the solar wind proton flux. However, this
trend is by no means universal; for example, for datasets~31 (panels~(g) and (u)), 39 (panels~(h)
and (v)), and 43 (panels~(i) and (w)), the intensity decreases with increasing solar wind proton
flux. Furthermore, some datasets show much stronger increases in SWCX intensity with increasing
solar wind proton flux than others; for example, compare datasets~16 (panel~(d)) and 49 (panel~(x))
with the other datasets in their respective panels.

The results in Figure~\ref{fig:Oxygen-vs-Protonflux} show that the solar wind proton flux alone is
not a good indicator of the amount of SWCX contamination in a SXRB spectra. This is not surprising
for a number of reasons. Firstly, the solar wind proton flux is measured in the vicinity of the
Earth, and so is insensitive to localized solar wind enhancements (such as coronal mass ejections)
moving across the line of sight far from the Earth \citep{koutroumpa07,henley08a}. Secondly,
a set of observations taken over several years may exhibit variation in the heliospheric
SWCX emission due to the solar cycle \citep{robertson03a,koutroumpa06}, which would be independent
of the near-Earth solar wind proton flux. Finally, although an increase
in the solar wind proton flux will tend to increase the overall brightness of the geocoronal
emission, the amount of geocoronal emission seen in a given observation will depend on which part
of the magnetosheath the sight-line passes through, with the brightest emission coming from the
sub-solar region \citep{robertson03b}. Because of \xmm's eccentric orbit, different observations of
the same direction can sample different parts of the magnetosheath \citep[e.g.,][]{kuntz08a}.

\begin{figure}
\centering
\plotone{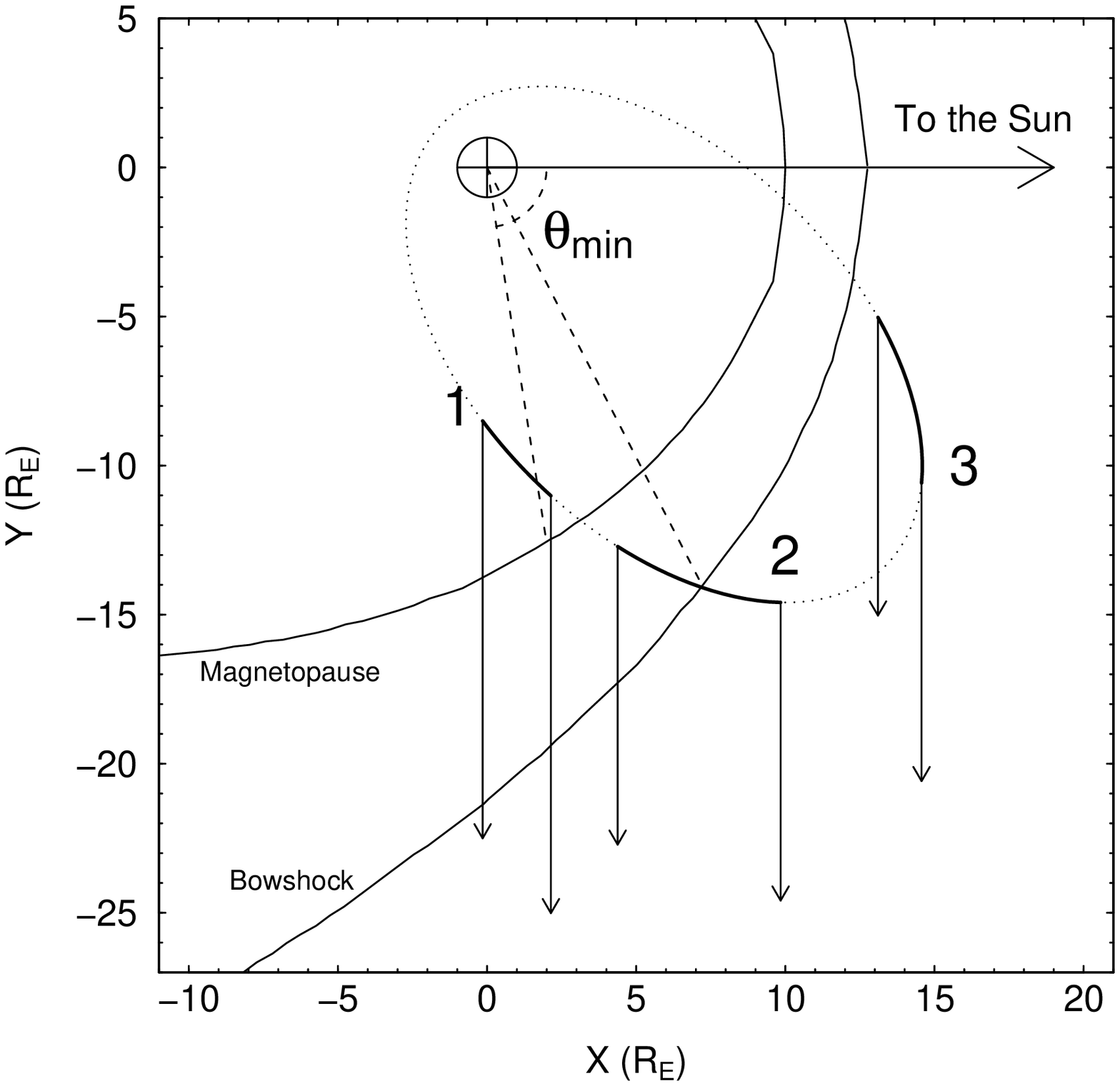}
\caption{Illustration of the \thetamin\ parameter for three different types of \xmm\ observation.
  The solid curves show the magnetopause and the bowshock (from \citealp{spreiter66}).  The dotted
  ellipse shows the \xmm\ orbit (note that the orbital orientation relative to the solar direction
  changes during the course of the year).  The solid sections of the ellipse show \xmm's position
  during three hypothetical observations of a direction indicated by the arrows.  For observation~1,
  \thetamin\ is defined throughout the observation.  For observation~2, \thetamin\ is defined for
  part of the observation.  For observation~3, \thetamin\ is undefined throughout the observation.
  See the text for more details.
  \label{fig:MagnetosheathDiagram}}
\end{figure}

We have investigated whether or not \xmm\ sightlines that pass close to or through the sub-solar
region of the magnetosheath lead to increased oxygen intensities. For each observation, we quantify
how close the \xmm\ sightline gets to the sub-solar region as follows. We use the orbital data file
to establish \xmm's position during the observation. For each time during the observation, we step
along the sightline, and for each point along the sightline that lies between the magnetopause and
the bowshock, we measure the Earth-centered angle $\theta$ between that point and the Earth-Sun
line. We use the minimum value of $\theta$, \thetamin, at each time during the observation to
quantify how close the \xmm\ sightline gets to the sub-solar region -- the smaller \thetamin\ is,
the closer the sightline is to the sub-solar region, and $\thetamin = 0\degr$ means that the sightline
crosses the Earth-Sun line in the magnetosheath. In general, \thetamin\ varies during the course
of an \xmm\ observation. Figure~\ref{fig:MagnetosheathDiagram} illustrates three different types
of observation relevant to this discussion. For observation~1, the sightline passes through the magnetosheath
throughout the observation, and so \thetamin\ is defined throughout. For observation~2, \thetamin\ is
defined for only part of the observation. For observation~3, the sightline never passes through the
magnetosheath, and \thetamin\ is undefined throughout the observation.

\begin{figure}
\centering
\plotone{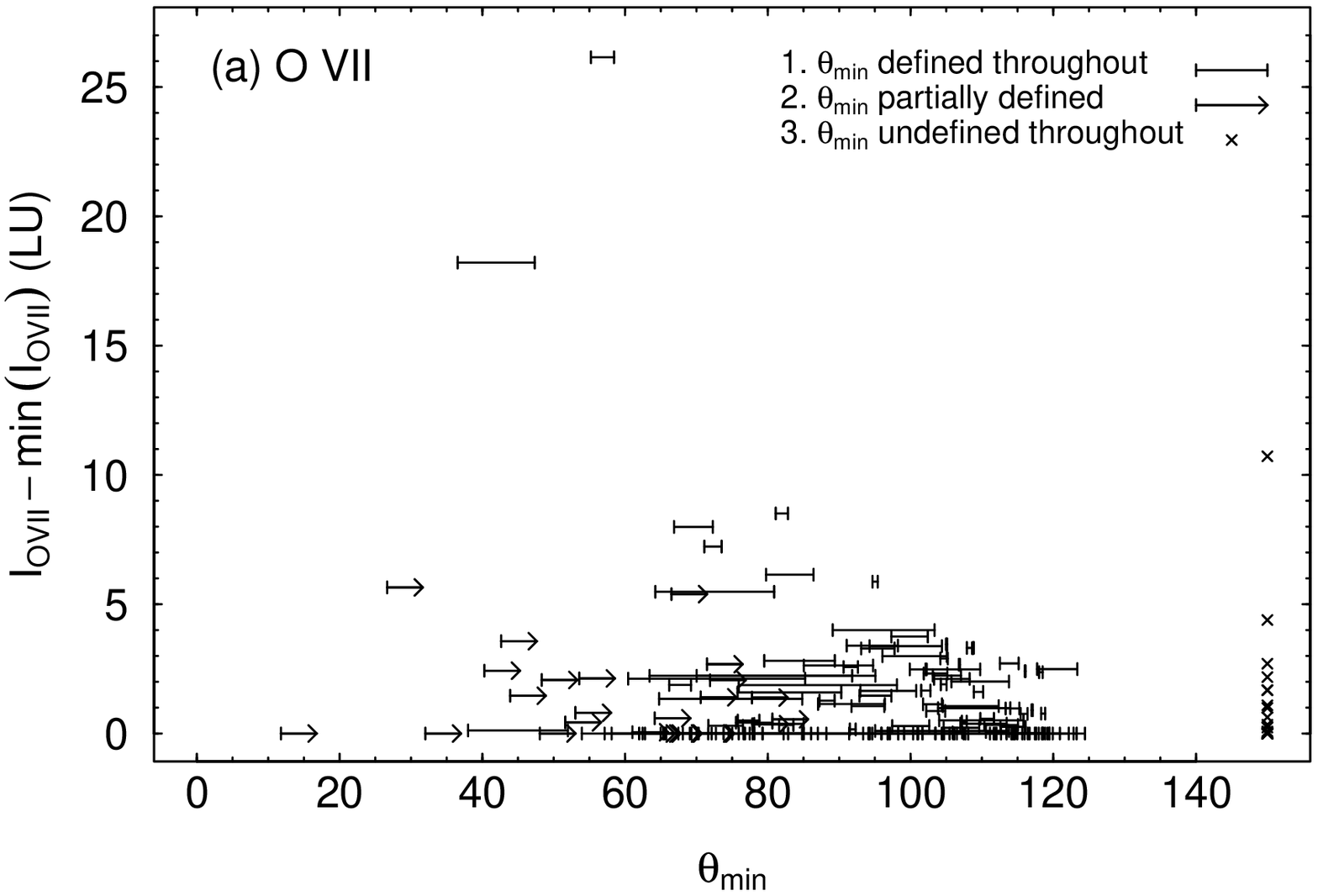}
\plotone{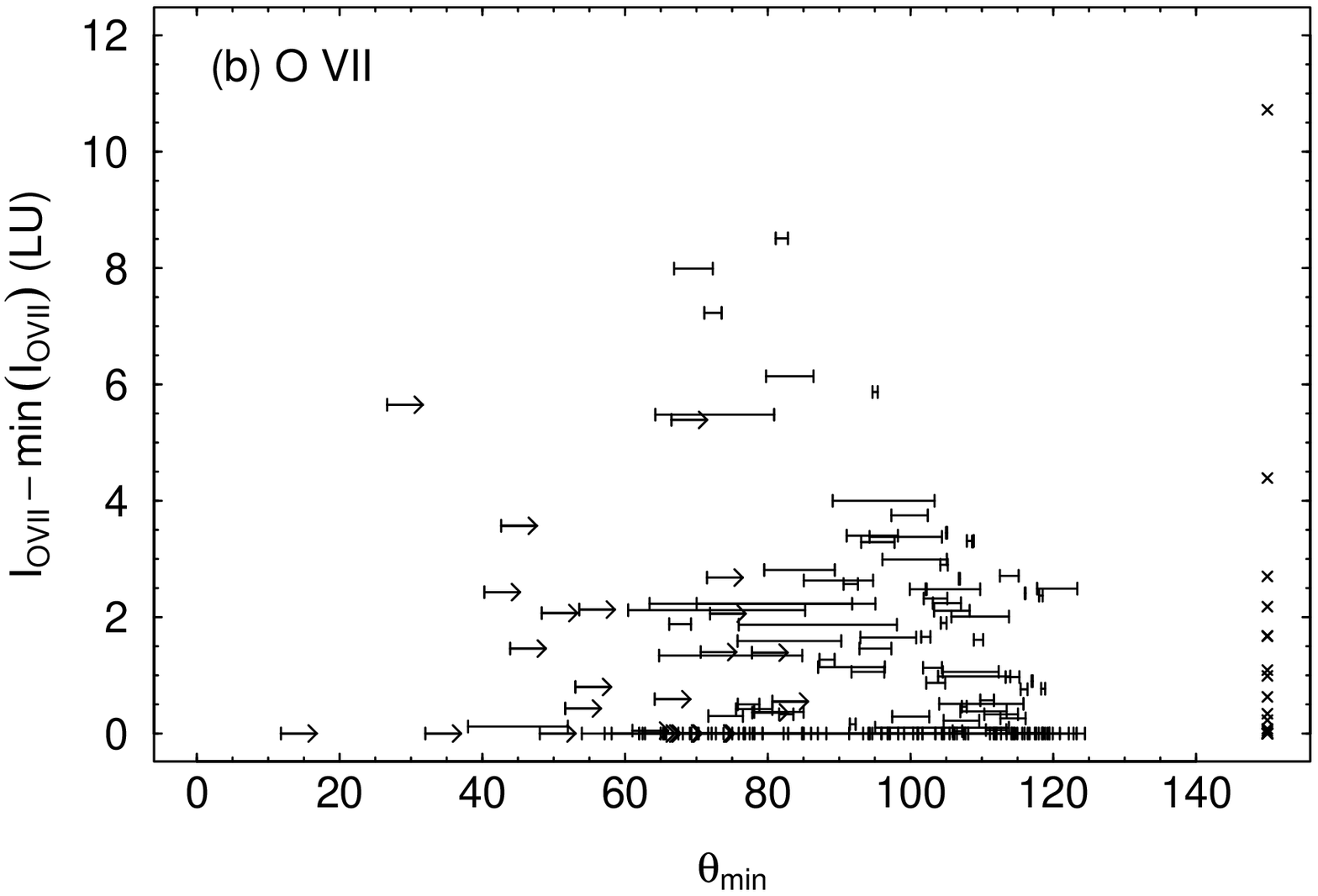}
\plotone{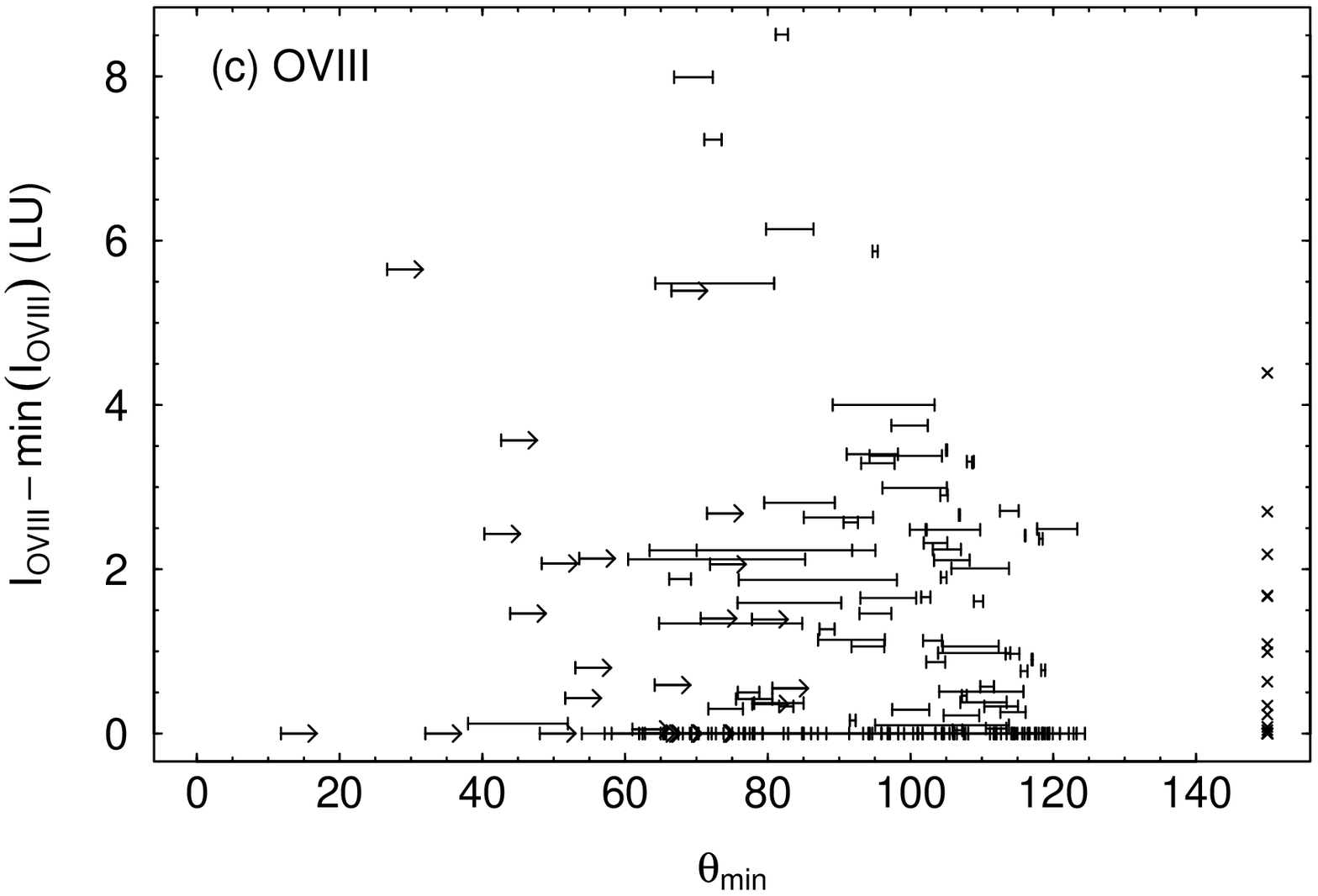}
\caption{$I-\min(I)$ against \thetamin\ for (a and b) \OVII\ and (c) \OVIII.  Panel (b)
  shows the same data as panel (a), but with a narrower $y$-axis range.  Observations for which
  \thetamin\ is defined throughout (observation~1 in Figure~\ref{fig:MagnetosheathDiagram}) are
  shown by horizontal bars indicating the range of \thetamin\ during the observation.  Observations
  for which \thetamin\ is defined for only part of the observation (observation~2 in
  Figure~\ref{fig:MagnetosheathDiagram}) are shown by arrows, the left-hand ends of which indicate
  the minimum value of \thetamin\ during the observation.  Observations for which \thetamin\ is
  undefined throughout (observation~3 in Figure~\ref{fig:MagnetosheathDiagram}) are shown by the
  crosses.
  \label{fig:IntensityMagnetosheath}}
\end{figure}

Figure~\ref{fig:IntensityMagnetosheath} shows $I - \min(I)$ for \OVII\ and \OVIII\ plotted against
\thetamin. As noted in the figure caption, the different symbols indicate the different types of
observation illustrated in Figure~\ref{fig:MagnetosheathDiagram}. Apart from the two observations
exhibiting the brightest \OVII\ enhancements, which have smaller-than-typical values of \thetamin,
there is no clear tendency for sightlines that pass closer to the sub-solar region of the magnetosheath
to produce larger oxygen intensity enhancements. Our results indicate that a single factor such as
the solar wind proton flux or the closeness of the sightline to the sub-solar region is usually not
sufficient for determining if an observation is likely to be SWCX contaminated.

\subsection{Oxygen Emission from the Galactic Halo}
\label{subsec:Halo}

In order to study the emission from the Galactic halo, we must first remove the foreground emission.
In Section~\ref{subsubsec:ForegroundRemoval} we apply various filters to our observations in order
to reduce the SWCX contamination. We then use \rosat\ shadowing data \citep{snowden00} to model the
foreground emission (due to SWCX and/or the LB) that remains after this filtering. We subtract this
foreground emission and calculate deabsorbed halo intensities. In Section~\ref{subsubsec:PlaneParallel}
we compare the halo intensities with a simple plane-parallel model for the halo, and in
Section~\ref{subsubsec:OxygenRatio} we look at the \OVII/\OVIII\ intensity ratio.

\subsubsection{Removing the Foreground Emission}
\label{subsubsec:ForegroundRemoval}

\begin{figure}
\centering
\plotone{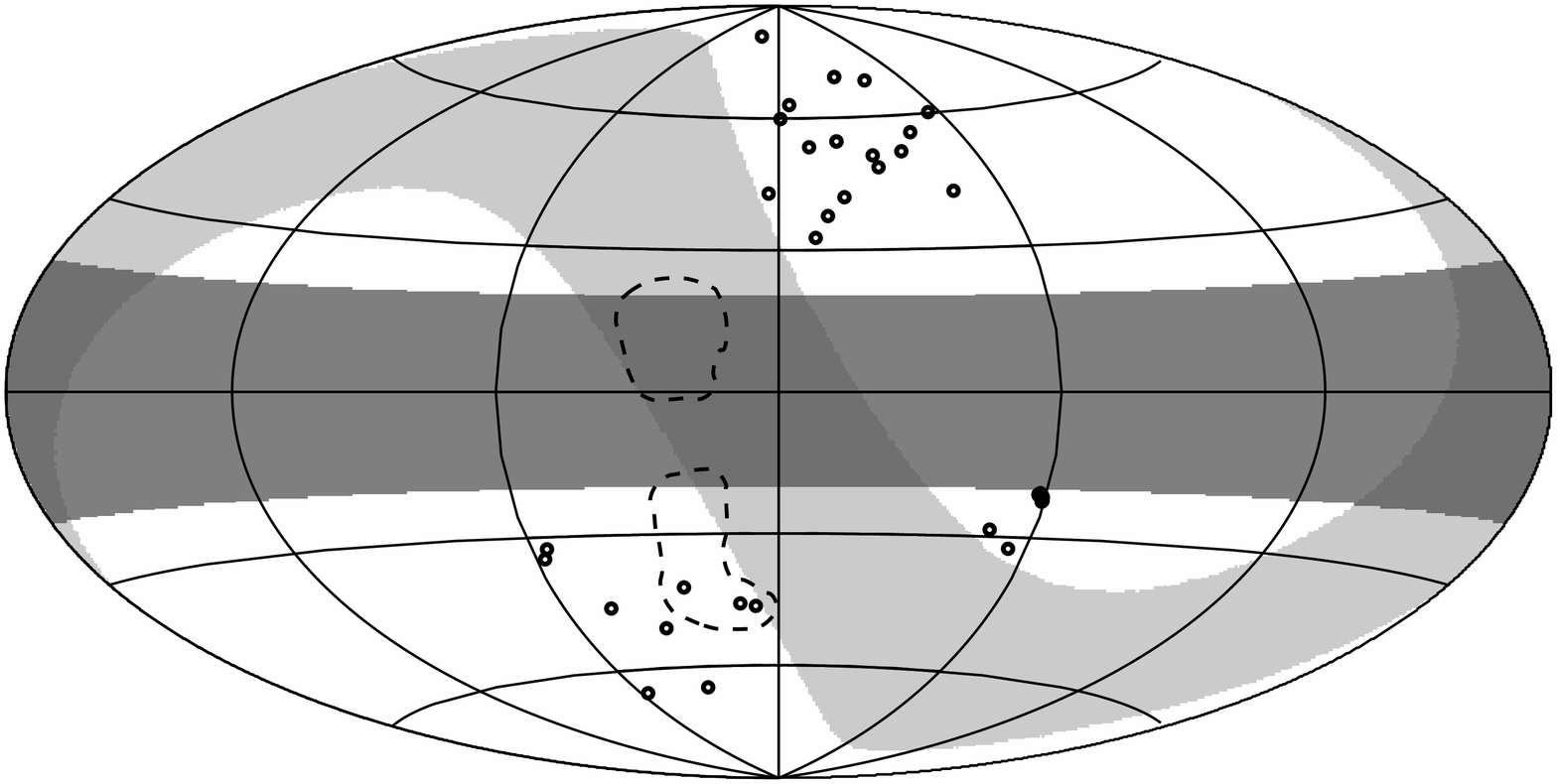}
\caption{All-sky Hammer-Aitoff projection in Galactic coordinates, centered on the Galactic
  Anticenter, showing the 39 observations that remain after the filters described in
  Section~\ref{subsec:Halo} are applied. The light gray band shows the region with ecliptic latitude
  $|\beta| \le 20\degr$, while the dark gray band shows the region with Galactic latitude $|b| \le
  20\degr$. The dashed lines outline two X-ray-bright regions: the Monogem Ring in the north and the
  Eridanus Enhancement in the south.
  \label{fig:ObservationLocation}}
\end{figure}

To reduce the SWCX contamination, we used only the results obtained with the proton flux filtering
described in Section~\ref{subsec:ProtonFluxFiltering}. As was noted in that section, the proton flux
filtering will only help reduce contamination from geocoronal SWCX emission and heliospheric SWCX
emission produced near the Earth. We therefore applied additional filters to our data to help further
reduce the heliospheric SWCX contamination. In particular, we removed observations of low ecliptic
latitudes ($|\beta| \le 20\degr$) and observations taken during high solar activity, as these
observations are expected to be more strongly contaminated by heliospheric SWCX (see
Section~\ref{sec:SWCX}). Although the transition from solar maximum to solar minimum is gradual, we
have taken 00:00UT on 2005 Jan 01 ($\mathrm{MJD} = 53371$) as the boundary between high and low
solar activity.\footnote{This date was estimated from sunspot data obtained from the National
  Geophyical Data Center (http://www.ngdc.noaa.gov/stp/SOLAR/ftpsunspotnumber.html).} At the time of
writing, we are still at solar minimum, so all observations with $\mathrm{MJD} > 53371$ are
considered to be at low solar activity. After we applied all these filters, just 43
observations remained. As we wish to study the halo, we removed a further 4 observations with $|b| \le
20\degr$.  The locations of the remaining 39 observations on the sky are shown in
Figure~\ref{fig:ObservationLocation}. Note that the region with $|\beta| \le 20\degr$ cuts
diagonally across the region with $120\degr \le l \le 240\degr$, so both Galactic hemispheres are
approximately equally sampled.

\begin{figure}
\centering
\plotone{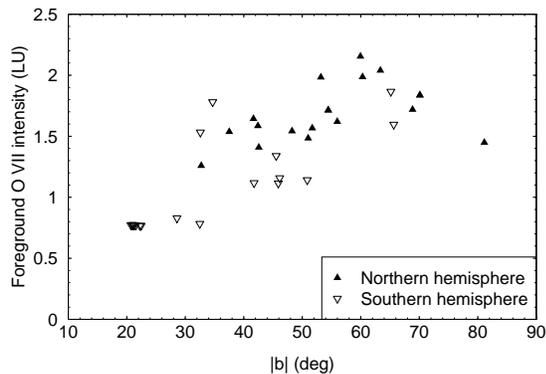}
\caption{Foreground \OVII\ intensities, calculated using SXRB shadow data from \citet{snowden00},
  against Galactic latitude.
\label{fig:ForegroundOxygen}}
\end{figure}

Despite the above filtering, some foreground oxygen emission (either from SWCX or the LB) may have remained
in our spectra. We modeled this foreground emission using the \citet{snowden00} catalog of SXRB
shadows.  This catalog contains foreground and background R12 (1/4~\kev) count-rates for 378 shadows
in the \rosat\ All-Sky Survey. For each of our \xmm\ observing directions, we found the 5 nearest
shadows in the catalog, and averaged their foreground count-rates, weighted by the inverses of their
distances from the \xmm\ pointing direction; i.e.,
\begin{equation}
  \mbox{Average foreground R12 count-rate} = \frac{\sum_i R_i / \theta_i}{\sum_i 1  / \theta_i},
  \label{eq:R12ave}
\end{equation}
where $R_i$ is the foreground R12 count-rate for the $i$th shadow, whose center is at an angular
distance $\theta_i$ from the \xmm\ pointing direction. Using a \raymondsmith\ model with
$T=10^{6.08}$~\K\ \citep{snowden00} to model the foreground emission, we converted the foreground
count-rates calculated above to emission measures.\footnote{We used a Raymond \& Smith model for this
  purpose because APEC is inaccurate in the 1/4~\kev\ band; see
  http://cxc.harvard.edu/atomdb/issues\_caveats.html.} We then used these emission measures with
line emissivity data from APEC to calculate foreground \OVII\ and \OVIII\ intensities, again
assuming $T=10^{6.08}$~\K. The foreground \OVII\ intensities calculated in this way are plotted in
Figure~\ref{fig:ForegroundOxygen}. The foreground intensity tends to increase with Galactic latitude.
The foreground \OVIII\ intensities follow the same trend, but are $\sim$100 times smaller.

Using this model to calculate the foreground oxygen intensities, \Ifg, we can calculate the deabsorbed
halo intensities, \Ihalo, from the observed intensities, \Iobs:
\begin{equation}
  \Ihalo = (\Iobs - \Ifg) \e^{\sigma \NH},
  \label{eq:Ihalo}
\end{equation}
where $\sigma$ is the photoelectric absorption cross-section and \NH\ is the hydrogen column
density. For \OVII\ we used $\sigma = 6.965 \times 10^{-22}~\cmsq$, calculated for $\Eovii =
0.57~\kev$, and for \OVIII\ we used $\sigma = 4.740 \times 10^{-22}~\cmsq$, calculated for $\Eoviii =
0.654~\kev$. These cross-sections were calculated using data from \citet{balucinska92}, with a
revised He cross-section from \citet{yan98}, and \citet{wilms00} interstellar abundances.
For \NH\ we used the \HI\ column density from the LAB survey \citep{kalberla05}.

The region with $120\degr \le l \le 240\degr$ includes two X-ray-bright regions: the Monogem Ring (a
supernova remnant; \citealp{plucinsky96}) and the Eridanus Enhancement (a superbubble;
\citealp{burrows93,snowden95b}). These features produce emission that is neither from the foreground
(LB and/or SWCX) nor from the halo. As a result, the above-described procedure will not yield
accurate halo intensities for observations within these features. The Monogem Ring lies within the
excluded $|\beta| \le 20\degr$ region, and so is not a problem. Three of the 39 observations shown
in Figure~\ref{fig:ObservationLocation}, however, are toward the Eridanus Enhancement. We therefore
removed these three observations from our subsequent analysis.

\subsubsection{A Plane-Parallel Model for the Halo Emission}
\label{subsubsec:PlaneParallel}

Here, we examine a simple plane-parallel model for the Galactic halo. In such a model, the intrinsic
emissivity, $\varepsilon$, of the halo gas is assumed to depend only on the height above the disk,
$z$. Models in which the density and temperature vary exponentially with height or are constant
within a given height range are subsets of the plane-parallel model. For such a model, the intrinsic
halo intensity for a given direction depends only on $\cosec |b|$; i.e.,
\begin{equation}
  \Ihalo(b) = I_{90} \cosec |b|,
  \label{eq:PlaneParallelModel}
\end{equation}
where $I_{90} = (1/4\pi) \int^\infty_0 \varepsilon(z) dz$ is the intrinsic halo intensity at $|b| = 90\degr$.

We fitted the above model to the deabsorbed halo intensities that we derived from our observations
using \eqref{eq:Ihalo}. We fitted the model to the northern and southern Galactic hemispheres
independently, using weighted least squares.  Although most of our measurements have asymmetrical
error bars, to simplify the fitting we assumed symmetrical errors, with the error on a given
intensity being equal to the larger of the positive and negative errors.

\begin{figure}
\centering
\plotone{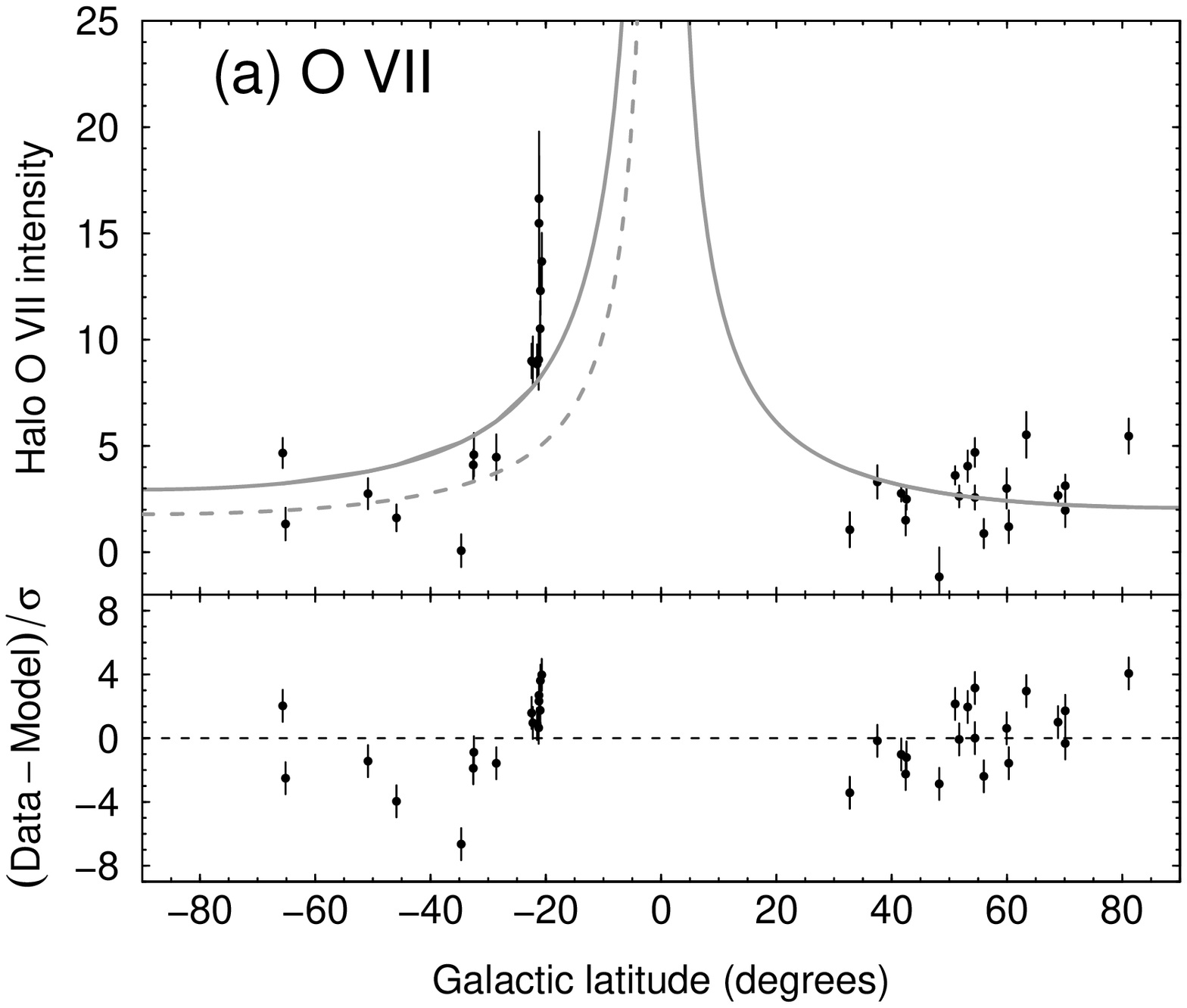}
\plotone{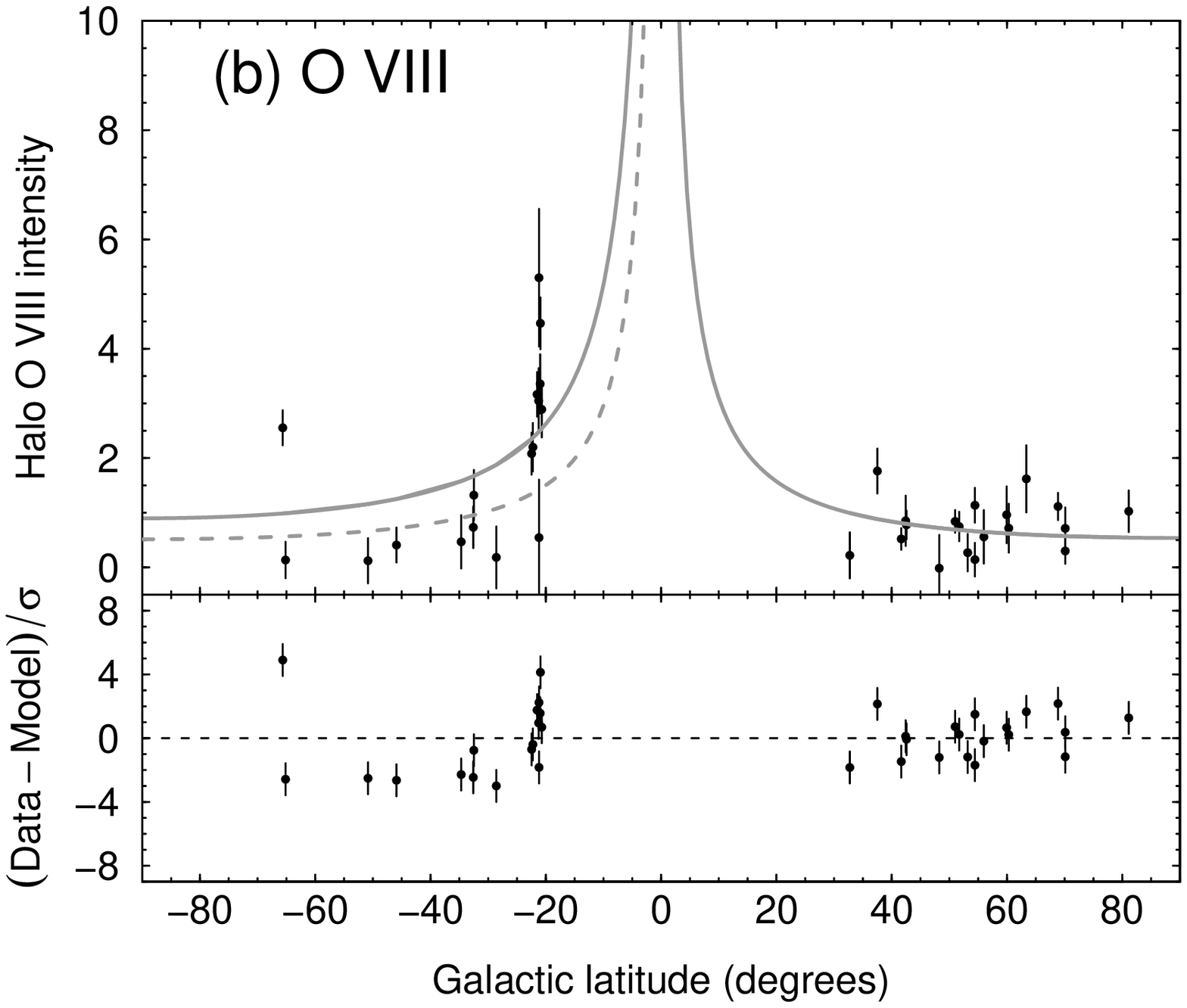}
\caption{Deabsorbed halo (a) \OVII\ and (b) \OVIII\ intensity against Galactic latitude, compared
  with a plane parallel model for the Galactic halo (\eqref{eq:PlaneParallelModel}). The solid gray
  lines show the best-fitting models obtained by fitting to each hemisphere independently. The
  residuals are for these models. The dashed gray lines show the best-fitting models obtained after
  the cluster of datapoints near $b = -20\degr$ was removed.
  \label{fig:PlaneParallelModel}}
\end{figure}

The deabsorbed halo intensities are shown in Figure~\ref{fig:PlaneParallelModel}, along with the
best-fitting plane-parallel halo models (shown with solid gray lines). For both lines, $I_{90}$ is
slightly larger in the southern hemisphere.  For \OVII, $I_{90} = 2.9 \pm 0.3~\LU$ in the south,
against $2.1 \pm 0.2~\LU$ in the north.  The corresponding values for \OVIII\ are $0.90 \pm
0.13~\LU$ in the south and $0.54 \pm 0.07~\LU$ in the north. The larger values of $I_{90}$ in the
southern hemisphere may be due to the cluster of datapoints near $b = -20\degr$.  These observations
are all near M31, and so may be contaminated by emission from M31's own halo. If we remove this
cluster of datapoints, we obtain the models shown by the dashed gray lines in
Figure~\ref{fig:PlaneParallelModel}. In this case, there is no significant difference between
the two hemispheres: the new values of $I_{90}$ in the south are $1.8 \pm 0.4~\LU$ for \OVII\
and $0.5 \pm 0.2~\LU$ for \OVIII.

There is some scatter in the halo intensity about the plane-parallel model. This suggests a
patchiness to the halo emission, as previously noted by \citet{yoshino09}. In addition, the
\OVII\ residuals are significantly correlated (at the 5\%\ level) with Galactic latitude in the
northern hemisphere, suggesting that the plane-parallel model may not be a good description of the
general distribution of emitting material in the halo. However, this correlation is dominated by the
two outermost datapoints, with $b = 32.7\degr$ and $81.1\degr$ -- if these two points are removed,
the correlation is no longer significant. We therefore cannot currently rule out the plane-parallel
halo model.  However, our completed survey, spanning the whole range of $l$, and ideally combined
with a SWCX model that would allow us to use a larger fraction of our observations, should allow us
to distinguish between different halo models (say, a plane-parallel model versus a Galactocentric
model).

\subsubsection{The Halo \OVII/\OVIII\ Ratio}
\label{subsubsec:OxygenRatio}

\begin{figure}
\centering
\plotone{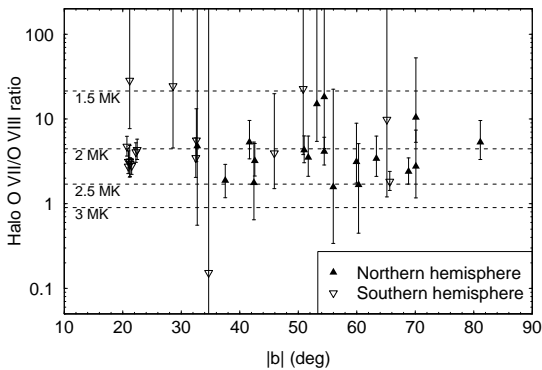}
\caption{Deabsorbed halo \OVII/\OVIII\ intensity ratio against Galactic latitude.  The dashed lines
  show the ratios expected for a plasma in collisional ionization equilibrium with $T = 1.5$, 2.0, 2.5, and $3.0 \times
  10^6$~\K\ (from top to bottom).
  \label{fig:O7O8ratio}}
\end{figure}

Figure~\ref{fig:O7O8ratio} shows the halo \OVII/\OVIII\ intensity ratio plotted against Galactic
latitude.  These ratios were calculated from the deabsorbed oxygen intensities, derived from the
observations using \eqref{eq:Ihalo}. Also shown in Figure~\ref{fig:O7O8ratio} are the
expected ratios for thermal plasmas in equilibrium at various temperatures, calculated using line
emissivity data from APEC. The expected \OVII/\OVIII\ intensity ratio decreases with
increasing temperature as the plasma becomes more highly ionized.

The \OVII/\OVIII\ ratios in Figure~\ref{fig:O7O8ratio} typically imply a halo temperature of
$\sim$2--$2.5 \times 10^6~\K$. Although there is some scatter in the datapoints in
Figure~\ref{fig:O7O8ratio}, the large error bars mean that we cannot tell whether or not there is
real variation in the temperature of the Galactic halo.

The halo temperature inferred from the \OVII/\OVIII\ ratios is in good agreement with the results of
studies of the Galactic halo with \xmm\ or \suzaku, which have obtained temperatures of $\sim$2--$3
\times 10^6~\K$, assuming (as we have implicitly done so) a single temperature for the halo
\citep{smith07a,galeazzi07,yoshino09,lei09,gupta09b}.  While some studies find that an isothermal
halo model is unable to explain all the available ultraviolet and X-ray data for the halo
\citep{yao07a,shelton07,lei09}, such a model is useful for characterizing the X-ray emission.
\citet{kuntz00} used a two-temperature model of the halo in their \rosat\ All-Sky Survey
analysis. The halo temperature inferred from our line intensity ratios lies between the
temperatures of their two components ($(1.1^{+0.6}_{-0.4}) \times 10^6$ and $(2.9^{+0.9}_{-0.5}) \times 10^6~\K$),
and is in reasonable agreement with the temperature of their hotter component.

As noted above, the \OVII/\OVIII\ intensity ratios, and hence the inferred halo temperatures, have
large error bars. Tighter constraints on the halo temperature can be obtained by fitting thermal
plasma models to the spectra, as this technique uses more of the information contained in the
spectra. In a forthcoming paper we will present such an analysis of our spectra, and also describe
the implications of the results for models of the hot halo (D.~B. Henley et al., in preparation).

\section{SUMMARY}
\label{sec:Summary}

We have presented measurements of the SXRB \OVII\ and \OVIII\ intensity between $l=120\degr$ and
$l=240\degr$, extracted from archival \xmm\ observations.  We have not restricted ourselves to
blank-sky observations -- if an observation target is not too bright or too extended, we excluded a
region surrounding the target, and extracted a SXRB spectrum from the remainder of the field of
view.

In an attempt to reduce SWCX contamination, we removed times of high solar wind proton flux from the data.
We measured oxygen intensities both with and without this proton flux filtering. Without the filtering, we
obtained measurements from 586 \xmm\ observations, and with the filtering from 303 observations. Four observations
appear in the latter set but not in the former (see Section~\ref{subsec:OxygenIntensitiesResults}), so
we have obtained measurements from a total of 590 \xmm\ observations.

We have found a very large range of oxygen intensities: 0.5 to 31.3~\LU\ for \OVII\ and 0.0 to
11.3~\LU\ for \OVIII.  For a total of 69 directions we have multiple observations, whose variation
in the oxygen line intensities can be used to constrain models of SWCX emission. Some observations
exhibit extremely bright SWCX emission, the brightest being an enhancement in the \OVII\ intensity
of $\sim$25~\LU\ over two other observations of the same direction. However, most SWCX enhancements
are $\la$4~\LU\ for \OVII\ and $\la$2~\LU\ for \OVIII.

For He-like \Kalpha\ emission due to SWCX, the forbidden line is expected to be the brightest
component, whereas for a hot collisionally excited plasma the resonance line is expected to be
brightest \citep{wargelin08}. However, for observations exhibiting enhanced emission due to SWCX, we
do not see a clear tendency for the \OVII\ centroid energy to shift toward that of the \OVII\ forbidden
line, apparently because the uncertainties in the measured \OVII\ centroids are too large.
We also find that enhanced SWCX emission is not univerally associated with increased solar wind flux; nor is
increased SWCX emission clearly associated with the sightline passing close to the sub-solar region
of the magnetosheath.

We have used our measurements to look at the oxygen emission from the Galactic halo. To this end, we
applied various filters to our results in an attempt to reduce SWCX contamination. As well as the
above-mentioned proton flux filtering, we removed observations from low ecliptic latitude and
observations that were taken around solar maximum. We also used \rosat\ shadowing data
\citep{snowden00} to model the remaining foreground emission. The deabsorbed halo intensities from
our culled dataset show some scatter about the predictions of a simple plane-parallel model,
indicating that the halo emission is patchy (as also noted by \citealp{yoshino09}). The halo
\OVII/\OVIII\ ratios for this filtered set of observations imply a temperature of $\sim$2.0--$2.5
\times 10^6~\K$, in good agreement with previous studies of the halo.

In a forthcoming paper we will present a more detailed analysis of the spectra that survive our
SWCX-reducing filtering, and will compare the results to the predictions of various physical models
of the hot halo (D.~B. Henley et al., in preparation). A future paper will expand our survey to
cover the whole sky, and incorporate EPIC-pn data to supplement the EPIC-MOS data that we have used
here. Subsequent papers will discuss the implications of our measurements for various topics of
interest, such as the structure and origin of the Galactic halo, the north-south asymmetry of the
halo seen with \rosat\ \citep{snowden98}, and the source of the diffuse X-ray emission in the Galactic
plane.

\acknowledgements
We would like to thank the anonymous referee, whose comments have helped improve this paper.
This research is based on observations obtained with \xmm, an ESA science mission with
instruments and contributions directly funded by ESA Member States and NASA. This research
was funded by NASA grant NNX08AJ47G, awarded through the Astrophysics Data Program.

\bibliography{references}

\clearpage
\LongTables
\begin{landscape}
\tabletypesize{\scriptsize}
\begin{deluxetable}{llrrccccrr@{,}lcrr@{,}lccc}
\tablewidth{0pt}
\tablecaption{\OVII\ and \OVIII\ Line Intensities (Without Proton Flux Filtering)\label{tab:OxygenIntensities1}}
\tablehead{
                                     &                                      &                                      &                                      & \multicolumn{2}{c}{MOS1}                                                    & \multicolumn{2}{c}{MOS2}                                                    \\
\colhead{Obs.~ID}                    & \colhead{Start}                      & \colhead{$l$}                        & \colhead{$b$}                        & \colhead{Exp.}                       & \colhead{$\Omega$}                   & \colhead{Exp.}                       & \colhead{$\Omega$}                   & \colhead{\Iovii}                     & \multicolumn{2}{c}{$\Delta\Iovii$}   & \colhead{\Eovii}                     & \colhead{\Ioviii}                    & \multicolumn{2}{c}{$\Delta\Ioviii$}  & \colhead{$\langle\fsw\rangle$}       & \colhead{$\frac{\Ftotal}{\Fexgal}$}  & \colhead{Data in}                    \\
                                     & \colhead{Date}                       & \colhead{(deg)}                      & \colhead{(deg)}                      & \colhead{(ks)}                       & \colhead{(arcmin$^2$)}               & \colhead{(ks)}                       & \colhead{(arcmin$^2$)}               & \colhead{(\LU)}                      & \multicolumn{2}{c}{(\LU)}            & \colhead{(\kev)}                     & \colhead{(\LU)}                      & \multicolumn{2}{c}{(\LU)}            & \colhead{($10^8~\pcmsq~\ps$)}        &                                      & \colhead{Table 2?}                   \\
\colhead{(1)}                        & \colhead{(2)}                        & \colhead{(3)}                        & \colhead{(4)}                        & \colhead{(5)}                        & \colhead{(6)}                        & \colhead{(7)}                        & \colhead{(8)}                        & \colhead{(9)}                        & \multicolumn{2}{c}{(10)}             & \colhead{(11)}                       & \colhead{(12)}                       & \multicolumn{2}{c}{(13)}             & \colhead{(14)}                       & \colhead{(15)}                       & \colhead{(16)}                       
}
\startdata
0153030101 & 2003-07-04 & 120.057 & $-37.320$ &             6.0 &             473 &             6.4 &             478 &  6.44 &  (5.30 &  7.66) & 0.5625 &  1.90 &  (1.32 &  2.45) &    1.40 &  1.86 &       Y \\
0300890101 & 2005-06-15 & 120.246 & $-58.654$ &            28.1 &             487 &            30.4 &             582 &  7.15 &  (6.49 &  7.48) & 0.5675 &  1.67 &  (1.43 &  1.95) &    1.71 &  1.70 &       Y \\
0402560701 & 2006-07-23 & 120.404 & $-22.166$ &            26.2 &             498 &            30.3 &             584 &  5.31 &  (4.58 &  5.55) & 0.5725 &  1.63 &  (1.29 &  1.96) &    3.32 &  1.63 &       N \\
0402560501 & 2006-07-20 & 120.554 & $-21.822$ &            53.9 &             378 &            55.0 &             498 &  5.90 &  (5.52 &  6.31) & 0.5675 &  1.28 &  (1.10 &  1.56) &    3.13 &  1.89 &       N \\
0304570101 & 2006-04-20 & 120.580 & $ 58.034$ &            10.6 &             469 &            10.5 &             484 &  6.91 &  (6.26 &  7.73) & 0.5575 &  0.67 &  (0.37 &  1.07) &    2.94 &  1.07 &       N \\
0402560801 & 2006-12-25 & 120.591 & $-22.242$ &            48.2 &             404 &            47.9 &             577 &  5.58 &  (5.14 &  6.23) & 0.5625 &  1.45 &  (1.28 &  1.74) &    0.95 &  1.50 &       Y \\
0112570301 & 2002-01-24 & 120.595 & $-22.245$ &            25.9 &             564 &            26.6 &             575 &  8.96 &  (8.27 &  9.22) & 0.5725 &  2.39 &  (2.20 &  2.73) &    2.62 &  1.99 &       N \\
0402560601 & 2006-07-28 & 120.742 & $-22.461$ &            31.2 &             474 &            32.0 &             505 &  5.86 &  (5.40 &  6.32) & 0.5675 &  1.42 &  (1.16 &  1.68) &    0.84 &  1.18 &       Y \\
\enddata
\end{deluxetable}

\clearpage
\end{landscape}

\clearpage
\begin{landscape}
\tabletypesize{\scriptsize}
\begin{deluxetable}{llrrccccrr@{,}lcrr@{,}lccc}
\tablewidth{0pt}
\tablecaption{\OVII\ and \OVIII\ Line Intensities (With Proton Flux Filtering)\label{tab:OxygenIntensities2}}
\tablehead{
                                     &                                      &                                      &                                      & \multicolumn{2}{c}{MOS1}                                                    & \multicolumn{2}{c}{MOS2}                                                    \\
\colhead{Obs.~ID}                    & \colhead{Start}                      & \colhead{$l$}                        & \colhead{$b$}                        & \colhead{Exp.}                       & \colhead{$\Omega$}                   & \colhead{Exp.}                       & \colhead{$\Omega$}                   & \colhead{\Iovii}                     & \multicolumn{2}{c}{$\Delta\Iovii$}   & \colhead{\Eovii}                     & \colhead{\Ioviii}                    & \multicolumn{2}{c}{$\Delta\Ioviii$}  & \colhead{$\langle\fsw\rangle$}       & \colhead{$\frac{\Ftotal}{\Fexgal}$}  & \colhead{Affected}                   \\
                                     & \colhead{Date}                       & \colhead{(deg)}                      & \colhead{(deg)}                      & \colhead{(ks)}                       & \colhead{(arcmin$^2$)}               & \colhead{(ks)}                       & \colhead{(arcmin$^2$)}               & \colhead{(\LU)}                      & \multicolumn{2}{c}{(\LU)}            & \colhead{(\kev)}                     & \colhead{(\LU)}                      & \multicolumn{2}{c}{(\LU)}            & \colhead{($10^8~\pcmsq~\ps$)}        &                                      & \colhead{by filtering?}              \\
\colhead{(1)}                        & \colhead{(2)}                        & \colhead{(3)}                        & \colhead{(4)}                        & \colhead{(5)}                        & \colhead{(6)}                        & \colhead{(7)}                        & \colhead{(8)}                        & \colhead{(9)}                        & \multicolumn{2}{c}{(10)}             & \colhead{(11)}                       & \colhead{(12)}                       & \multicolumn{2}{c}{(13)}             & \colhead{(14)}                       & \colhead{(15)}                       & \colhead{(16)}                       
}
\startdata
0153030101 & 2003-07-04 & 120.057 & $-37.320$ &             6.0 &             473 &             6.4 &             478 &  6.44 &  (5.30 &  7.66) & 0.5625 &  1.90 &  (1.31 &  2.45) &    1.40 &  1.86 &       N \\
0300890101 & 2005-06-15 & 120.246 & $-58.654$ &            26.2 &             487 &            28.5 &             582 &  7.09 &  (6.31 &  7.36) & 0.5725 &  1.60 &  (1.25 &  1.80) &    1.60 &  1.64 &       Y \\
0402560801 & 2006-12-25 & 120.591 & $-22.242$ &            48.2 &             404 &            47.9 &             577 &  5.58 &  (5.14 &  6.23) & 0.5625 &  1.45 &  (1.28 &  1.74) &    0.95 &  1.50 &       N \\
0402560601 & 2006-07-28 & 120.742 & $-22.461$ &            31.2 &             474 &            32.0 &             505 &  5.86 &  (5.40 &  6.32) & 0.5675 &  1.42 &  (1.16 &  1.68) &    0.84 &  1.18 &       N \\
0402560301 & 2006-07-01 & 120.784 & $-21.514$ &            46.4 &             410 &            47.6 &             512 &  5.09 &  (4.64 &  5.46) & 0.5675 &  1.95 &  (1.73 &  2.20) &    1.52 &  1.75 &       N \\
0402560901 & 2006-12-26 & 121.001 & $-21.239$ &            44.5 &             348 &            44.8 &             515 &  4.75 &  (4.13 &  5.02) & 0.5675 &  1.75 &  (1.59 &  2.09) &    1.03 &  1.70 &       N \\
0109270701 & 2002-01-05 & 121.429 & $-21.264$ &            55.3 &             573 &            55.0 &             570 &  8.44 &  (8.03 &  8.72) & 0.5675 &  3.56 &  (3.33 &  3.74) &    1.20 &  1.93 &       N \\
0402561301 & 2007-01-03 & 121.597 & $-20.708$ &            27.4 &             486 &            26.8 &             582 &  5.55 &  (5.08 &  5.96) & 0.5675 &  1.42 &  (1.17 &  1.66) &    1.83 &  1.20 &       Y \\
\enddata
\end{deluxetable}

\clearpage
\end{landscape}

\clearpage
\tabletypesize{\small}
\begin{deluxetable}{lccccc}
\tablewidth{0pt}
\tablecaption{Ranges and Quartiles of the Oxygen Intensities\label{tab:RangeMedianMode}}
\tablehead{
\colhead{Line}	& \colhead{Proton flux}	& \colhead{Range}	& \colhead{Lower quartile}	& \colhead{Median}	& \colhead{Upper quartile}	\\
		& \colhead{filtering?}	& \colhead{(\LU)}	& \colhead{(\LU)}		& \colhead{(\LU)}	& \colhead{(\LU)}
}
\startdata
\OVII		& N			& 0.5--31.2		& 3.79 (3.63,3.93)		& 5.22 (4.99,5.39)	& 6.72 (6.51,7.02)		\\
\OVII		& Y			& 0.5--17.6		& 3.62 (3.46,3.77)		& 4.89 (4.62,5.14)	& 6.24 (5.96,6.50)		\\
\OVIII		& N			& 0.0--11.3		& 0.64 (0.58,0.71)		& 1.06 (1.01,1.16)	& 1.75 (1.68,1.90)		\\
\OVIII		& Y			& 0.0--6.0		& 0.60 (0.49,0.69)		& 1.01 (0.94,1.06)	& 1.62 (1.49,1.73)
\enddata
\tablecomments{The numbers in parentheses are the 90\%\ confidence intervals, calculated by bootstrapping.}
\end{deluxetable}

\begin{deluxetable}{lcccccc}
\tablewidth{0pt}
\tablecaption{Correlation Coefficients for Intensity against Latitude\label{tab:Correlation}}
\tablehead{
\colhead{Line}	& \colhead{Proton flux}	& \multicolumn{2}{c}{North}                                           && \multicolumn{2}{c}{South}                      \\
\cline{3-4} \cline{6-7}
                & \colhead{filtering?}	& \colhead{$b>0\degr$}             & \colhead{$b>20\degr$}            && \colhead{$b<0\degr$}  & \colhead{$b<-20\degr$}
}
\startdata
\OVII		& N			& 0.21 ($2.9 \times 10^{-7}$)       & 0.14 ($7.4 \times 10^{-4}$)       && \nodata               & 0.18 ($3.8 \times 10^{-5}$)  \\
\OVII		& Y			& 0.23 ($3.5 \times 10^{-5}$)       & 0.20 ($6.7 \times 10^{-4}$)       && \nodata               & 0.19 ($1.1 \times 10^{-3}$)  \\
\OVIII		& N			& \nodata                          & \nodata                          && \nodata               & 0.17 ($9.0 \times 10^{-5}$)  \\
\OVIII		& Y			& \nodata                          & \nodata                          && \nodata               & 0.17 ($4.5 \times 10^{-3}$)  \\
\enddata
\tablecomments{Correlation coefficients are Kendall's $\tau$ for the relevant intensity against $b$.
Values in parentheses are $p$-values (i.e., the probabilities of obtaining correlation coefficients
at least as large as the observed values, assuming that the null hypothesis is true).
Only correlations that are significant at the 5\%\ level are included.}
\end{deluxetable}

\tabletypesize{\scriptsize}
\begin{deluxetable}{cclrrrr@{,}lrr@{,}lcrr@{,}lrr@{,}l}
\tablewidth{0pt}
\tablecaption{Oxygen Line Intensities from Directions with Multiple Observations\label{tab:MultipleObs}}
\tablehead{
&&&&& \multicolumn{6}{c}{Without proton flux filtering}&& \multicolumn{6}{c}{With proton flux filtering} \\
\cline{6-11} \cline{13-18}
\colhead{Dataset}                    & \colhead{$N_\mathrm{obs}$}           & \colhead{Obs.~ID}                    & \colhead{$l$}                        & \colhead{$b$}                        & \colhead{\Iovii}                     & \multicolumn{2}{c}{$\Delta\Iovii$}   & \colhead{\Ioviii}                    & \multicolumn{2}{c}{$\Delta\Ioviii$}  && \colhead{\Iovii}                     & \multicolumn{2}{c}{$\Delta\Iovii$}   & \colhead{\Ioviii}                    & \multicolumn{2}{c}{$\Delta\Ioviii$}  \\
                                     &                                      &                                      & \colhead{(deg)}                      & \colhead{(deg)}                      & \colhead{(\LU)}                      & \multicolumn{2}{c}{(\LU)}            & \colhead{(\LU)}                      & \multicolumn{2}{c}{(\LU)}            && \colhead{(\LU)}                      & \multicolumn{2}{c}{(\LU)}            & \colhead{(\LU)}                      & \multicolumn{2}{c}{(\LU)}            \\
\colhead{(1)}                        & \colhead{(2)}                        & \colhead{(3)}                        & \colhead{(4)}                        & \colhead{(5)}                        & \colhead{(6)}                        & \multicolumn{2}{c}{(7)}              & \colhead{(8)}                        & \multicolumn{2}{c}{(9)}              && \colhead{(10)}                       & \multicolumn{2}{c}{(11)}             & \colhead{(12)}                       & \multicolumn{2}{c}{(13)}             \\
}
\startdata
 1 &  2 & 0112570301 & 120.595 & $-22.245$ &     8.96 &        (8.27 &        9.22) &     2.39 &        (2.20 &        2.73) & &  \nodata & \multicolumn{2}{c}{\nodata} &  \nodata & \multicolumn{2}{c}{\nodata}\\
   &    & 0402560801 & 120.591 & $-22.242$ &     5.58 &        (5.14 &        6.23) &     1.45 &        (1.28 &        1.74) & &     5.58 &        (5.14 &        6.23) &     1.45 &        (1.28 &        1.74)\\
 2 &  2 & 0410582001 & 120.820 & $-21.565$ &     5.21 &        (4.28 &        5.68) &     1.52 &        (1.25 &        2.03) & &  \nodata & \multicolumn{2}{c}{\nodata} &  \nodata & \multicolumn{2}{c}{\nodata}\\
   &    & 0402560301 & 120.784 & $-21.514$ &     5.09 &        (4.64 &        5.46) &     1.95 &        (1.73 &        2.20) & &     5.09 &        (4.64 &        5.46) &     1.95 &        (1.73 &        2.20)\\
 3 &  2 & 0109270301 & 121.707 & $-20.938$ &     9.04 &        (8.56 &        9.50) &     2.73 &        (2.48 &        2.97) & &     9.04 &        (8.56 &        9.50) &     2.73 &        (2.48 &        2.97)\\
   &    & 0402561501 & 121.703 & $-20.934$ &     5.04 &        (4.65 &        5.43) &     2.18 &        (1.95 &        2.41) & &     5.04 &        (4.65 &        5.43) &     2.18 &        (1.95 &        2.41)\\
 4 &  2 & 0403530501 & 121.807 & $-21.183$ &     6.37 &        (5.23 &        7.24) &     0.28 &        (0.00 &        0.81) & &     6.37 &        (5.23 &        7.24) &     0.28 &        (0.00 &        0.81)\\
   &    & 0403530401 & 121.808 & $-21.183$ &     6.79 &        (5.65 &        7.93) &     2.66 &        (2.04 &        3.29) & &     6.79 &        (5.65 &        7.93) &     2.66 &        (2.04 &        3.29)\\
\enddata
\end{deluxetable}

\clearpage
\tabletypesize{\small}
\begin{deluxetable}{clcccccc}
\tablewidth{0pt}
\tablecaption{Directions with the Brightest SWCX Emission\label{tab:BrightSWCX}}
\tablehead{
					&			& \multicolumn{2}{c}{Faintest}				&& \multicolumn{2}{c}{Brightest}			&				\\
\cline{3-4} \cline{6-7}
\colhead{Dataset\tablenotemark{a}}	& \colhead{Line}	& \colhead{Obs. ID}	& \colhead{$I$}			&& \colhead{Obs. Id}	& \colhead{$I$}			& \colhead{Difference}		\\
					&			&			& \colhead{(\LU)}		&&			& \colhead{(\LU)}		& \colhead{(\LU)}
}
\startdata
12					& \OVII\		& 0400560301		& $ 4.11^{+0.30}_{-0.38}$	&& 0059140901		& $14.83^{+0.91}_{-0.91}$	& $10.72^{+0.99}_{-0.96}$	\\
16					& \OVII\		& 0112520901		& $ 5.14^{+0.86}_{-0.77}$	&& 0112520601		& $31.29^{+1.31}_{-1.12}$	& $26.15^{+1.52}_{-1.41}$	\\
26					& \OVII\		& 0143150301		& $ 3.26^{+0.74}_{-0.76}$	&& 0143150601		& $21.47^{+0.63}_{-0.85}$	& $18.21^{+0.99}_{-1.13}$	\\
49					& \OVIII\		& 0302352201		& $ 0.58^{+0.40}_{-0.45}$	&& 0203362201		& $ 8.78^{+0.34}_{-0.31}$	& $ 8.20^{+0.56}_{-0.51}$	\\

\enddata
\tablenotetext{a}{The dataset number from Table~\ref{tab:MultipleObs}.}
\end{deluxetable}

\end{document}